\RequirePackage{ifpdf}
\documentclass{JHEP3}
\pdfoutput=1
\usepackage{graphics}
\usepackage{graphicx}
\usepackage{amsmath}
\usepackage[normalem]{ulem}
\usepackage{cite}
\usepackage{dsfont}
\newcommand{\loopint}[1]{\int \!\!\! \frac{d^D #1}{\left(2\pi\right)^D}\!}

\newcommand{\pFq}[5]{\, \! _{#1} F_{#2}\left( #3 \, ; \, #4 \, ; \, #5 \right)}

\def\be{\begin{equation}}
\def\ee{\end{equation}}
\def\bea{\begin{eqnarray}}
\def\eea{\end{eqnarray}}
\def\eps{\epsilon}
\def\nnb{\nonumber}
\def\ep{\epsilon}
\def\eps{\epsilon}

\def\zc{z_f}
\def\ub{\bar u}

\preprint{OUTP-14-15P\\
SI-HEP-2014-24\\
QFET-2014-18}
\title{Master integrals for the two-loop penguin contribution in non-leptonic $B$-decays}

\author{Guido Bell$^{a}$ and Tobias Huber$^{b}$\\
$^a$ Rudolf Peierls Centre for Theoretical Physics, University of Oxford,\\
1 Keble Road, Oxford OX1 3NP, United Kingdom\\
$^b$ Theoretische Physik 1, Naturwissenschaftlich-Technische Fakult\"at,\\
Universit\"at Siegen, Walter-Flex-Stra{\ss}e 3, D-57068 Siegen, Germany\\

\email{guido.bell@physics.ox.ac.uk \\ huber@tp1.physik.uni-siegen.de}}

\abstract{
We compute the master integrals that arise in the calculation of the leading penguin amplitudes in non-leptonic $B$-decays at two-loop order. The application of differential equations in a canonical basis enables us to give analytic results for all master integrals in terms of iterated integrals with rational weight functions. It is the first application of this method to the case of two different internal masses.
}


\begin{document}


\newpage
\section{Introduction}
\label{sec-intro}

The study of flavour-changing quark transitions provides an important indirect probe to search for new heavy particles as well as to test the CKM mechanism of flavour mixing and CP violation. One prominent class of such transitions are non-leptonic $B$-meson decays, which offer a rich and interesting phenomenology including many CP-violating asymmetries. Non-leptonic two-body decays therefore play a central role at current and future $B$-physics experiments. The extraction of the underlying decay amplitudes is, however, complicated by the strong-interaction dynamics of the purely hadronic environment. A systematic formalism to compute the hadronic matrix elements arises in the heavy-quark limit~\cite{Beneke:1999br,Beneke:2000ry,Beneke:2001ev}. Schematically, 
\begin{align}
\langle M_1 M_2 | Q_i | \bar{B} \rangle
&\,\simeq\, 
F^{B M_1} \,
\int du \;T_{i}^I(u) \, \phi_{M_2}(u)
\nonumber \\
&\quad\:\,+\,  
\int d\omega \,dv \,du \; T_{i}^{II}(\omega,v,u)
\, \phi_B(\omega) \, \phi_{M_1}(v) \,  \phi_{M_2}(u)\,,
\label{factorisation}
\end{align}
where $M_{1,2}$ are light (charmless) pseudo-scalar or vector mesons and $Q_i$ is a generic operator of the effective weak Hamiltonian. The hadronic dynamics in the above factorisation formula is encoded in a form factor $F$ and in light-cone distribution amplitudes $\phi$. The hard-scattering kernels $T$, on the other hand, can be computed to all orders in perturbation theory in a partonic calculation. In the last few years, the perturbative corrections have been worked out to next-to-next-to-leading order (NNLO) accuracy. While the full set of ${\cal O}(\alpha_s^2)$ corrections to the spectator-scattering kernels $T_i^{II}$ is known~\cite{Beneke:2005vv,Kivel:2006xc,Beneke:2006mk,Jain:2007dy,Pilipp:2007mg}, NNLO corrections to the kernels $T^{I}_i$ have to date only been determined for the topological tree amplitudes~\cite{Bell:2007tv,Bell:2009nk,Beneke:2009ek}. 

The missing NNLO ingredient consists of a two-loop calculation of the hard-scattering kernels $T^{I}_i$ in the penguin sector. The calculation involves various types of operator insertions, for details we refer to a future publication~\cite{BBHL}. The one-loop contribution of the magnetic dipole operator has been computed in~\cite{Kim:2011jm}. The most difficult part of the calculation consists in the computation of massive two-loop penguin diagrams like the ones shown in Fig.~\ref{fig:diagrams}. Whereas the integrals that entered the two-loop tree calculation~\cite{Bell:2006tz,Huber:2009se} can be expressed in terms of Harmonic Polylogarithms (HPLs)~\cite{Remiddi:1999ew}, the massive propagator in the penguin loop introduces an additional scale and complicates the calculation. In the present paper we give analytic results for the master integrals that arise in this calculation.

A convenient technique for the calculation of multi-scale integrals is the method of differential equations~\cite{Kotikov:1990kg,Kotikov:1991pm,Remiddi:1997ny}. In combination with integration-by-parts identities~\cite{Tkachov:1981wb,Chetyrkin:1981qh} and Laporta's reduction algorithm~\cite{Laporta:2001dd}, the master integrals are computed by solving a set of differential equations where the derivatives are taken with respect to the external scales of the process. It has recently been pointed out that the solution simplifies considerably if the basis of master integrals is chosen appropriately~\cite{Henn:2013pwa}. We will discuss the properties of such a \emph{canonical basis} in detail below. The method has been successfully applied to compute various massless as well as massive two-loop and three-loop integrals~\cite{Henn:2013tua,Henn:2013woa,Henn:2013nsa,Argeri:2014qva,Henn:2014lfa,Gehrmann:2014bfa,Caola:2014lpa,Hoschele:2014qsa,DiVita:2014pza,vonManteuffel:2014mva}. The present calculation is the first application of the method in which the integrals have two different internal masses. 

Our paper is organised as follows. We first discuss the kinematics of the process and introduce a generalisation of the HPLs in Section~\ref{sec-def}. The canonical basis of master integrals is defined in Section~\ref{sec-basis}, and analytic results for all master integrals are given in Section~\ref{sec-results}. We comment on several cross-checks of our calculation in Section~\ref{sec-checks}, before we conclude in Section~\ref{sec-conclusion}. The paper is complemented by three appendices with various technical details, as well as an electronic file that contains the analytic results of all master integrals and is attached to the arXiv submission of the present work.

\FIGURE[t]{
\includegraphics[width=0.29\textwidth]{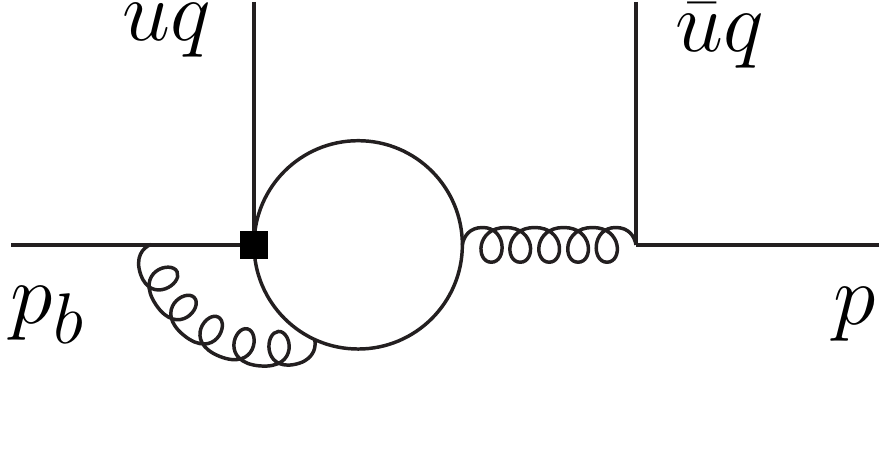}
\hspace{6mm}
\includegraphics[width=0.29\textwidth]{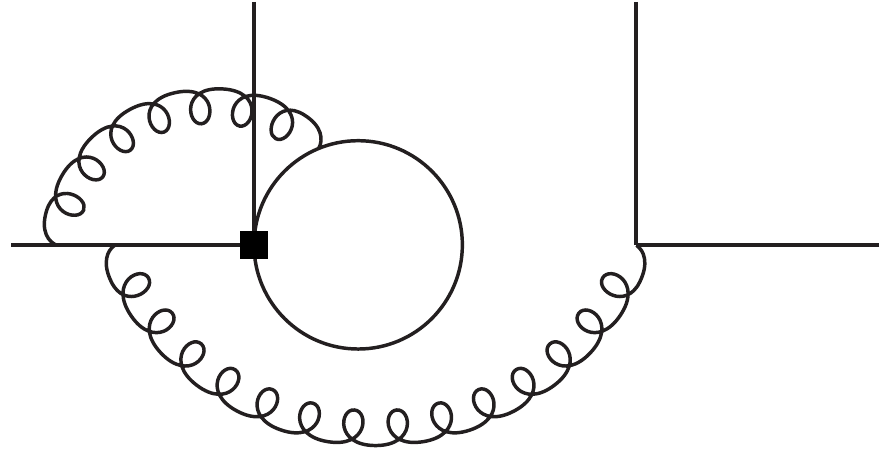}
\hspace{6mm}
\includegraphics[width=0.29\textwidth]{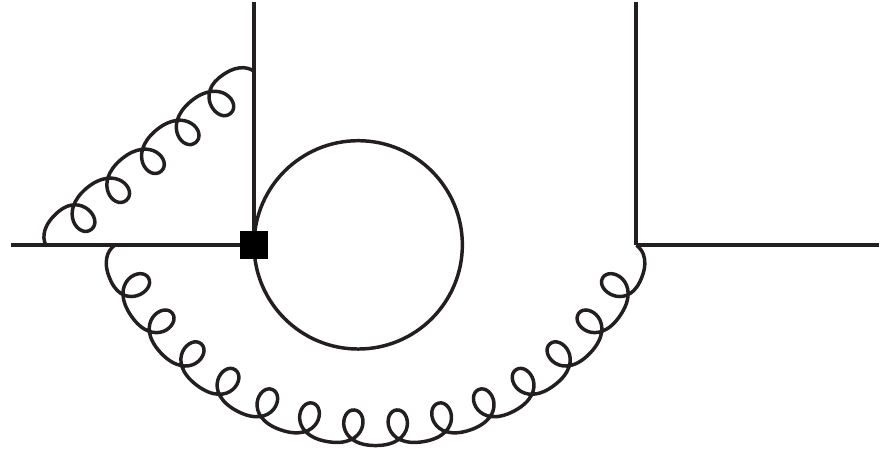}
\caption{Sample diagrams that arise in the two-loop calculation of the leading penguin amplitudes. The black square denotes an insertion of an operator from the effective weak Hamiltonian. The line to the left of the square is the incoming $b$-quark with momentum $p_b=q+p$. The quark in the penguin loop can either be massless (up, down, strange) or massive (charm, bottom). The momenta of the massless final state quarks are outgoing.}\label{fig:diagrams}}


\section{Definitions and Notation}
\label{sec-def}

\subsection{Kinematics}
\label{sec-kinematics}

The kinematics of the process is depicted in Fig.~\ref{fig:diagrams}. We write $p_b = q+p$ with $p_b^2=m_b^2$ and $p^2=q^2=0$. The momentum $q$ of the emitted final state meson is split up into two parallel momenta $q_1=uq$ and $q_2=(1-u)q \equiv \ub q$ of the quark and anti-quark, respectively, where $u\in[0,1]$ is the convolution variable that enters the first term of Eq.~(\ref{factorisation}). The quark in the penguin loop can either be massless in the case of up, down and strange quarks, or massive of mass $m_c$ or $m_b$ in the case of charm or bottom. For massless quarks, the master integrals are already known from the calculation of the two-loop tree amplitudes in~\cite{Bell:2006tz,Huber:2009se}. We therefore only consider the situation with a massive quark in the penguin loop in the following. The problem  then depends on two dimensionless variables, which we choose as the momentum fraction $\ub$ of the anti-quark and the mass ratio $\zc \equiv m_f^2/m_b^2$, with $f=c,b$. The analytic continuation is done via $\zc \to \zc-i\eta$, with infinitesimally small $\eta>0$.

In order to express the solution to the master integrals in terms of iterated integrals with rational weights, it will be convenient to trade the variables $\ub$ and $\zc$ for other sets of variables. Our default choice is the set $(r,s)$ with 
\begin{align} \label{eq:rs}
r \equiv & \sqrt{1-4\zc} \, ,\qquad\quad
s \equiv  \sqrt{1-\frac{4\zc}{\ub}}\,, 
\end{align}
which, when solved for the original variables, implies
\begin{align}
\ub = & \frac{1-r^2}{1-s^2} \, , \qquad\quad
\zc =  \frac{1-r^2}{4} \, .
\end{align}
Let us have a look at the possible values of $s$. When $\ub$ runs from $0 \to 1$, the variable $s$ for $4 \zc>1$ runs from $+i\infty \to r$ along the imaginary axis. For $4\zc<1$, $s$ runs from $+i\infty \to 0$ along the imaginary axis, followed by $0 \to r$ along the real axis. In this case the threshold at $\ub = 4 \zc$ is mapped onto $s=0$.

Another convenient choice of variables will be the set $(r,s_1)$, with
\begin{align} \label{eq:s1}
s_1 \equiv & \sqrt{1-\frac{4z_b}{\ub}}
\end{align}
and $z_b=1-i\eta$. The variable $s_1$ runs from $+i\infty \to +i\sqrt{3}$ along the imaginary axis once we let $\ub$ run from $0 \to 1$. 

A third choice of variables consists of the set $(r,p)$ with
\begin{align} \label{eq:p}
p \equiv & \frac{1-\sqrt{u^2+4\ub\zc}}{\ub} \, .
\end{align}
When solved for the original variable $\ub$ one obtains
\begin{align}
\ub =& \frac{r^2+1-2 p}{1-p^2}\; .
\end{align}
When $\ub$ runs from $0 \to 1$, the variable $p$ runs from $1-2\zc \to 1-2\sqrt{\zc}$.

\subsection{Iterated integrals}

One of the classical examples of iterated integrals are HPLs~\cite{Remiddi:1999ew}. They are generalisations of ordinary polylogarithms and appear in many calculations of higher-order corrections in perturbative Quantum Field Theory. The HPLs are defined by
\begin{eqnarray}\label{eq:defHPL}
H_{a_1 , a_2 , \ldots, a_n }(x) = \int_0^x dt \; f_{a_1}(t) \, H_{a_2, \ldots, a_n }(t)  \,, 
\end{eqnarray}
where the parameters $a_i$ can take the values $0$ or $\pm1$, and $n$ is called the \emph{weight} of the HPL. In the special case that all indices are zero, one defines $H_{\vec0_n}(x) = \frac{1}{n!} \ln^n (x)$. The weight functions $f_{a_i}(x)$ are given by
\begin{align}\label{eq:weights}
{f_{1}(x) = \frac{1}{1-x}}\,,\qquad  {f_{0}(x) = \frac{1}{x}}\,, \qquad {f_{-1}(x) = \frac{1}{1+x}}\,.
\end{align}
In addition one assigns the weight $k$ to numbers like $\pi^k$, $\ln^k(2)$ and the Riemann zeta function $\zeta_k$, and one uses that the product of two expressions of weights $k_1$ and $k_2$ has weight $k_1+k_2$.

These definitions were generalised in~\cite{Maitre:2007kp} by introducing linear combinations of $f_{1}(x)$ and $f_{-1}(x)$, the so-called ``$+$'' and ``$-$''-weights, according to
\begin{align}\label{eq:pm}
f_{+}(x) = & f_{1}(x) + f_{-1}(x) = \frac{2}{1-x^2}\,, \\
f_{-}(x) = & f_{1}(x) - f_{-1}(x) = \frac{2x}{1-x^2}\,.
\end{align}

In the present work we further generalise the weights by allowing more generic expressions to appear in the weight functions. For any expression $w \neq 0$ we define
\begin{align}\label{eq:genweights}
{f_{w}(x) = \frac{1}{w-x}}\,, \qquad {f_{-w}(x) = \frac{1}{w+x}}\, ,
\end{align}
and accordingly
\begin{align}
f_{w^+}(x) = & f_{w}(x) + f_{-w}(x) = \frac{2w}{w^2-x^2}\,, \\
f_{w^-}(x) = & f_{w}(x) - f_{-w}(x) = \frac{2x}{w^2-x^2}\,. \label{eq:genpm}
\end{align}
Also with these newly introduced weight functions we define a general HPL by means of Eq.~(\ref{eq:defHPL}), but we also allow the weights~(\ref{eq:genweights})~--~(\ref{eq:genpm}) to enter the integrand. In the current calculation, we encounter the following expressions for $w$,
\begin{align}
w_1 = & 1\, , \qquad\qquad\qquad w_4 = 
1 + \sqrt{1-r^2}\, , \nnb \\
w_2 = & r\, , \qquad\qquad\qquad w_5 = 
1 - \sqrt{1-r^2}\, . \nnb \\
w_3 = & \frac{r^2+1}{2} \, ,
\end{align}
We will refer to $w_1$~--~$w_5$ as {\emph{rational weights}}, since any of the $w_i$ is rational either in $r$ or $m_f$, given that $\sqrt{1-r^2} = 2 \sqrt{\zc} = 2 m_f/m_b$ is free of any square roots.

As a matter of fact, the generalised HPLs are closely related to Goncharov poly\-logarithms~\cite{Goncharov:1998kja}, which are defined by
\begin{align}
G_{a_1 , a_2 , \ldots, a_n }(x) = \int_0^x \frac{dt}{t-a_1} \, G_{a_2, \ldots, a_n }(t) \,  
\label{eq:defGonch}
\end{align}
and $G_{\vec0_n}(x) = H_{\vec0_n}(x)$. We can therefore always write a generalised HPL as a linear combination of Goncharov polylogarithms, for example
\begin{align}
H_{w_2^+}(x) = & G_{-r}(x) - G_{r}(x) \, ,
\end{align}
and similarly for higher weights.

The structure of the differential equations in the subsequent sections reveals that the results of the master integrals are most compactly written in terms of HPLs with generalised weights. For their numerical evaluation described in Section~\ref{sec-checks}, however, we prefer the notation in terms of Goncharov polylogarithms.


\section{Canonical Basis}
\label{sec-basis}

Within dimensional regularisation where space-time is analytically continued to $D=4-2\ep$ dimensions, integration-by-parts identities~\cite{Tkachov:1981wb,Chetyrkin:1981qh} provide non-trivial relations between different loop integrals. It has now become a standard tool to use automated reduction algorithms to express complicated multi-loop calculations in terms of a much smaller set of irreducible master integrals. The choice of the master integrals is, however, not unique. Henn recently conjectured that the set $\vec M$ of master integrals can always be chosen in a way such that the set of differential equations assumes the form~\cite{Henn:2013pwa}
\begin{align}
\frac{\partial}{\partial x_m} \, \vec M (\ep,x_n) = & \ep \, A_m(x_n) \,  \vec M (\ep,x_n) \, , 
\label{eq:Hennbasis}
\end{align}
where $x_n$ are dimensionless kinematic variables and $A_m(x_n)$ is a matrix which does not depend on $\ep$. In this form the system of differential equations decouples order-by-order in the $\ep$-expansion. The system~(\ref{eq:Hennbasis}) can be written as a total differential,
\begin{align}
d \, \vec M (\ep,x_n) = & \ep \, d \tilde A(x_n) \,  \vec M (\ep,x_n) \; . \label{eq:totaldifferential}
\end{align}
The matrix $\tilde A$ contains the relevant information about the structure of the occurring weight functions. Together with suitably chosen boundary conditions, this entirely fixes the solution. As an additional feature, the solutions to the master integrals contain functions that are of uniform weight at each order in $\ep$, and the weight increases by unit steps as one goes from one power to the next one in the $\ep$-expansion. As a consequence, by assigning the weight $-1$ to $\ep$ and multiplying the master integrals by an appropriate power of $\ep$, one can achieve that the total weight of each master integral is zero to all orders in $\ep$. Integrals with the latter property and a system of differential equations of the form~(\ref{eq:totaldifferential}) will be referred to as a \emph{canonical basis}.

\FIGURE[p!]{
\includegraphics[width=1\textwidth]{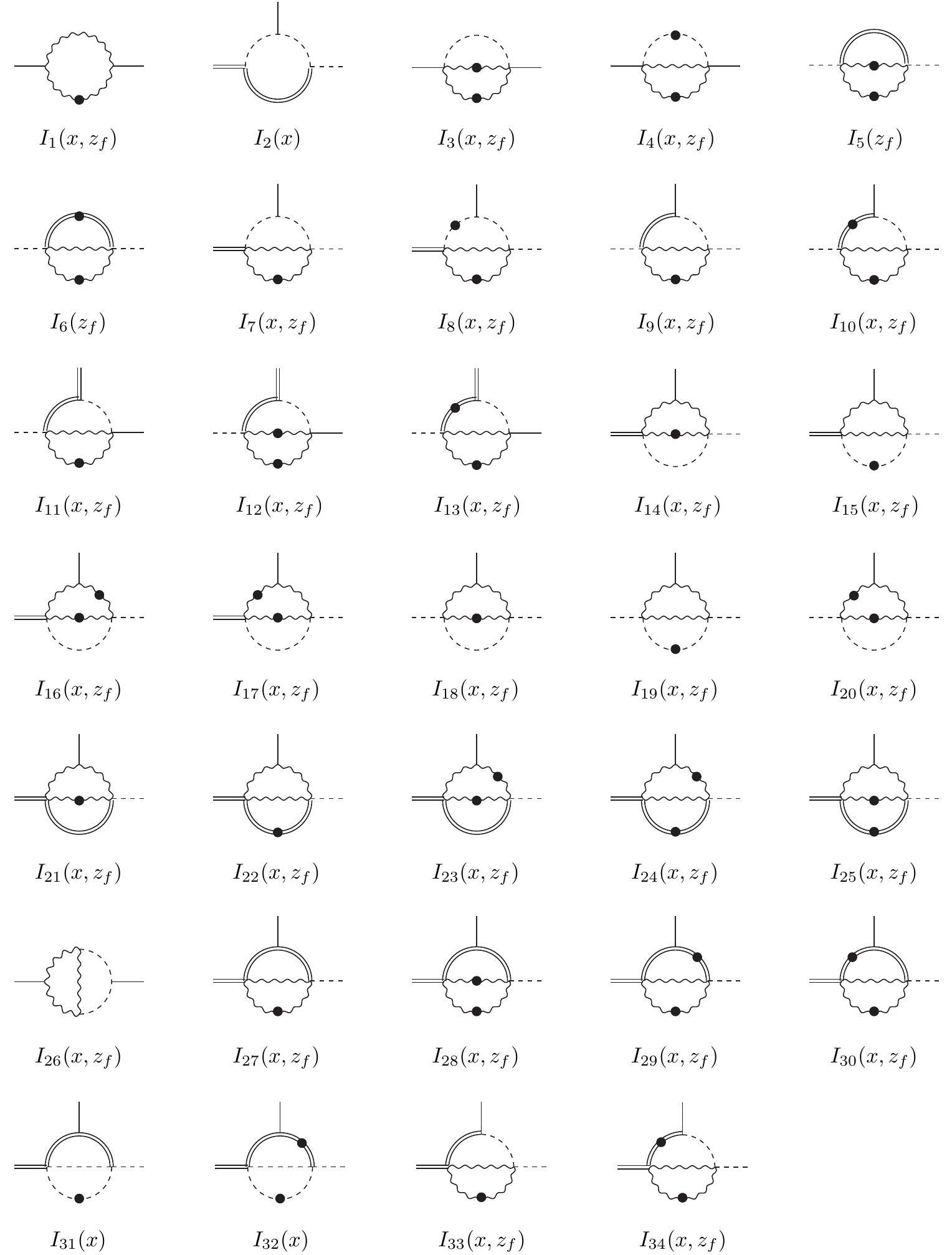}
\caption{Integrals required to define the basis integrals in \eqref{eq:M1}-\eqref{eq:M29}. Dashed/wavy/double internal lines denote propagators with mass $0/\sqrt{\zc}m_b/m_b$. Dashed/solid/double external lines correspond to virtualities $0/x m_b^2/m_b^2$. Dotted propagators are taken to be squared.}\label{fig:integrals}
\vspace{-0.5cm}
}

At present there does not exist a systematic algorithm to find a canonical basis of master integrals. The construction therefore requires some level of experimentation, for some guidelines cf.~the discussions in~\cite{Henn:2013tua, Argeri:2014qva, Gehrmann:2014bfa, Hoschele:2014qsa, DiVita:2014pza}. In the current calculation we mainly used explicit integral representations to find the canonical basis. The basis consists of 29 master integrals which we denote by $M_{1-29}$. In terms of the integrals $I_{1-34}$ defined in Fig.~\ref{fig:integrals}, they are given by
\allowdisplaybreaks{
\begin{align} \label{eq:M1}
M_1(r,s) &= \ep \, \ub \, s \, I_1(\ub,z_f) \,, \\[0.6em]
M_2(\ub) &= \ep^2 u \, I_2(\ub) \,, \\[0.6em]
M_3(r,s) &= \ep^2 \ub \, I_3(\ub,z_f) \,, \\[0.3em]
M_4(r,s) &= \ep^2 \ub \, s \, 
\Big( I_3(\ub,z_f) + 2 I_4(\ub,z_f) \Big)\,, \\[0.2em]
M_5(r) &= \ep^2 r \, 
\Big( I_5(z_f) + 2 I_6(z_f) \Big)\,, \\[0.3em]
M_6(r,s) &= \ep^3 \ub \, I_7(u,z_f) \,, \\[0.1em]
M_7(r,s) &= \frac{\ep^2 \ub \, s}{2 m_b^2} \, 
\Big( 2 u  m_b^2 \,I_8(u,z_f) - I_3(1,z_f)  
- 2 I_4(1,z_f) \Big)\,, \\[0.1em]
M_8(r,s) &= \ep^3 u \, I_9(u,z_f) \,, \\[0.1em]
M_9(r,s) &= \frac{\ep^2 \ub \, s}{2m_b^2} \, 
\Big(  2 u m_b^2 \,I_{10}(u,z_f) - I_5(z_f)  
- 2I_6(z_f) \Big)\,, \\[0.1em]
M_{10}(r,s) &= \ep^3 u \, I_{11}(\ub,z_f) \,, \\[0.3em]
M_{11}(r,s) &= \ep^2  \ub \, s \, 
\Big( I_{12}(\ub,z_f) + 2 I_{13}(\ub,z_f) \Big)\,, \\[0.3em]
M_{12}(r,s) &= \ep^3 u \, I_{14}(\ub,z_f) \,, \\[0.6em]
M_{13}(r,s) &= \ep^3 u \, I_{15}(\ub,z_f) \,, \\[0.0em]
M_{14}(r,s) &= \frac{\ep^2 s}{(1 + r^2)m_b^2} \, 
\Big\{ 4 z_f m_b^2 \big(1- \ub +\ub \zc \big)
\big(I_{16}(\ub,z_f) + I_{17}(\ub,z_f)\big) 
 \nnb \\[0.0em]
&  
+ 3 I_3(1,z_f) + 2 \ep\big(1- \ub +2\ub\zc\big)
\big(I_{15}(\ub,z_f) + 2I_{14}(\ub,z_f)\big)\Big\}\,, 
\\[0.3em]
M_{15}(r,s) &= \ep^3 \ub \, I_{18}(\ub,z_f) \,, \\[0.6em]
M_{16}(r,s) &= \ep^3 \ub \, I_{19}(\ub,z_f) \,, \\[0.1em]
M_{17}(r,s) &= \frac{\ep^2 \ub \, s}{m_b^2} \, 
\Big(  2 z_f  m_b^2 \,I_{20}(\ub,z_f) + \ep\,I_{19}(\ub,z_f) 
+ 2\ep\,I_{18}(\ub,z_f)  \Big)\,, \\[0.1em]
M_{18}(r,s) &= \ep^3 u \, I_{21}(\ub,z_f) \,, \\[0.6em]
M_{19}(r,s) &= \ep^3 u \, I_{22}(\ub,z_f) \,, \\[0.1em]
M_{20}(r,s) &= - \frac{\ep^2 \ub \, s}{2m_b^2} \, 
\Big\{  u m_b^2 \big( I_{23}(\ub,z_f) + I_{24}(\ub,z_f) \big) 
+ I_{5}(z_f) + 2I_{6}(z_f)  \Big\}\,, \\[-0.1em]
M_{21}(r,s) &= \frac{\ep^2}{\ub m_b^2} \, 
\Big\{  2 m_b^2  \big((1 + \ub)^2 \zc - \ub^2\big) I_{25}(\ub, \zc)  
+2 \zc (1 + \ub)  \big(I_5'(\zc) + 2 I_4'(\zc)\big) 
\nnb\\[0.1em]
& + \big(\ub^2 - 2 (1 + \ub) \zc\big) \big(I_5(\zc) + 2 I_6(\zc)\big) + 
 2 \ep  \, u \ub \,\big(I_{21}(\ub, \zc) + I_{22}(\ub, \zc)\big)
\nnb\\[0.1em]
& - \, \ub  m_b^2  (1 + \ub)(\ub - 4 \zc)   
\big(I_{23}(\ub, \zc) + I_{24}(\ub, \zc) \big)
+2 \ub \, I_4'(\zc)
\Big\}\,, \\[0.1em]
M_{22}(r,s) &=  \ep^3 (1-2\eps) \, \ub \, I_{26}(\ub,z_f) \,, \\[0.6em]
M_{23}(r,s_1) &= \ep^3 \, u \, I_{27}(\ub,z_f) \,, \\[0.0em]
M_{24}(r,s_1) &= \frac{2\ep^2(1+s_1)}{(1-s_1)^2 \, m_b^2} \; \sqrt{1+\frac{8\zc(1-s_1)}{(1+s_1)^2}}\,
\Big\{m_b^2 \, (1-s_1)\, \big(I_{28}(\ub,z_f) + 2\,I_{29}(\ub,z_f)\big) \nnb \\[0.1em]
 &-2m_b^2\,(1+s_1) \, I_{30}(\ub,z_f) +(1-s_1)\, \big(I_5'(\zc) + 2 I_4'(\zc)\big)\Big\} \, ,\\[0.0em]
M_{25}(r,s_1) &= \frac{2\ep^2(1-s_1)}{(1+s_1)^2 \, m_b^2} \; \sqrt{1+\frac{8\zc(1+s_1)}{(1-s_1)^2}}\,
\Big\{m_b^2 \, (1+s_1)\, \big(I_{28}(\ub,z_f) + 2\,I_{29}(\ub,z_f)\big) \nnb \\[0.1em]
 &-2m_b^2\,(1-s_1) \, I_{30}(\ub,z_f) +(1+s_1)\, \big(I_5'(\zc) + 2 I_4'(\zc)\big)\Big\} \, ,\\[0.1em]
M_{26}(s_1) &= \ep^3 u \, I_{31}(\ub) \,, \\[0.0em]
M_{27}(s_1) &= -  \frac{2\ep^2 s_1}{(1 - s_1^2)m_b^2} \,
\Big( m_b^2 \, I_{32}(\ub) + 3\ep \, I_{31}(\ub) \Big)  \,, \\[0.1em]
M_{28}(r,p) &= \ep^3 u \, I_{33}(\ub,\zc) \, , \\[0.0em]
M_{29}(r,p) &= \frac{\ep^2}{2m_b^2}
\Big\{ 2u \big(1-\ub p\big) m_b^2 I_{34}(\ub,\zc) - 
\big( \ub p -1 + 2 \sqrt{\zc} \big)  \big(I_5'(\zc) + 2 I_4'(\zc)\big) 
\Big\} \,.
\label{eq:M29}
\end{align}
}

\noindent
The variables $r$, $s$, $s_1$ and $p$ have been introduced in Section~\ref{sec-kinematics}, and the definition of the integrals $I_{4,5}'(\zc)$ can be found in Appendix~\ref{app:auxints}. In addition there are seven auxiliary integrals, labeled $M_{1-7}'$, which are already known from previous calculations but which are needed in order to close the system of differential equations.

In the given integral basis the system of differential equations takes the form (\ref{eq:totaldifferential}). Instead of one large matrix $\tilde A$, we solve each topology separately and in turn get several smaller matrices $\tilde A_k$. We give the solution to the basis integrals $M_{1-29}$ in the next section, together with the relevant boundary conditions. The solution to the auxiliary integrals $M_{1-7}'$ can be found in Appendix~\ref{app:auxints}.


\section{Results}
\label{sec-results}

We write the results for the master integrals in the form
\begin{align}
M =& \; i^L \, S_\Gamma^L \, \left(m_b^2\right)^{L\,D/2-N} \, \tilde M \, , \label{eq:mastergeneric}
\end{align}
with the number of loops $L$ and an integer $N$ which denotes the sum of all propagator powers. The integral $\tilde M$ is therefore dimensionless. Our integration measure per loop is $\int d^Dk/(2\pi)^D$ and the pre-factor $S_\Gamma$ reads
\begin{align}
S_\Gamma =& \frac{1}{\left(4\pi\right)^{D/2} \, \Gamma(1-\ep)} \; .
\end{align}

Once the differential equations are set up, the only missing ingredient are the boundary conditions. It turns out that the following conditions -- almost all of which describe the vanishing of an integral in a particular kinematic point -- are sufficient to write down the entire solution to an integral. We find that $M_{1,3,4,6,7,9,11,14-17,20-22}(r,s)$ and $M_{27}(s_1)$ vanish in $\ub=0$, corresponding to $s=+i\infty$ or $s_1=+i\infty$. Furthermore, $M_{8,10,12,13,18,19}(r,s)$, $M_2(\ub)$, $M_{23}(r,s_1)$, $M_{26}(s_1)$, and $M_{28,29}(r,p)$ vanish in $\ub=1$, corresponding to $s=r$, $s_1=+i\sqrt{3}$ or $p=1-2\sqrt{\zc}$. Moreover, $M_{5}(r)$ vanishes in $r=0$. Finally, the integrals $M_{24,25}(r,s_1)$ fulfill
\begin{align}
\tilde M_{24,25}(r,s_1=+i\infty) & = 4 \, \tilde M_{23}(r,s_1=+i\infty) \, - \, 4 \, \tilde M_{4}'(\zc) \, ,
\end{align}
which can be derived using the Laporta reduction algorithm~\cite{Laporta:2001dd}. All these considerations lead to the full set of solutions which we list below.

\subsection{$M_{1}$}
\label{sec-M01}
As a warm-up exercise and to demonstrate how the method of differential equations in the canonical basis works, we consider the one-loop integral 
\begin{align}
M_1(r,s) &= \loopint{k} \; \frac{\ep\, \ub\, s}{[(k + p -u q)^2 - z_f m_b^2] [k^2 - z_f m_b^2]^2}\,.
\end{align}
The auxiliary integral 
\begin{align}
M_{1}'(\zc) &= \loopint{k} \; \frac{\ep}{[k^2 - z_f m_b^2]^2}
\end{align}
appears as a subtopology and has to be taken into account in order to make the system of differential equations complete. The solution to the auxiliary integral $M_{1}'(\zc)$ is elementary and can be found in Appendix~\ref{app:auxints}.

In terms of the variables $r$ and $s$, the system of differential equations becomes
\begin{align}
\frac{\partial \tilde M_1(r,s)}{\partial s} = & -\frac{2\, \ep \, \tilde M_{1}(r,s)}{s \, (1-s^2)} + \frac{2\,\ep \, \tilde M_{1}'(\zc)}{1-s^2} \,,  
\label{eq:M01s}\\
\frac{\partial \tilde M_{1}'(\zc)}{\partial s} = & 0 \,,  \\
\intertext{and}
\frac{\partial \tilde M_1(r,s)}{\partial r} = & \frac{2\, \ep \, r \, \tilde M_{1}(r,s)}{1-r^2} \,, 
\label{eq:M01r}\\
\frac{\partial \tilde M_{1}'(\zc)}{\partial r} = & \frac{2\,\ep \, r \, \tilde M_{1}'(\zc)}{1-r^2} \, .
\end{align}
The system of differential equations can be brought into the canonical form \eqref{eq:totaldifferential}, with $\vec{M} = \{\tilde M_1(r,s), \tilde M_{1}'(\zc)\}$ and
\begin{align}
\tilde A_1(r,s) =& \left(\begin{array}{cc}
 \ln(1-s^2)-2 \ln(s)-\ln(1-r^2) \quad& \ln\left(\displaystyle\frac{1+s}{1-s}\textstyle\right) \\[1em]
 0 & -\ln(1-r^2)
\end{array}\right) \, . \label{eq:AtildeM01}
\end{align}
Solving Eqs.~\eqref{eq:M01s} and \eqref{eq:M01r} together with the aforementioned boundary condition gives
\begin{align}
\tilde M_{1}(r,s)= & 
\,z_f^{-\ep}\, \Big\{ \ep\,[H_{w_1^+}(s) - i\pi] \nnb\\
+& \ep^2\,[\pi^2 + 2\,i\pi\,H_{0}(s) + 
   i\pi\,H_{w_1^-}(s) - 2\,H_{0, w_1^+}(s)-H_{w_1^-, w_1^+}(s) + 
   2\,i\pi\,\ln(2)] \nnb\\+& \ep^3\,[\frac{i\pi^3}{6}  - 2\,\pi^2\,H_{0}(s) - 
   \pi^2\,H_{w_1^-}(s) + \frac{\pi^2}{6}\,H_{w_1^+}(s) - 4\,i\pi\,H_{0, 0}(s) - 
   2\,i\pi\,H_{0, w_1^-}(s) \nnb\\-& 2\,i\pi\,H_{w_1^-, 0}(s) - 
   i\pi\,H_{w_1^-, w_1^-}(s) + 4\,H_{0, 0, w_1^+}(s) + 
   2\,H_{0, w_1^-, w_1^+}(s) + 2\,H_{w_1^-, 0, w_1^+}(s) \nnb\\+& 
   H_{w_1^-, w_1^-, w_1^+}(s) - 2\,\pi^2\,\ln(2) - 4\,i\pi\,H_{0}(s)\,\ln(2) - 
   2\,i\pi\,H_{w_1^-}(s)\,\ln(2) - 2\,i\pi\,\ln^2(2)] \nnb\\+& {\cal O}(\ep^4)\Big\} \,. \label{eq:M01res}
\end{align}
The solution can also be obtained from the following closed form,
\begin{align}
\tilde M_{1}(r,s)= & 
\, z_f^{-\ep}\; \frac{2 \, \ep \, s \, \Gamma(1-\ep) \, \Gamma(1+\ep)}{s^2-1} \;
  \pFq{2}{1}{1,1+\ep}{\frac{3}{2}}{\frac{1}{1-s^2}} \; ,\label{eq:M01exact}
\end{align}
by expanding the hypergeometric function e.g.~with {\tt HypExp}~\cite{Huber:2005yg,Huber:2007dx}.

\subsection{$M_{2}$}
\label{sec-M02}
From now on, we will not give the explicit form of the differential equations anymore, but only the corresponding matrices $\tilde A_i$ and the final solution to the integrals. The integral $M_2$ only depends on one kinematic variable, which we choose to be $\ub$. The set of integrals is now given by $\vec{M} = \left\{\tilde M_2(\ub),\tilde M_{1}'(\zc=1),\tilde M_{2}'(\ub)\right\}$, and we have
\begin{align}
\tilde A_2(\ub) =& \left(\begin{array}{ccc}
  2\ln(1-\ub)-2\ln(\ub)\quad &-\ln(\ub) \quad& -\ln(\ub)\\
   0 & 0 & 0\\
  0 & 0 & -\ln(\ub) \\
\end{array}\right) \, . \label{eq:AtildeM02}
\end{align}
The solution reads
\begin{align}
\tilde M_{2}(\ub)=&\ep^2\,[i\pi\,H_{0}(\ub) - H_{0, 0}(\ub)] \nnb\\ +& 
 \ep^3\,[-\frac{i\pi^3}{3} - \frac{2}{3}\,\pi^2\,H_{0}(\ub) - 3\,i\pi\,H_{0, 0}(\ub) - 
   i\pi\,H_{w_1^-, 0}(\ub) - i\pi\,H_{w_1^+, 0}(\ub) + 
   3\,H_{0, 0, 0}(\ub) \nnb \\+& H_{w_1^-, 0, 0}(\ub) + 
   H_{w_1^+, 0, 0}(\ub) - 2\,\zeta_3]+ {\cal O}(\ep^4) \; . \label{eq:M02res}
\end{align}
Also in this case the solution can be obtained from an expression containing hypergeometric functions,
\begin{align}
\tilde M_{2}(\ub)= & \frac{(1-\ub) \, \ep \, \Gamma(1-\ep) \, \Gamma(1+\ep)}{\Gamma(2-2\ep)} \,
  \big\{\Gamma(1-2\ep) \, \pFq{2}{1}{1,1}{2-2\ep}{1-\ub} \nnb \\
  &  -\Gamma^2(1-\ep) \, e^{i\pi\ep} \, \ub^{-\ep} \, \pFq{2}{1}{1,1-\ep}{2-2\ep}{1-\ub} \big\}\; .\label{eq:M02exact}
\end{align}

\subsection{$M_{3}$ and $M_{4}$}
\label{sec-M03}
In this topology we have the set of integrals $\vec{M}= \big\{\tilde M_{3}(r,s),\tilde M_{4}(r,s),\big[\tilde M_{1}'(\zc)\big]^2\big\}$, together with the corresponding matrix $\tilde A_{3,4}(r,s)$. Since the expressions for the matrices $\tilde A_k$ become more and more involved, we from now on relegate them to Appendix~\ref{app:Atilde}. The solution to $M_{3}(r,s)$ and $M_{4}(r,s)$ reads
\allowdisplaybreaks{
\begin{align}
\tilde M_{3}(r,s)= & 
\,z_f^{-2\ep}\, 
\Big\{
\ep^2\,[-\pi^2 - 2\,i\pi\,H_{w_1^+}(s) + 2\,H_{w_1^+, w_1^+}(s)] \nnb \\
  +& \ep^3\,[-\pi^2\,H_{w_1^-}(s)+6\,\pi^2\,H_{w_1^+}(s)-
2\,i\pi\,H_{w_1^-,w_1^+}(s)+12\,i\pi\,H_{w_1^+,0}(s) \nnb \\+&
8\,i\pi\,H_{w_1^+,w_1^-}(s)+2\,H_{w_1^-,w_1^+,w_1^+}(s)-
12\,H_{w_1^+,0,w_1^+}(s)-8\,H_{w_1^+,w_1^-,w_1^+}(s) \nnb \\-&
2\,\pi^2\,\ln(2)+16\,i\pi\,H_{w_1^+}(s)\,\ln(2)-21\,\zeta_3]+{\cal O}(\ep^4) \Big\} \, ,  \\[0.5em]
\tilde M_{4}(r,s)= &
\,z_f^{-2\ep}\, \Big\{
  \ep\,[2\,i\pi - 2\,H_{w_1^+}(s)] \nnb \\
  +& \ep^2\,[12\,H_{0,w_1^+}(s)+8\,H_{w_1^-,w_1^+}(s)-12\,i\pi\,H_{0}(s)-
8\,i\pi\,H_{w_1^-}(s)-16\,i\pi\,\ln(2)-6\,\pi^2] \nnb \\
  +& \ep^3\,[-4\,i\pi^3+36\,\pi^2\,H_{0}(s)+
24\,\pi^2\,H_{w_1^-}(s)-\frac{20}{3}\,\pi^2\,H_{w_1^+}(s)+
72\,i\pi\,H_{0,0}(s) \nnb \\+& 48\,i\pi\,H_{0,w_1^-}(s)+
48\,i\pi\,H_{w_1^-,0}(s)+32\,i\pi\,H_{w_1^-,w_1^-}(s)-
12\,i\pi\,H_{w_1^+,w_1^+}(s)\nnb \\-& 72\,H_{0,0,w_1^+}(s)-
48\,H_{0,w_1^-,w_1^+}(s)-48\,H_{w_1^-,0,w_1^+}(s)-
32\,H_{w_1^-,w_1^-,w_1^+}(s)\nnb \\+& 12\,H_{w_1^+,w_1^+,w_1^+}(s)+
48\,\pi^2\,\ln(2)+96\,i\pi\,H_{0}(s)\,\ln(2)\nnb \\+& 64\,i\pi\,H_{w_1^-}(s)\,
\ln(2)+64\,i\pi\,\ln^2(2)] + {\cal O} (\ep^4) \Big\} \, .
\end{align}
}
A closed form of these integrals is given by
\begin{align}
\tilde M_{3}(r,s)= & 
\,z_f^{-2\ep}\, \frac{\Gamma^2(1-\ep) \, \Gamma^2(1+\ep)}{2 \, s^2 \, (\ep-1)} \;
\bigg\{ (\ep-1) \, (s^2+1) \nnb \\
+& (1-2\ep) \, (3-s^2) \, \pFq{3}{2}{1,\ep,2\ep}{2-\ep,\frac{1}{2}+\ep}{\frac{1}{1-s^2}} \nnb \\
-& (2-3\ep) \, (1-s^2) \, \pFq{3}{2}{1,\ep,2\ep-1}{2-\ep,\frac{1}{2}+\ep}{\frac{1}{1-s^2}} \bigg\} \, ,\label{eq:M03exact} \\[0.5em]
\tilde M_{4}(r,s)= &  
\,z_f^{-2\ep}\, \frac{\Gamma^2(1-\ep) \, \Gamma(1+\ep)\, \Gamma(\ep-1)}{4 \, s^3} \;
\bigg\{ (\ep-1) \left[\ep (4 s^4-6 s^2+6)-(s^2-1)^2\right] \nnb \\
-& (1-2 \ep) (3+s^2) \left[\ep \left(4 s^2-6\right)-s^2+1\right] \, \pFq{3}{2}{1,\ep,2\ep}{2-\ep,\frac{1}{2}+\ep}{\frac{1}{1-s^2}} \nnb \\
-& (2-3 \ep) (1-s^2) \left[\ep \left(4 s^2+6\right)-s^2-1\right] \, \pFq{3}{2}{1,\ep,2\ep-1}{2-\ep,\frac{1}{2}+\ep}{\frac{1}{1-s^2}} \!\!\bigg\} \, .
\label{eq:M04exact}
\end{align}

\subsection{$M_{5}$}
\label{sec-M05}
In this case the set of integrals consists of $\vec{M} =\big\{\tilde M_{5}(r),\big[\tilde M_{1}'(\zc)\big]^2,\tilde M_{1}'(\zc) \, \tilde M_{1}'(\zc=1)\big\}$. The matrix $\tilde A_{5}(r)$ can be found in Appendix~\ref{app:Atilde}, and the solution becomes
\allowdisplaybreaks{
\begin{align}
\tilde M_{5}(r)= & \ep^2\,[-2\,H_{w_1^+, w_1^-}(r) - 4\,H_{w_1^+}(r)\,\ln(2)]\nnb \\
 +& \ep^3\,[4\,H_{0, w_1^+, w_1^-}(r) - 6\,H_{w_1^+, w_1^-, w_1^-}(r) + 
    8\,H_{0, w_1^+}(r)\,\ln(2) - 12\,H_{w_1^+, w_1^-}(r)\,\ln(2) \nnb \\
    -& 12\,H_{w_1^+}(r)\,\ln^2(2)] +{\cal O} (\ep^4) \; ,
\end{align}
}
which can also be obtained from the expansion of
\begin{align}
\tilde M_{5}(r)= & \frac{4^{1+\ep} \, \ep \, r \, \Gamma^2(1-\ep)\Gamma^2(1+\ep)}{1+2\ep} \nnb \\
  \times & \bigg\{(1-r^2)^{-\ep} \, \pFq{2}{1}{1,\frac{1}{2}}{\frac{3}{2}+\ep}{r^2} - 4^\ep \, (1-r^2)^{-2\ep} \, \pFq{2}{1}{1,\frac{1}{2}-\ep}{\frac{3}{2}+\ep}{r^2} \bigg\} \; .
\end{align}

\subsection{$M_{6}$ and $M_{7}$}
\label{sec-M06}
Here the topology consists of six integrals
\begin{equation}
\vec{M} = \left\{\tilde M_{6}(r,s),\tilde M_{7}(r,s),\tilde M_{3}(r,s=r),\tilde M_{4}(r,s=r),\tilde M_{1}'(\zc)\tilde M_{2}'(u),\big[\tilde M_{1}'(\zc)\big]^2\right\} \,,
\end{equation}
and the corresponding matrix is $\tilde A_{6,7}(r,s)$. The solutions to the integrals reads
\allowdisplaybreaks{
\begin{align}
\tilde M_{6}(r,s)= &
\ep^3\,[-\frac{i\pi^3}{2} + \pi^2\,H_{0}(r) + \frac{\pi^2}{2}\,H_{w_1^-}(s) + 
    i\pi\,H_{w_1^-}(s)\,H_{w_1^+}(r) - \frac{\pi^2}{2}\,H_{w_2^-}(s) \nnb \\
    -& i\pi\,H_{w_1^+}(r)\,H_{w_2^-}(s) + 2\,i\pi\,H_{0, w_1^+}(r) - 
    H_{w_1^-}(s)\,H_{w_1^+, w_1^+}(r) + H_{w_2^-}(s)\,
     H_{w_1^+, w_1^+}(r) \nnb \\
    +& i\pi\,H_{w_1^+, w_1^+}(s) - 
    2\,H_{0}(r)\,H_{w_1^+, w_1^+}(s) - H_{w_1^-}(r)\,
     H_{w_1^+, w_1^+}(s) - i\pi\,H_{w_1^+, w_2^+}(s) \nnb \\
     +& H_{w_1^+}(r)\,H_{w_1^+, w_2^+}(s) - 2\,H_{0, w_1^+, w_1^+}(r) - 
    H_{w_1^+, w_1^+, w_1^-}(s) + H_{w_1^+, w_1^+, w_2^-}(s)\nnb \\
     - & 2\,H_{w_1^+, w_1^+}(s)\,\ln(2) + \frac{7}{2}\,\zeta_3]+ {\cal O}(\ep^4) \, , \\[0.5em]
\tilde M_{7}(r,s)= & \ep^2\,[-i\pi\,H_{w_1^+}(s) + 2\,H_{0}(r)\,H_{w_1^+}(s) + 
    H_{w_1^-}(r)\,H_{w_1^+}(s) + i\pi\,H_{w_2^+}(s) \nnb \\
     -& H_{w_1^+}(r)\,H_{w_2^+}(s) + H_{w_1^+, w_1^-}(s) - 
    H_{w_1^+, w_2^-}(s) + 2\,H_{w_1^+}(s)\,\ln(2)] \nnb \\
    +& \ep^3\,[\frac{13}{6}\,\pi^2\,H_{w_1^+}(s) + 2\,i\pi\,H_{0}(r)\,H_{w_1^+}(s) - 
    i\pi\,H_{w_1^-}(r)\,H_{w_1^+}(s) - 3\,\pi^2\,H_{w_2^+}(s) \nnb \\
    +& 3\,i\pi\,H_{w_1^+}(r)\, H_{w_1^+}(s)  - 6\,i\pi\,H_{0}(r)\,
     H_{w_2^+}(s) - 2\,i\pi\,H_{w_1^-}(r)\,H_{w_2^+}(s)+ H_{w_2^-, w_1^+, w_1^-}(s)\nnb \\
     -& 4\,H_{w_1^+}(s)\,H_{0, 0}(r) + 2\,H_{w_1^+}(s)\,H_{0, w_1^-}(r) + 
    6\,H_{w_2^+}(s)\,H_{0, w_1^+}(r) + 4\,i\pi\,H_{0, w_1^+}(s) \nnb \\
    -& 8\,H_{0}(r)\,H_{0, w_1^+}(s) - 4\,H_{w_1^-}(r)\,H_{0, w_1^+}(s) - 
    4\,i\pi\,H_{0, w_2^+}(s) + 4\,H_{w_1^+}(r)\,H_{0, w_2^+}(s) \nnb \\
    +& 2\,H_{w_1^+}(s)\,H_{w_1^-, 0}(r) + 3\,H_{w_1^+}(s)\,
     H_{w_1^-, w_1^-}(r) + 2\,H_{w_2^+}(s)\,H_{w_1^-, w_1^+}(r) \nnb \\
     -& H_{w_1^+, w_2^-, w_2^-}(s) + 3\,i\pi\,H_{w_1^-, w_1^+}(s) - 6\,H_{0}(r)\,H_{w_1^-, w_1^+}(s) - 
    3\,H_{w_1^-}(r)\,H_{w_1^-, w_1^+}(s) \nnb \\
    -& 3\,i\pi\,H_{w_1^-, w_2^+}(s) + 3\,H_{w_1^+}(r)\,H_{w_1^-, w_2^+}(s) - 2\,H_{w_2^+}(s)\,
     H_{w_1^+, w_1^-}(r) + i\pi\,H_{w_1^+, w_1^-}(s) \nnb \\
     -& 2\,H_{0}(r)\,H_{w_1^+, w_1^-}(s) + H_{w_1^-}(r)\,
     H_{w_1^+, w_1^-}(s) - 3\,H_{w_1^+}(s)\,H_{w_1^+, w_1^+}(r) - 
    i\pi\,H_{w_1^+, w_2^-}(s) \nnb \\
    +& 2\,H_{0}(r)\,H_{w_1^+, w_2^-}(s) - 
    H_{w_1^-}(r)\,H_{w_1^+, w_2^-}(s) - i\pi\,H_{w_2^-, w_1^+}(s) + 
    2\,H_{0}(r)\,H_{w_2^-, w_1^+}(s) \nnb \\
     +& H_{w_1^-}(r)\,H_{w_2^-, w_1^+}(s) + i\pi\,H_{w_2^-, w_2^+}(s) - 
    H_{w_1^+}(r)\,H_{w_2^-, w_2^+}(s) - 4\,H_{0, w_1^+, w_1^-}(s) \nnb \\
    +& 4\,H_{0, w_1^+, w_2^-}(s) - 3\,H_{w_1^-, w_1^+, w_1^-}(s) + 
    3\,H_{w_1^-, w_1^+, w_2^-}(s) - H_{w_1^+, w_1^-, w_1^-}(s) \nnb \\
    +& H_{w_1^+, w_1^-, w_2^-}(s) + H_{w_1^+, w_2^-, w_1^-}(s) - 
    H_{w_2^-, w_1^+, w_2^-}(s) - 2\,i\pi\,H_{w_1^+}(s)\,\ln(2)\nnb \\
    +& 4\,H_{0}(r)\,H_{w_1^+}(s)\,\ln(2) + 6\,H_{w_1^-}(r)\,H_{w_1^+}(s)\,
     \ln(2) - 4\,i\pi\,H_{w_2^+}(s)\,\ln(2) \nnb \\
     -& 4\,H_{w_1^+}(r)\,H_{w_2^+}(s)\,\ln(2) - 8\,H_{0, w_1^+}(s)\,\ln(2) - 
    6\,H_{w_1^-, w_1^+}(s)\,\ln(2) \nnb \\
    +& 2\,H_{w_1^+, w_1^-}(s)\,\ln(2) - 
    2\,H_{w_1^+, w_2^-}(s)\,\ln(2) + 2\,H_{w_2^-, w_1^+}(s)\,\ln(2) \nnb \\
    +& 6\,H_{w_1^+}(s)\,\ln^2(2)]+{\cal O}(\ep^4) \, .
\end{align}
}

\subsection{$M_{8}$ and $M_{9}$}
\label{sec-M08}
Also here the topology consists of six integrals, namely
\begin{equation}
\vec{M} = \left\{\tilde M_{8}(r,s),\tilde M_{9}(r,s),\tilde M_{5}(r),\tilde M_{1}'(\zc)\tilde M_{3}'(u),\big[\tilde M_{1}'(\zc)\big]^2,\tilde M_{1}'(\zc)\tilde M_{1}'(\zc=1)\right\} \, ,
\end{equation}
and the matrix $\tilde A_{8,9}(r,s)$. Owing to simple boundary conditions, the result is quite short,
\allowdisplaybreaks{
\begin{align}   
\tilde M_{8}(r,s)=&\ep^3\,[-H_{w_1^+, w_1^+, w_1^-}(r) + H_{w_1^+, w_1^+, w_1^-}(s) - 
    2\,H_{w_1^+, w_1^+}(r)\,\ln(2) \nnb \\ +& 2\,H_{w_1^+, w_1^+}(s)\,\ln(2)] +{\cal O}(\ep^4) \, ,\\[0.5em]    
\tilde M_{9}(r,s)=&  \ep^2\,[H_{w_1^+, w_1^-}(s) + 2\,H_{w_1^+}(s)\,\ln(2)] \nnb \\
 +& \ep^3\,[-2\,H_{w_1^+}(s)\,H_{0, w_1^-}(r) - 
    H_{w_1^+}(s)\,H_{w_1^-, w_1^-}(r) + H_{w_2^+}(s)\,
     H_{w_1^+, w_1^-}(r)  \nnb \\
     +& 2\,H_{w_1^-}(r)\,H_{w_1^+, w_1^-}(s) + 
    H_{w_1^-}(r)\,H_{w_1^+, w_2^-}(s) - 4\,H_{0, w_1^+, w_1^-}(s) - 
    3\,H_{w_1^-, w_1^+, w_1^-}(s)\nnb \\
    - &H_{w_1^+, w_1^-, w_1^-}(s) - 
    H_{w_1^+, w_2^-, w_1^-}(s) - H_{w_2^-, w_1^+, w_1^-}(s) + 
    4\,H_{w_1^-}(r)\,H_{w_1^+}(s)\,\ln(2)\nnb \\
    +& 2\,H_{w_1^+}(r)\,
     H_{w_2^+}(s)\,\ln(2) - 8\,H_{0, w_1^+}(s)\,\ln(2) - 
    6\,H_{w_1^-, w_1^+}(s)\,\ln(2) \nnb \\
    +& 2\,H_{w_1^+, w_1^-}(s)\,\ln(2) - 
    2\,H_{w_2^-, w_1^+}(s)\,\ln(2) + 6\,H_{w_1^+}(s)\,\ln^2(2)]+{\cal O}(\ep^4) \, .
\end{align}
}

\subsection{$M_{10}$ and $M_{11}$}
\label{sec-M11}
This topology consists of seven integrals
\begin{equation}
\vec{M}  = \left\{\tilde M_{10}(r,s),\tilde M_{11}(r,s),\tilde M_{3}(r,s),\tilde M_{4}(r,s),\tilde M_{5}(r),\big[\tilde M_{1}'(\zc)\big]^2,\tilde M_{1}'(\zc)\tilde M_{1}'(\zc=1)\right\} \,,
\end{equation}
and the matrix $\tilde A_{10,11}(r,s)$. The result is rather long since we need functions up to weight four in $M_{10}(r,s)$,
\allowdisplaybreaks{
\begin{align}
\tilde M_{10}(r,s)=& \ep^3\,[-\frac{\pi^2}{2}\,H_{w_1^-}(r) + \frac{\pi^2}{2}\,H_{w_1^-}(s) - 
    i\,\pi\,H_{w_1^-}(r)\,H_{w_1^+}(s) + i\,\pi\,H_{w_1^-, w_1^+}(s) \nnb\\ +& 
    i\,\pi\,H_{w_1^+, w_1^-}(s) + H_{w_1^-}(r)\,H_{w_1^+, w_1^+}(s) - 
    H_{w_1^-, w_1^+, w_1^+}(s) - H_{w_1^+, w_1^-, w_1^+}(s) \nnb\\-& 
    H_{w_1^+, w_1^+, w_1^-}(r) - 2\,H_{w_1^+, w_1^+}(r)\,\ln(2) + 
    2\,H_{w_1^+, w_1^+}(s)\,\ln(2)]  \nnb\\+& 
  \ep^4\,[3\,\pi^2\,H_{w_1^-}(r)\,H_{w_1^+}(s) + \pi^2\,H_{w_1^-}(r)\,
     H_{w_2^-}(s) - 4\,i \pi\,H_{w_1^+}(s)\,H_{0, w_1^-}(r) \nnb\\-& 
    \frac{3}{2}\,\pi^2\,H_{w_1^-, w_1^-}(r) - 5\,i \pi\,H_{w_1^+}(s)\,
     H_{w_1^-, w_1^-}(r) + \frac{3}{2}\,\pi^2\,H_{w_1^-, w_1^-}(s) -
    3\,\pi^2\,H_{w_1^-, w_1^+}(s) \nnb\\-& \pi^2\,H_{w_1^-, w_2^-}(r) + 
    6\,i \pi\,H_{w_1^-}(r)\,H_{w_1^+, 0}(s) - 3\,\pi^2\,H_{w_1^+, w_1^-}(s) \nnb\\+&
    5\,i \pi\,H_{w_1^-}(r)\,H_{w_1^+, w_1^-}(s) + 
    3\,H_{w_1^-, w_1^-}(r)\,H_{w_1^+, w_1^+}(s) +
    2\,i \pi\,H_{w_1^-}(r)\,H_{w_1^+, w_2^-}(s) \nnb\\+& 
    2\,H_{w_1^+, w_1^-}(r)\,H_{w_1^+, w_2^+}(s) -
    \pi^2\,H_{w_2^-, w_1^-}(s) + 2\,i \pi\,H_{w_1^-}(r)\,
     H_{w_2^-, w_1^+}(s) \nnb\\+& 4\,i \pi\,H_{0, w_1^-, w_1^+}(r)+
    4\,i \pi\,H_{0, w_1^+, w_1^-}(r) + 2\,i \pi\,H_{w_1^-, w_1^-, w_1^+}(r) + 
    3\,i \pi\,H_{w_1^-, w_1^-, w_1^+}(s) \nnb\\-& 6\,i \pi\,H_{w_1^-, w_1^+, 0}(s) + 
    2\,i \pi\,H_{w_1^-, w_1^+, w_1^-}(r) - 
    2\,i \pi\,H_{w_1^-, w_1^+, w_1^-}(s) - 
    2\,i \pi\,H_{w_1^-, w_1^+, w_2^-}(r) \nnb\\-& 
    2\,i \pi\,H_{w_1^-, w_2^-, w_1^+}(r) + 4\,i \pi\,H_{w_1^+, 0, w_1^-}(r) - 
    6\,i \pi\,H_{w_1^+, 0, w_1^-}(s) \nnb\\-& 6\,H_{w_1^-}(r)\,
     H_{w_1^+, 0, w_1^+}(s) - 6\,i \pi\,H_{w_1^+, w_1^-, 0}(s) + 
    2\,i \pi\,H_{w_1^+, w_1^-, w_1^-}(r) \nnb\\-& 
    7\,i \pi\,H_{w_1^+, w_1^-, w_1^-}(s) - 5\,H_{w_1^-}(r)\,
     H_{w_1^+, w_1^-, w_1^+}(s) - 2\,i \pi\,H_{w_1^+, w_1^-, w_2^-}(r) \nnb\\-& 
    2\,H_{w_1^-}(s)\,H_{w_1^+, w_1^+, w_1^-}(r) + 
    2\,H_{w_2^-}(s)\,H_{w_1^+, w_1^+, w_1^-}(r) - 
    2\,i \pi\,H_{w_1^+, w_2^-, w_1^-}(s)\nnb\\ -& 2\,H_{w_1^-}(r)\,
     H_{w_1^+, w_2^-, w_1^+}(s) - 2\,i \pi\,H_{w_2^-, w_1^-, w_1^+}(s) - 
    2\,i \pi\,H_{w_2^-, w_1^+, w_1^-}(s) \nnb\\-& 2\,H_{w_1^-}(r)\,
     H_{w_2^-, w_1^+, w_1^+}(s) - 3\,H_{w_1^-, w_1^-, w_1^+, w_1^+}(s) + 
    6\,H_{w_1^-, w_1^+, 0, w_1^+}(s) \nnb\\+& 2\,H_{w_1^-, w_1^+, w_1^-, w_1^+}(s) - 
    H_{w_1^-, w_1^+, w_1^+, w_1^-}(r) + 
    2\,H_{w_1^-, w_1^+, w_2^-, w_1^+}(r) \nnb\\+& 
    2\,H_{w_1^-, w_2^-, w_1^+, w_1^+}(r) + 
    6\,H_{w_1^+, 0, w_1^-, w_1^+}(s) + 6\,H_{w_1^+, 0, w_1^+, w_1^-}(r) + 
    6\,H_{w_1^+, w_1^-, 0, w_1^+}(s) \nnb\\+& 7\,H_{w_1^+, w_1^-, w_1^-, w_1^+}(s) + 
    4\,H_{w_1^+, w_1^-, w_1^+, w_1^-}(r) - 
    2\,H_{w_1^+, w_1^-, w_1^+, w_2^+}(r) \nnb\\+& 
    2\,H_{w_1^+, w_1^-, w_2^-, w_1^+}(r) + H_{w_1^+, w_1^+, w_1^-, w_1^-}(r) - 2\,H_{w_1^+, w_1^+, w_1^-, w_2^-}(r)\nnb\\-& 
    4\,H_{w_1^+, w_1^+, w_1^-, w_2^+}(r) - 
    2\,H_{w_1^+, w_1^+, w_2^-, w_1^-}(r) - 
    4\,H_{w_1^+, w_1^+, w_2^+, w_1^-}(r) \nnb\\+& 
    2\,H_{w_1^+, w_2^-, w_1^-, w_1^+}(s) - 
    2\,H_{w_1^+, w_2^+, w_1^+, w_1^-}(r) + 
    2\,H_{w_2^-, w_1^-, w_1^+, w_1^+}(s) \nnb\\+& 
    2\,H_{w_2^-, w_1^+, w_1^-, w_1^+}(s) - 3\,\pi^2\,H_{w_1^-}(r)\,\ln(2) + 
    3\,\pi^2\,H_{w_1^-}(s)\,\ln(2) \nnb\\+& 4\,i \pi\,H_{w_1^-}(r)\,H_{w_1^+}(s)\,
     \ln(2) - 4\,i \pi\,H_{w_1^-, w_1^+}(s)\,\ln(2) - 
    4\,i \pi\,H_{w_1^+, w_1^-}(s)\,\ln(2) \nnb\\-& 4\,H_{w_1^-}(s)\,
     H_{w_1^+, w_1^+}(r)\,\ln(2) + 4\,H_{w_2^-}(s)\,H_{w_1^+, w_1^+}(r)\,
     \ln(2) \nnb\\+& 6\,H_{w_1^-}(r)\,H_{w_1^+, w_1^+}(s)\,\ln(2) + 
    4\,H_{w_1^+}(r)\,H_{w_1^+, w_2^+}(s)\,\ln(2) - 
    2\,H_{w_1^-, w_1^+, w_1^+}(r)\,\ln(2) \nnb\\+& 12\,H_{w_1^+, 0, w_1^+}(r)\,
     \ln(2) - 12\,H_{w_1^+, 0, w_1^+}(s)\,\ln(2) + 
    8\,H_{w_1^+, w_1^-, w_1^+}(r)\,\ln(2) \nnb\\-& 10\,H_{w_1^+, w_1^-, w_1^+}(s)\,
     \ln(2) - 2\,H_{w_1^+, w_1^+, w_1^-}(r)\,\ln(2) - 
    4\,H_{w_1^+, w_1^+, w_2^-}(r)\,\ln(2) \nnb\\-& 8\,H_{w_1^+, w_1^+, w_2^+}(r)\,
     \ln(2) - 4\,H_{w_1^+, w_2^-, w_1^+}(s)\,\ln(2) - 
    4\,H_{w_1^+, w_2^+, w_1^+}(r)\,\ln(2) \nnb\\-& 4\,H_{w_2^-, w_1^+, w_1^+}(s)\,
     \ln(2) - 6\,H_{w_1^+, w_1^+}(r)\,\ln^2(2) + 6\,H_{w_1^+, w_1^+}(s)\,
     \ln^2(2) \nnb\\-& \frac{21}{2}\,H_{w_1^-}(r)\,\zeta_3 + \frac{21}{2}\,H_{w_1^-}(s)\,\zeta_3]
     +{\cal O}(\ep^5) \, , \\[2.0em]
\tilde M_{11}(r,s)=& \ep\,[H_{w_1^+}(s)-i\pi] \nnb\\+& 
  \ep^2\,[3\,\pi^2 + 6\,i\pi\,H_{0}(s) - i\,\pi\,H_{w_1^-}(r) + 
    3\,i\pi\,H_{w_1^-}(s) + H_{w_1^-}(r)\,H_{w_1^+}(s) \nnb\\-& 
    6\,H_{0, w_1^+}(s) - 3\,H_{w_1^-, w_1^+}(s) + 4\,i\pi\,\ln(2) + 
    2\,H_{w_1^+}(s)\,\ln(2)] \nnb\\+& \ep^3\,[2\,i\pi^3 - 18\,\pi^2\,H_{0}(s) + 
    3\,\pi^2\,H_{w_1^-}(r) + 6\,i\pi\,H_{0}(s)\,H_{w_1^-}(r) - 
    9\,\pi^2\,H_{w_1^-}(s) \nnb\\+& 3\,i\pi\,H_{w_1^-}(r)\,H_{w_1^-}(s) + 
    \frac{10}{3}\,\pi^2\,H_{w_1^+}(s) - 2\,i\pi\,H_{w_1^-}(r)\,H_{w_2^-}(s) - 
    36\,i\pi\,H_{0, 0}(s) \nnb\\+& 4\,i\pi\,H_{0, w_1^-}(r) - 
    18\,i\pi\,H_{0, w_1^-}(s) - 6\,H_{w_1^-}(r)\,H_{0, w_1^+}(s) - 
    18\,i\pi\,H_{w_1^-, 0}(s) \nnb\\+& i\,\pi\,H_{w_1^-, w_1^-}(r) + 
    H_{w_1^+}(s)\,H_{w_1^-, w_1^-}(r) - 9\,i\pi\,H_{w_1^-, w_1^-}(s) - 
    3\,H_{w_1^-}(r)\,H_{w_1^-, w_1^+}(s) \nnb\\-& 2\,H_{w_2^+}(s)\,
     H_{w_1^+, w_1^-}(r) + 6\,i\pi\,H_{w_1^+, w_1^+}(s) + 
    2\,i\pi\,H_{w_2^-, w_1^-}(s) + 2\,H_{w_1^-}(r)\,
     H_{w_2^-, w_1^+}(s) \nnb\\+& 36\,H_{0, 0, w_1^+}(s) + 
    18\,H_{0, w_1^-, w_1^+}(s) + 18\,H_{w_1^-, 0, w_1^+}(s) + 
    9\,H_{w_1^-, w_1^-, w_1^+}(s) \nnb\\-& 6\,H_{w_1^+, w_1^+, w_1^+}(s) - 
    2\,H_{w_2^-, w_1^-, w_1^+}(s) - 12\,\pi^2\,\ln(2) - 
    24\,i\pi\,H_{0}(s)\,\ln(2) \nnb\\+& 4\,i\pi\,H_{w_1^-}(r)\,\ln(2) - 
    12\,i\pi\,H_{w_1^-}(s)\,\ln(2) + 2\,H_{w_1^-}(r)\,H_{w_1^+}(s)\,
     \ln(2) \nnb\\-& 4\,H_{w_1^+}(r)\,H_{w_2^+}(s)\,\ln(2) -
    12\,H_{0, w_1^+}(s)\,\ln(2) - 6\,H_{w_1^-, w_1^+}(s)\,\ln(2) \nnb\\+& 
    4\,H_{w_2^-, w_1^+}(s)\,\ln(2) - 8\,i\pi\,\ln^2(2) + 
    2\,H_{w_1^+}(s)\,\ln^2(2)] + {\cal O}(\ep^4) \, .
\end{align}
}

\subsection{$M_{12}$ -- $M_{14}$}
\label{sec-M12}
Again we need seven integrals to complete the system of differential equations. They are
\begin{equation}
\vec{M} = \left\{\tilde M_{12}(r,s),\tilde M_{13}(r,s),\tilde M_{14}(r,s),\tilde M_{3}(r,r),\tilde M_{4}(r,r),\big[\tilde M_{1}'(\zc)\big]^2,\tilde M_{1}(r,s)\tilde M_{1}'(\zc)\right\} \, ,
\end{equation}
together with the matrix $\tilde A_{12-14}(r,s)$. The results are
\allowdisplaybreaks{
\begin{align}
\tilde M_{12}(r,s)=& \ep^3\,[\pi^2\,H_{w_1^-}(r) - \pi^2\,H_{w_1^-}(s) - 2\,\pi^2\,H_{w_1^+}(r) - 
    2\,i\pi\,H_{w_1^-}(s)\,H_{w_1^+}(r) + 2\,\pi^2\,H_{w_1^+}(s) \nnb \\+& 
    4\,i\pi\,H_{0}(r)\,H_{w_1^+}(s) + 2\,i\pi\,H_{w_1^-}(r)\,
     H_{w_1^+}(s) - \frac{3}{4}\,\pi^2\,H_{w_3^-}(r) +
    \frac{3}{4}\,\pi^2\,H_{w_3^-}(s) \nnb \\+& 2\,i\pi\,H_{w_1^+}(r)\,
     H_{w_3^-}(s) + \pi^2\,H_{w_3^+}(r) - \pi^2\,H_{w_3^+}(s) -
    2\,i\pi\,H_{0}(r)\,H_{w_3^+}(s) \nnb \\-& i\pi\,H_{w_1^-}(r)\,
     H_{w_3^+}(s) - 4\,i\pi\,H_{0, w_1^+}(r) + 
    2\,i\pi\,H_{0, w_3^+}(r) + i\pi\,H_{w_1^-, w_3^+}(r) \nnb \\-& 
    4\,i\pi\,H_{w_1^+, 0}(r) - i\pi\,H_{w_1^+, w_1^-}(r) + 
    i\pi\,H_{w_1^+, w_1^-}(s) + 2\,H_{w_1^-}(s)\,H_{w_1^+, w_1^+}(r) \nnb \\-& 
    2\,H_{w_3^-}(s)\,H_{w_1^+, w_1^+}(r) - 
    2\,i\pi\,H_{w_1^+, w_3^-}(r) + 2\,i\pi\,H_{w_1^+, w_2^-}(r) - 
    2\,i\pi\,H_{w_1^+, w_2^-}(s) \nnb \\-& 2\,i\pi\,H_{w_1^+, w_2^+}(r) + 
    2\,i\pi\,H_{w_1^+, w_2^+}(s) - 2\,H_{w_1^+}(r)\,H_{w_1^+, w_2^+}(s) - 
    \frac{3}{2}\,i\pi\,H_{w_3^-, w_1^+}(r) \nnb \\-& \frac{1}{2}\,i\pi\,H_{w_3^-, w_1^+}(s) + 
    2\,i\pi\,H_{w_3^+, 0}(r) + \frac{3}{2}\,i\pi\,H_{w_3^+, w_1^-}(r) - 
    \frac{1}{2}\,i\pi\,H_{w_3^+, w_1^-}(s) \nnb \\-& i\pi\,H_{w_3^+, w_2^-}(r) + 
    i\pi\,H_{w_3^+, w_2^-}(s) + i\pi\,H_{w_3^+, w_2^+}(r) - 
    i\pi\,H_{w_3^+, w_2^+}(s) \nnb \\+& H_{w_1^+}(r)\,H_{w_3^+, w_2^+}(s) - 
    2\,H_{w_1^-, w_1^+, w_1^+}(r) - H_{w_1^+, w_1^-, w_1^+}(r) - 
    H_{w_1^+, w_1^-, w_1^+}(s) \nnb \\-& 2\,H_{w_1^+, w_1^+, w_1^-}(r) + 
    2\,H_{w_1^+, w_1^+, w_3^-}(r) + 4\,H_{w_1^+, w_1^+, w_2^+}(r) + 
    2\,H_{w_1^+, w_3^-, w_1^+}(r) \nnb \\-& H_{w_1^+, w_3^+, w_2^+}(r) - 
    2\,H_{w_1^+, w_2^-, w_1^+}(r) + 2\,H_{w_1^+, w_2^-, w_1^+}(s) + 
    2\,H_{w_1^+, w_2^+, w_1^+}(r) \nnb \\+& \frac{3}{2}\,H_{w_3^-, w_1^+, w_1^+}(r) + 
    \frac{1}{2}\,H_{w_3^-, w_1^+, w_1^+}(s) - \frac{1}{2}\,H_{w_3^+, w_1^-, w_1^+}(r) + 
    \frac{1}{2}\,H_{w_3^+, w_1^-, w_1^+}(s) \nnb \\-& H_{w_3^+, w_1^+, w_2^+}(r) + 
    H_{w_3^+, w_2^-, w_1^+}(r) - H_{w_3^+, w_2^-, w_1^+}(s) - 
    H_{w_3^+, w_2^+, w_1^+}(r) \nnb \\-& 2\,i\pi\,H_{w_1^+}(r)\,\ln(2) + 
    2\,i\pi\,H_{w_1^+}(s)\,\ln(2) + i\pi\,H_{w_3^+}(r)\,\ln(2) - 
    i\pi\,H_{w_3^+}(s)\,\ln(2)] \nnb \\+& {\cal O}(\ep^4) \, ,\\[2.0em]
\tilde M_{13}(r,s)=& \ep^2\,[i\pi\,H_{w_1^+}(r) - i\pi\,H_{w_1^+}(s) - H_{w_1^+, w_1^+}(r) + 
    H_{w_1^+, w_1^+}(s)] \nnb \\ +& \ep^3\,[-\frac{1}{2}\,\pi^2\,H_{w_1^-}(r) + 
    \frac{1}{2}\,\pi^2\,H_{w_1^-}(s) - \pi^2\,H_{w_1^+}(r) + \pi^2\,H_{w_1^+}(s) + \frac{3}{4}\,\pi^2\,H_{w_3^-}(r)\nnb \\-&
    2\,i\pi\,H_{w_1^-}(r)\,H_{w_1^+}(s)  - 
    \frac{3}{4}\,\pi^2\,H_{w_3^-}(s) - 2\,i\pi\,H_{w_1^+}(r)\,
     H_{w_3^-}(s) - \pi^2\,H_{w_3^+}(r) \nnb \\+& \pi^2\,H_{w_3^+}(s) + 
    2\,i\pi\,H_{0}(r)\,H_{w_3^+}(s) + i\pi\,H_{w_1^-}(r)\,
     H_{w_3^+}(s) + 2\,i\pi\,H_{w_1^+}(r)\,H_{w_2^-}(s) \nnb \\-& 
    2\,i\pi\,H_{0, w_3^+}(r) + i\pi\,H_{w_1^-, w_1^+}(r) + 
    i\pi\,H_{w_1^-, w_1^+}(s) - i\pi\,H_{w_1^-, w_3^+}(r) - 
    2\,i\pi\,H_{w_1^+, 0}(r) \nnb \\+& 2\,i\pi\,H_{w_1^+, 0}(s) + 
    i\pi\,H_{w_1^+, w_1^-}(r) + i\pi\,H_{w_1^+, w_1^-}(s) + 
    2\,H_{w_3^-}(s)\,H_{w_1^+, w_1^+}(r) \nnb \\-& 2\,H_{w_2^-}(s)\,
     H_{w_1^+, w_1^+}(r) + 2\,H_{w_1^-}(r)\,H_{w_1^+, w_1^+}(s) + 
    2\,i\pi\,H_{w_1^+, w_3^-}(r) - 2\,i\pi\,H_{w_1^+, w_2^-}(r) \nnb \\+& 
    \frac{3}{2}\,i\pi\,H_{w_3^-, w_1^+}(r) + \frac{1}{2}\,i\pi\,H_{w_3^-, w_1^+}(s) - 
    2\,i\pi\,H_{w_3^+, 0}(r) - \frac{3}{2}\,i\pi\,H_{w_3^+, w_1^-}(r) \nnb \\+& 
    \frac{1}{2}\,i\pi\,H_{w_3^+, w_1^-}(s) + i\pi\,H_{w_3^+, w_2^-}(r) - 
    i\pi\,H_{w_3^+, w_2^-}(s) - i\pi\,H_{w_3^+, w_2^+}(r) \nnb \\+& 
    i\pi\,H_{w_3^+, w_2^+}(s) - H_{w_1^+}(r)\,H_{w_3^+, w_2^+}(s) - 
    2\,i\pi\,H_{w_2^-, w_1^+}(s) - H_{w_1^-, w_1^+, w_1^+}(r) \nnb \\-& 
    H_{w_1^-, w_1^+, w_1^+}(s) + 2\,H_{w_1^+, 0, w_1^+}(r) -
    2\,H_{w_1^+, 0, w_1^+}(s) - H_{w_1^+, w_1^-, w_1^+}(r) \nnb \\-& 
    H_{w_1^+, w_1^-, w_1^+}(s) - 2\,H_{w_1^+, w_1^+, w_1^-}(r) -
    2\,H_{w_1^+, w_1^+, w_3^-}(r) + 2\,H_{w_1^+, w_1^+, w_2^-}(r) \nnb \\-& 
    2\,H_{w_1^+, w_3^-, w_1^+}(r) + H_{w_1^+, w_3^+, w_2^+}(r) + 
    2\,H_{w_1^+, w_2^-, w_1^+}(r) - \frac{3}{2}\,H_{w_3^-, w_1^+, w_1^+}(r) \nnb \\-& 
    \frac{1}{2}\,H_{w_3^-, w_1^+, w_1^+}(s) + \frac{1}{2}\,H_{w_3^+, w_1^-, w_1^+}(r) - 
    \frac{1}{2}\,H_{w_3^+, w_1^-, w_1^+}(s) + H_{w_3^+, w_1^+, w_2^+}(r) \nnb \\-& 
    H_{w_3^+, w_2^-, w_1^+}(r) + H_{w_3^+, w_2^-, w_1^+}(s) + 
    H_{w_3^+, w_2^+, w_1^+}(r) + 2\,H_{w_2^-, w_1^+, w_1^+}(s) \nnb \\+& 
    2\,i\pi\,H_{w_1^+}(r)\,\ln(2) - 2\,i\pi\,H_{w_1^+}(s)\,\ln(2) - 
    i\pi\,H_{w_3^+}(r)\,\ln(2) + i\pi\,H_{w_3^+}(s)\,\ln(2) \nnb \\-& 
    4\,H_{w_1^+, w_1^+}(r)\,\ln(2) + 4\,H_{w_1^+, w_1^+}(s)\,\ln(2)]
    + {\cal O}(\ep^4)  \, ,
\end{align}
}

\vspace{-7mm}
\allowdisplaybreaks{
\begin{align}
\tilde M_{14}(r,s)=& \ep^2\,[-2\,\pi^2 - 4\,i\pi\,H_{0}(r) - 2\,i\pi\,H_{w_1^-}(r) - 
    i\pi\,H_{w_1^-}(s) + 2\,i\pi\,H_{w_2^-}(s) - 2\,i\pi\,H_{w_2^+}(s) \nnb \\+& 
    2\,H_{w_1^+}(r)\,H_{w_2^+}(s) + H_{w_1^-, w_1^+}(s) - 
    2\,H_{w_2^-, w_1^+}(s) - 2\,i\pi\,\ln(2)] \nnb \\+& 
  \ep^3\,[-\frac{11}{3}\,i\pi^3 - 2\,\pi^2\,H_{-1}(r^2) + 12\,\pi^2\,H_{0}(r) + 
    4\,\pi^2\,H_{0}(s) + 8\,i\pi\,H_{0}(r)\,H_{0}(s) \nnb \\+& 
    4\,i\pi\,H_{0}(s)\,H_{w_1^-}(r) + 5\,\pi^2\,H_{w_1^-}(s) + 
    8\,i\pi\,H_{0}(r)\,H_{w_1^-}(s) + 2\,i\pi\,H_{w_1^-}(r)\,
     H_{w_1^-}(s) \nnb \\-& \frac{3}{2}\,\pi^2\,H_{w_1^+}(s) - 
    4\,i\pi\,H_{w_1^+}(r)\,H_{w_1^+}(s) + \pi^2\,H_{w_3^-}(s) + 
    2\,i\pi\,H_{0}(r)\,H_{w_3^-}(s) \nnb \\+& i\pi\,H_{w_1^-}(r)\,
     H_{w_3^-}(s) - \frac{3}{4}\,\pi^2\,H_{w_3^+}(s) - 
    2\,i\pi\,H_{w_1^+}(r)\,H_{w_3^+}(s) - 6\,\pi^2\,H_{w_2^-}(s) \nnb \\-& 
    8\,i\pi\,H_{0}(r)\,H_{w_2^-}(s) + 6\,\pi^2\,H_{w_2^+}(s) + 
    12\,i\pi\,H_{0}(r)\,H_{w_2^+}(s) + 4\,i\pi\,H_{w_1^-}(r)\,
     H_{w_2^+}(s)  \nnb \\-& 2\,i\pi\,H_{-1, 0}(r^2) - 2\,i\pi\,H_{-1, 1}(r^2) + 
    16\,i\pi\,H_{0, 0}(r) + 2\,i\pi\,H_{0, w_1^-}(s) \nnb \\-& 
    12\,H_{w_2^+}(s)\,H_{0, w_1^+}(r) - 4\,i\pi\,H_{0, w_2^-}(s) + 
    4\,i\pi\,H_{0, w_2^+}(s) - 4\,H_{w_1^+}(r)\,H_{0, w_2^+}(s) \nnb \\-& 
    4\,i\pi\,H_{w_1^-, 0}(r) + 2\,i\pi\,H_{w_1^-, 0}(s) - 
    6\,i\pi\,H_{w_1^-, w_1^-}(r) + 3\,i\pi\,H_{w_1^-, w_1^-}(s) \nnb \\-& 
    4\,H_{w_2^+}(s)\,H_{w_1^-, w_1^+}(r) + 2\,H_{w_1^-}(r)\,
     H_{w_1^-, w_1^+}(s) - 4\,i\pi\,H_{w_1^-, w_2^-}(s) + 
    4\,i\pi\,H_{w_1^-, w_2^+}(s) \nnb \\-& 4\,H_{w_1^+}(r)\,H_{w_1^-, w_2^+}(s) + 
    4\,H_{w_2^+}(s)\,H_{w_1^+, w_1^-}(r) + 4\,H_{w_1^+}(s)\,
     H_{w_1^+, w_1^+}(r) \nnb \\+& 2\,H_{w_3^+}(s)\,H_{w_1^+, w_1^+}(r) + 
    i\pi\,H_{w_1^+, w_1^+}(s) + \frac{1}{2}\,i\pi\,H_{w_3^-, w_1^-}(s) - 
    i\pi\,H_{w_3^-, w_2^-}(s) \nnb \\+& i\pi\,H_{w_3^-, w_2^+}(s) - 
    H_{w_1^+}(r)\,H_{w_3^-, w_2^+}(s) + \frac{1}{2}\,i\pi\,H_{w_3^+, w_1^+}(s) -
    4\,i\pi\,H_{w_2^-, 0}(s) \nnb \\-& 4\,i\pi\,H_{w_2^-, w_1^-}(s) - 
    4\,H_{w_1^-}(r)\,H_{w_2^-, w_1^+}(s) + 
    4\,i\pi\,H_{w_2^-, w_2^-}(s) - 4\,i\pi\,H_{w_2^-, w_2^+}(s) \nnb \\+& 
    4\,H_{w_1^+}(r)\,H_{w_2^-, w_2^+}(s) - 2\,H_{0, w_1^-, w_1^+}(s) + 
    4\,H_{0, w_2^-, w_1^+}(s) - 2\,H_{w_1^-, 0, w_1^+}(s) \nnb \\-& 
    3\,H_{w_1^-, w_1^-, w_1^+}(s) + 4\,H_{w_1^-, w_2^-, w_1^+}(s) - 
    H_{w_1^+, w_1^+, w_1^+}(s) - \frac{1}{2}\,H_{w_3^-, w_1^-, w_1^+}(s) \nnb \\+& 
    H_{w_3^-, w_2^-, w_1^+}(s) - \frac{1}{2}\,H_{w_3^+, w_1^+, w_1^+}(s) + 
    4\,H_{w_2^-, 0, w_1^+}(s) + 4\,H_{w_2^-, w_1^-, w_1^+}(s) \nnb \\-& 
    4\,H_{w_2^-, w_2^-, w_1^+}(s) + 2\,\pi^2\,\ln(2) - 
    2\,i\pi\,H_{-1}(r^2)\,\ln(2) + 4\,i\pi\,H_{0}(s)\,\ln(2) \nnb \\-& 
    6\,i\pi\,H_{w_1^-}(r)\,\ln(2) + 2\,i\pi\,H_{w_1^-}(s)\,\ln(2) + 
    i\pi\,H_{w_3^-}(s)\,\ln(2) + 8\,i\pi\,H_{w_2^+}(s)\,\ln(2) \nnb \\+& 
    8\,H_{w_1^+}(r)\,H_{w_2^+}(s)\,\ln(2) + 4\,H_{w_1^-, w_1^+}(s)\,
     \ln(2) - 8\,H_{w_2^-, w_1^+}(s)\,\ln(2) - 2\,i\pi\,\ln^2(2)]\nnb \\
    +& {\cal O}(\ep^4) \, .
\end{align}
}

\subsection{$M_{15}$ -- $M_{17}$}
\label{sec-M15}
The integrals in this topology only depend on one non-trivial scale ratio, and their solution can be written in terms of ordinary HPLs. The topology involves five integrals,\begin{equation}
\vec{M} = \left\{\tilde M_{15}(r,s),\tilde M_{16}(r,s),\tilde M_{17}(r,s),\big[\tilde M_{1}'(\zc)\big]^2,\tilde M_{1}(r,s)\tilde M_{1}'(\zc)\right\} \,,
\end{equation}
and the matrix $\tilde A_{15-17}(r,s)$. The result reads
 \allowdisplaybreaks{
\begin{align}
\tilde M_{15}(r,s) = & 
z_f^{-2\ep} \, \Big\{\ep^3\,[-i\pi\,H_{w_1^+,w_1^-}(s)+
H_{w_1^+,w_1^-,w_1^+}(s)-2\,i\pi\,H_{w_1^+}(s)\, \ln(2)- 7\,\zeta_3] + {\cal O}(\ep^4) \Big\} \, ,\\[0.5em]
\tilde M_{16}(r,s) = & 
z_f^{-2\ep} \,\Big\{ \ep^2\,[\frac{\pi^2}{2}+i\pi\,H_{w_1^+}(s)-H_{w_1^+,w_1^+}(s)] \nnb \\
+&
\ep^3\,[-\frac{\pi^2}{2}\,H_{w_1^-}(s)-\pi^2\,H_{w_1^+}(s)-
i\pi\,H_{w_1^-,w_1^+}(s)-2\,i\pi\,H_{w_1^+,0}(s)-
i\pi\,H_{w_1^+,w_1^-}(s)\nnb \\+& H_{w_1^-,w_1^+,w_1^+}(s)+
2\,H_{w_1^+,0,w_1^+}(s)+H_{w_1^+,w_1^-,w_1^+}(s)-\pi^2\,\ln(2)-
2\,i\pi\,H_{w_1^+}(s)\,\ln(2)\nnb \\+& \frac{21}{2}\,\zeta_3] + {\cal O}(\ep^4) \Big\} \, , 
\\[0.5em]
\tilde M_{17}(r,s) = & 
z_f^{-2\ep} \,\Big\{ \ep^2\,[i\pi\,H_{w_1^-}(s)-H_{w_1^-,w_1^+}(s)+2\,i\pi\,\ln(2)]\nnb \\+&
\ep^3\,[\frac{i\pi^3}{6}-\pi^2\,H_{w_1^-}(s)-\frac{\pi^2}{2}\,H_{w_1^+}(s)-
2\,i\pi\,H_{0,w_1^-}(s)-2\,i\pi\,H_{w_1^-,0}(s)\nnb \\-&
3\,i\pi\,H_{w_1^-,w_1^-}(s)-i\pi\,H_{w_1^+,w_1^+}(s)+
2\,H_{0,w_1^-,w_1^+}(s)+2\,H_{w_1^-,0,w_1^+}(s)\nnb \\+&
3\,H_{w_1^-,w_1^-,w_1^+}(s)+H_{w_1^+,w_1^+,w_1^+}(s)-
2\,\pi^2\,\ln(2)-4\,i\pi\,H_{0}(s)\,\ln(2)\nnb \\-& 6\,i\pi\,H_{w_1^-}(s)\,
\ln(2)-6\,i\pi\,\ln^2(2)] + {\cal O}(\ep^4) \Big\} \, .
\end{align}
}

\subsection{$M_{18}$ -- $M_{21}$}
\label{sec-M18}

This is the largest topology with eleven integrals,
\begin{multline}
\vec{M} = \left\{\tilde M_{18}(r,s),\tilde M_{19}(r,s),\tilde M_{20}(r,s),\tilde M_{21}(r,s),\tilde M_{5}(r),\big[\tilde M_{1}'(\zc)\big]^2, \tilde M_{1}'(\zc)\tilde M_{1}'(\zc=1),  \right. \\
\left. \tilde M_{1}'(\zc)\tilde M_{1}(r,s),\tilde M_{1}'(\zc=1)\tilde M_{1}(r,s),\tilde M_{4}'(\zc),\tilde M_{5}'(\zc)\right\}\,,
\end{multline}
and the matrix $\tilde A_{18-21}(r,s)$. It turns out that we need the combination $\tilde M_{18}(r,s)+\tilde M_{19}(r,s)$ up to functions of weight four. This very coefficient fills several pages and is relegated to Appendix~\ref{app:M1819ep4}. The results up to functions of weight three are
\allowdisplaybreaks{
\begin{align}
\tilde M_{18}(r,s)=& \ep^3\,[-\frac{\pi^2}{6}\,H_{w_1^-}(r) + \frac{\pi^2}{6}\,H_{w_1^-}(s) - 
    \frac{\pi^2}{12}\,H_{w_3^-}(r) + \frac{\pi^2}{12}\,H_{w_3^-}(s) - 
    i\pi\,H_{w_1^-}(r)\,H_{w_3^+}(s) \nnb\\ +& H_{w_1^-}(s)\,
     H_{w_1^-, w_1^-}(r) - H_{w_3^-}(s)\,H_{w_1^-, w_1^-}(r) + 
    i\pi\,H_{w_1^-, w_3^+}(r) - i\pi\,H_{w_1^+, w_1^-}(r) \nnb\\+& 
    i\pi\,H_{w_1^+, w_1^-}(s) - \frac{1}{2}\,i\pi\,H_{w_3^-, w_1^+}(r) + 
    \frac{1}{2}\,i\pi\,H_{w_3^-, w_1^+}(s) + \frac{1}{2}\,i\pi\,H_{w_3^+, w_1^-}(r) \nnb\\+& 
    \frac{1}{2}\,i\pi\,H_{w_3^+, w_1^-}(s) + H_{w_1^-}(r)\,
     H_{w_3^+, w_1^+}(s) - 3\,H_{w_1^-, w_1^-, w_1^-}(r) + 
    H_{w_1^-, w_1^-, w_3^-}(r) \nnb\\+& H_{w_1^-, w_3^-, w_1^-}(r) - 
    H_{w_1^-, w_3^+, w_1^+}(r) + H_{w_1^+, w_1^-, w_1^+}(r) - 
    H_{w_1^+, w_1^-, w_1^+}(s) \nnb\\+& H_{w_3^-, w_1^-, w_1^-}(r) + 
    \frac{1}{2}\,H_{w_3^-, w_1^+, w_1^+}(r) - \frac{1}{2}\,H_{w_3^-, w_1^+, w_1^+}(s) - 
    \frac{1}{2}\,H_{w_3^+, w_1^-, w_1^+}(r) \nnb\\-& \frac{1}{2}\,H_{w_3^+, w_1^-, w_1^+}(s) - 
    H_{w_3^+, w_1^+, w_1^-}(r) + 2\,H_{w_1^-}(r)\,H_{w_1^-}(s)\,
     \ln(2) - 2\,i\pi\,H_{w_1^+}(r)\,\ln(2) \nnb\\+& 2\,i\pi\,H_{w_1^+}(s)\,\ln(2) - 
    2\,H_{w_1^-}(r)\,H_{w_3^-}(s)\,\ln(2) + i\pi\,H_{w_3^+}(r)\,
     \ln(2) \nnb\\-& i\pi\,H_{w_3^+}(s)\,\ln(2) - 4\,H_{w_1^-, w_1^-}(r)\,
     \ln(2) + 2\,H_{w_1^-, w_3^-}(r)\,\ln(2) + 
    2\,H_{w_3^-, w_1^-}(r)\,\ln(2) \nnb\\-& 2\,H_{w_3^+, w_1^+}(r)\,\ln(2) + 
    2\,H_{w_3^+, w_1^+}(s)\,\ln(2) - 2\,H_{w_1^-}(r)\,\ln^2(2) + 
    2\,H_{w_1^-}(s)\,\ln^2(2) \nnb\\+& 2\,H_{w_3^-}(r)\,\ln^2(2) - 
    2\,H_{w_3^-}(s)\,\ln^2(2) - H_{w_1^-}(r)\,{\rm Li}_2(1-\zc) + 
    H_{w_1^-}(s)\,{\rm Li}_2(1-\zc) \nnb\\+& H_{w_3^-}(r)\,
     {\rm Li}_2(1-\zc) - H_{w_3^-}(s)\,{\rm Li}_2(1-\zc)] + {\cal O}(\ep^4) \, , \\[1.5em]
\tilde M_{19}(r,s) = &\ep^3\,[-\frac{\pi^2}{3}\,H_{w_1^-}(r) + \frac{\pi^2}{3}\,H_{w_1^-}(s) - 
    i\pi\,H_{w_1^-}(r)\,H_{w_1^+}(s) + \frac{\pi^2}{12}\,H_{w_3^-}(r) - 
    \frac{\pi^2}{12}\,H_{w_3^-}(s) \nnb \\+& i\pi\,H_{w_1^-}(r)\,H_{w_3^+}(s) - 
    H_{w_1^-}(s)\,H_{w_1^-, w_1^-}(r) + H_{w_3^-}(s)\,
     H_{w_1^-, w_1^-}(r) + i\pi\,H_{w_1^-, w_1^+}(s) \nnb \\-& 
    i\pi\,H_{w_1^-, w_3^+}(r) + i\pi\,H_{w_1^+, w_1^-}(r) + 
    H_{w_1^-}(r)\,H_{w_1^+, w_1^+}(s) + \frac{1}{2}\,i\pi\,H_{w_3^-, w_1^+}(r)
     \nnb \\-& \frac{1}{2}\,i\pi\,H_{w_3^-, w_1^+}(s) - \frac{1}{2}\,i\pi\,H_{w_3^+, w_1^-}(r)
     - \frac{1}{2}\,i\pi\,H_{w_3^+, w_1^-}(s) - H_{w_1^-}(r)\,
     H_{w_3^+, w_1^+}(s) \nnb \\+& 3\,H_{w_1^-, w_1^-, w_1^-}(r) - 
    H_{w_1^-, w_1^-, w_3^-}(r) - H_{w_1^-, w_1^+, w_1^+}(s) - 
    H_{w_1^-, w_3^-, w_1^-}(r) \nnb \\+& H_{w_1^-, w_3^+, w_1^+}(r) - 
    H_{w_1^+, w_1^-, w_1^+}(r) - H_{w_1^+, w_1^+, w_1^-}(r) - 
    H_{w_3^-, w_1^-, w_1^-}(r) \nnb \\-& \frac{1}{2}\,H_{w_3^-, w_1^+, w_1^+}(r) + 
    \frac{1}{2}\,H_{w_3^-, w_1^+, w_1^+}(s) + \frac{1}{2}\,H_{w_3^+, w_1^-, w_1^+}(r) + 
    \frac{1}{2}\,H_{w_3^+, w_1^-, w_1^+}(s) \nnb \\+& H_{w_3^+, w_1^+, w_1^-}(r) - 
    2\,H_{w_1^-}(r)\,H_{w_1^-}(s)\,\ln(2) + 2\,i\pi\,H_{w_1^+}(r)\,\ln(2) \nnb \\-&
    2\,i\pi\,H_{w_1^+}(s)\,\ln(2) + 2\,H_{w_1^-}(r)\,H_{w_3^-}(s)\,
     \ln(2) - i\pi\,H_{w_3^+}(r)\,\ln(2) \nnb \\+& i\pi\,H_{w_3^+}(s)\,\ln(2) +
    4\,H_{w_1^-, w_1^-}(r)\,\ln(2) - 2\,H_{w_1^-, w_3^-}(r)\,\ln(2) - 
    2\,H_{w_1^+, w_1^+}(r)\,\ln(2) \nnb \\+& 2\,H_{w_1^+, w_1^+}(s)\,\ln(2) - 
    2\,H_{w_3^-, w_1^-}(r)\,\ln(2) + 2\,H_{w_3^+, w_1^+}(r)\,\ln(2) \nnb \\-& 
    2\,H_{w_3^+, w_1^+}(s)\,\ln(2) + 2\,H_{w_1^-}(r)\,\ln^2(2) - 
    2\,H_{w_1^-}(s)\,\ln^2(2) - 2\,H_{w_3^-}(r)\,\ln^2(2) \nnb \\+& 
    2\,H_{w_3^-}(s)\,\ln^2(2) + H_{w_1^-}(r)\,{\rm Li}_2(1-\zc) - 
    H_{w_1^-}(s)\,{\rm Li}_2(1-\zc) \nnb \\-& H_{w_3^-}(r)\,
     {\rm Li}_2(1-\zc) + H_{w_3^-}(s)\,{\rm Li}_2(1-\zc)]
    + {\cal O}(\ep^4) \, , \\[1.5em]
\tilde M_{20}(r,s) = &  \ep^2\,[-i\pi\,H_{w_1^-}(r) + i\pi\,H_{w_1^-}(s) + 
    H_{w_1^-}(r)\,H_{w_1^+}(s) - H_{w_1^-, w_1^+}(s) + 
    2\,H_{w_1^+}(s)\,\ln(2)] \nnb \\+& \ep^3\,[\pi^2\,H_{w_1^-}(r) + 
    2\,i\pi\,H_{0}(s)\,H_{w_1^-}(r) - \pi^2\,H_{w_1^-}(s) + 
    3\,i\pi\,H_{w_1^-}(r)\,H_{w_1^-}(s) \nnb \\-& 2\,i\pi\,H_{w_1^-}(r)\,
     H_{w_5^-}(s) - \frac{\pi^2}{6}\,H_{w_5^+}(s) - 
    2\,i\pi\,H_{w_1^-}(r)\,H_{w_4^-}(s) +
    \frac{11}{6}\,\pi^2\,H_{w_4^+}(s) \nnb \\-& \frac{2}{3}\,\pi^2\,H_{w_1^+}(s) + 
    i\pi\,H_{w_1^-}(r)\,H_{w_3^-}(s) - \frac{\pi^2}{12}\,H_{w_3^+}(s) + 
    2\,i\pi\,H_{w_1^-}(r)\,H_{w_2^-}(s) \nnb \\-& 2\,i\pi\,H_{-1, 1}(r^2) + 
    4\,i\pi\,H_{0, w_1^-}(r) - 2\,i\pi\,H_{0, w_1^-}(s) - 
    2\,H_{w_1^-}(r)\,H_{0, w_1^+}(s) \nnb \\-& 2\,i\pi\,H_{w_1^-, 0}(s) - 
    3\,i\pi\,H_{w_1^-, w_1^-}(r) - H_{w_5^+}(s)\,
     H_{w_1^-, w_1^-}(r) - H_{w_4^+}(s)\,H_{w_1^-, w_1^-}(r) \nnb \\+& 
    2\,H_{w_1^+}(s)\,H_{w_1^-, w_1^-}(r) + H_{w_3^+}(s)\,
     H_{w_1^-, w_1^-}(r) - 3\,i\pi\,H_{w_1^-, w_1^-}(s) \nnb \\-& 
    3\,H_{w_1^-}(r)\,H_{w_1^-, w_1^+}(s) + 
    2\,i\pi\,H_{w_5^-, w_1^-}(s) + 2\,H_{w_1^-}(r)\,
     H_{w_5^-, w_1^+}(s) + 2\,i\pi\,H_{w_5^+, w_1^+}(s) \nnb \\+& 
    2\,i\pi\,H_{w_4^-, w_1^-}(s) + 2\,H_{w_1^-}(r)\,
     H_{w_4^-, w_1^+}(s) + 2\,i\pi\,H_{w_4^+, w_1^+}(s) \nnb \\-& 
    2\,H_{w_5^+}(s)\,H_{w_1^+, 0}(\sqrt{\zc}) + 
    2\,H_{w_4^+}(s)\,H_{w_1^+, 0}(\sqrt{\zc}) + 
    2\,H_{w_2^+}(s)\,H_{w_1^+, w_1^-}(r) \nnb \\-& i\pi\,H_{w_1^+, w_1^+}(s) - 
    \frac{1}{2}\,i\pi\,H_{w_3^-, w_1^-}(s) - H_{w_1^-}(r)\,
     H_{w_3^-, w_1^+}(s) - \frac{1}{2}\,i\pi\,H_{w_3^+, w_1^+}(s) \nnb \\-& 
    2\,i\pi\,H_{w_2^-, w_1^-}(s) - 2\,H_{w_1^-}(r)\,
     H_{w_2^-, w_1^+}(s) + 2\,H_{0, w_1^-, w_1^+}(s) + 
    2\,H_{w_1^-, 0, w_1^+}(s) \nnb \\+& 3\,H_{w_1^-, w_1^-, w_1^+}(s) - 
    2\,H_{w_5^-, w_1^-, w_1^+}(s) - 2\,H_{w_5^+, w_1^+, w_1^+}(s) - 
    2\,H_{w_4^-, w_1^-, w_1^+}(s) \nnb \\-& 2\,H_{w_4^+, w_1^+, w_1^+}(s) + 
    H_{w_1^+, w_1^+, w_1^+}(s) + \frac{1}{2}\,H_{w_3^-, w_1^-, w_1^+}(s) + 
    \frac{1}{2}\,H_{w_3^+, w_1^+, w_1^+}(s) \nnb \\+& 2\,H_{w_2^-, w_1^-, w_1^+}(s) - 
    2\,i\pi\,H_{-1}(r^2)\,\ln(2) - 2\,i\pi\,H_{w_1^-}(r)\,\ln(2) \nnb \\-& 
    2\,H_{w_1^-}(r)\,H_{w_5^+}(s)\,\ln(2) - 
    2\,H_{w_1^-}(r)\,H_{w_4^+}(s)\,\ln(2) + 
    4\,H_{w_1^-}(r)\,H_{w_1^+}(s)\,\ln(2) \nnb \\+& i\pi\,H_{w_3^-}(s)\,
     \ln(2) + 2\,H_{w_1^-}(r)\,H_{w_3^+}(s)\,\ln(2) + 
    4\,H_{w_1^+}(r)\,H_{w_2^+}(s)\,\ln(2) \nnb \\-& 4\,H_{0, w_1^+}(s)\,\ln(2) - 
    6\,H_{w_1^-, w_1^+}(s)\,\ln(2) + 4\,H_{w_5^-, w_1^+}(s)\,\ln(2) + 
    4\,H_{w_4^-, w_1^+}(s)\,\ln(2) \nnb \\-& 2\,H_{w_3^-, w_1^+}(s)\,\ln(2) - 
    4\,H_{w_2^-, w_1^+}(s)\,\ln(2) - 2\,H_{w_5^+}(s)\,\ln^2(2) - 
    2\,H_{w_4^+}(s)\,\ln^2(2) \nnb \\+& 4\,H_{w_1^+}(s)\,\ln^2(2) + 
    2\,H_{w_3^+}(s)\,\ln^2(2) - H_{w_5^+}(s)\,{\rm Li}_2(1-\zc) 
    - H_{w_4^+}(s)\,{\rm Li}_2(1-\zc)
    \nnb \\-& 
     H_{w_1^+}(s)\,
     {\rm Li}_2(1-\zc) + H_{w_3^+}(s)\,{\rm Li}_2(1-\zc)]
    + {\cal O}(\ep^4) \, , \\[1.5em]
\tilde M_{21}(r,s) = &  \ep^2\,[-\pi^2 - 2\,i\pi\,H_{w_1^+}(s) + 2\,H_{w_1^+, w_1^+}(s)] \nnb\\+& 
  \ep^3\,[-4\,\pi^2\,H_{0}\left(\textstyle\frac{1}{1+2\,\sqrt{\zc}}\displaystyle\right) - \pi^2\,H_{w_1^-}(r) + 
    \frac{8}{3}\,\pi^2\,H_{w_1^-}(s) + \frac{\pi^2}{3}\,H_{w_5^-}(s) \nnb\\+& 
    4\,i\pi\,H_{w_1^-}(r)\,H_{w_5^+}(s) - 
    \frac{11}{3}\,\pi^2\,H_{w_4^-}(s) + 4\,i\pi\,H_{w_1^-}(r)\,
     H_{w_4^+}(s) + 2\,\pi^2\,H_{w_1^+}(s) \nnb\\-& 
    8\,i\pi\,H_{w_1^-}(r)\,H_{w_1^+}(s) + \frac{\pi^2}{6}\,H_{w_3^-}(s) - 
    2\,i\pi\,H_{w_1^-}(r)\,H_{w_3^+}(s) - 2\,H_{w_1^-}(s)\,
     H_{w_1^-, w_1^-}(r) \nnb\\+& 2\,H_{w_5^-}(s)\,H_{w_1^-, w_1^-}(r) +
    2\,H_{w_4^-}(s)\,H_{w_1^-, w_1^-}(r) - 
    2\,H_{w_3^-}(s)\,H_{w_1^-, w_1^-}(r) \nnb\\+& 
    6\,i\pi\,H_{w_1^-, w_1^+}(s) - 4\,i\pi\,H_{w_5^-, w_1^+}(s) - 
    4\,i\pi\,H_{w_5^+, w_1^-}(s) - 4\,H_{w_1^-}(r)\,
     H_{w_5^+, w_1^+}(s) \nnb\\-& 4\,i\pi\,H_{w_4^-, w_1^+}(s) - 
    4\,i\pi\,H_{w_4^+, w_1^-}(s) - 4\,H_{w_1^-}(r)\,
     H_{w_4^+, w_1^+}(s) + 4\,i\pi\,H_{w_1^+, 0}(s) \nnb\\+& 
    4\,H_{w_5^-}(s)\,H_{w_1^+, 0}(\sqrt{\zc}) - 
    4\,H_{w_4^-}(s)\,H_{w_1^+, 0}(\sqrt{\zc}) + 
    6\,i\pi\,H_{w_1^+, w_1^-}(s) \nnb\\+& 8\,H_{w_1^-}(r)\,H_{w_1^+, w_1^+}(s) + 
    i\pi\,H_{w_3^-, w_1^+}(s) + i\pi\,H_{w_3^+, w_1^-}(s) + 
    2\,H_{w_1^-}(r)\,H_{w_3^+, w_1^+}(s) \nnb\\+& 
    16\,H_{0, w_1^+, 0}\left(\textstyle\frac{1}{1+2\,\sqrt{\zc}}\displaystyle\right) - 
    8\,H_{0, w_1^+, w_1^-}(1-2\,\sqrt{\zc}) + 
    8\,H_{0, w_1^+, w_1^-}\left(\textstyle\frac{1}{1+2\,\sqrt{\zc}}\displaystyle\right) \nnb\\+& 
    2\,H_{0, w_1^+, w_1^-}(1-2\,\zc) - 
    8\,H_{0, w_1^+, w_1^+}(1-2\,\sqrt{\zc}) + 
    8\,H_{0, w_1^+, w_1^+}\left(\textstyle\frac{1}{1+2\,\sqrt{\zc}}\displaystyle\right) \nnb\\+& 
    2\,H_{0, w_1^+, w_1^+}(1-2\,\zc) - 6\,H_{w_1^-, w_1^+, w_1^+}(s) + 
    4\,H_{w_5^-, w_1^+, w_1^+}(s) + 4\,H_{w_5^+, w_1^-, w_1^+}(s) \nnb\\+& 
    4\,H_{w_4^-, w_1^+, w_1^+}(s) + 4\,H_{w_4^+, w_1^-, w_1^+}(s) - 
    4\,H_{w_1^+, 0, w_1^+}(s) - 6\,H_{w_1^+, w_1^-, w_1^+}(s)\nnb\\-& 
    H_{w_3^-, w_1^+, w_1^+}(s) - H_{w_3^+, w_1^-, w_1^+}(s) - 
    3\,\pi^2\,\ln(2) - 4\,H_{w_1^-}(r)\,H_{w_1^-}(s)\,\ln(2) \nnb\\+& 
    4\,H_{w_1^-}(r)\,H_{w_5^-}(s)\,\ln(2) + 
    4\,H_{w_1^-}(r)\,H_{w_4^-}(s)\,\ln(2) - 
    4\,i\pi\,H_{w_1^+}(s)\,\ln(2) \nnb\\-& 4\,H_{w_1^-}(r)\,H_{w_3^-}(s)\,
     \ln(2) - 2\,i\pi\,H_{w_3^+}(s)\,\ln(2) - 
    16\,H_{0, w_1^+}(1-2\,\sqrt{\zc})\,\ln(2) \nnb\\+& 
    16\,H_{0, w_1^+}\left(\textstyle\frac{1}{1+2\,\sqrt{\zc}}\displaystyle\right)\,\ln(2) + 
    4\,H_{0, w_1^+}(1-2\,\zc)\,\ln(2) \nnb\\-& 8\,H_{w_5^+, w_1^+}(s)\,\ln(2) - 
    8\,H_{w_4^+, w_1^+}(s)\,\ln(2) + 16\,H_{w_1^+, w_1^+}(s)\,\ln(2) \nnb\\+& 
    4\,H_{w_3^+, w_1^+}(s)\,\ln(2) - 4\,H_{w_1^-}(s)\,\ln^2(2) + 
    4\,H_{w_5^-}(s)\,\ln^2(2) + 4\,H_{w_4^-}(s)\,\ln^2(2) \nnb\\-& 
    4\,H_{w_3^-}(s)\,\ln^2(2) - 2\,H_{w_1^-}(s)\,{\rm Li}_2(1-\zc) + 
    2\,H_{w_5^-}(s)\,{\rm Li}_2(1-\zc) \nnb\\+& 2\,H_{w_4^-}(s)\,
     {\rm Li}_2(1-\zc) - 2\,H_{w_3^-}(s)\,{\rm Li}_2(1-\zc) + 
    14\,\zeta_3] + {\cal O}(\ep^4) \, .
\end{align}
}

\subsection{$M_{22}$}
\label{sec-M22}

This is the only integral with five lines. However, since it is essentially a one-scale integral its result can be written in terms of ordinary HPLs. The topology consists of seven integrals,
\begin{multline}
\vec{M} =\left\{\tilde M_{22}(r,s),\tilde M_{3}(r,s),\tilde M_{4}(r,s),\tilde M_{1}'(\zc)\tilde M_{2}'(\ub),\big[\tilde M_{1}'(\zc)\big]^2, \right. \\
\left. \tilde M_{1}(r,s)\tilde M_{1}'(\zc),\tilde M_{1}(r,s)\tilde M_{2}'(\ub)\right\},
\end{multline}
and the matrix $\tilde A_{22}(r,s)$. The result reads
\begin{align}
\tilde M_{22}(r,s) = & 
z_f^{-2\ep}\,\Big\{ \ep^3 \, 
  [-\frac{i\pi^3}{2} - \pi^2 \, H_{w_1^-}(s) + \pi^2 \, H_{w_1^+}(s) - 
   2 \, i\pi \, H_{w_1^-, w_1^+}(s) + i\pi \, H_{w_1^+, w_1^-}(s)
   \nnb \\
   +&  i\pi \, H_{w_1^+, w_1^+}(s) + 2 \, H_{w_1^-, w_1^+, w_1^+}(s) - 
   H_{w_1^+, w_1^-, w_1^+}(s)  - 2 \, \pi^2 \, \ln(2) - H_{w_1^+, w_1^+, w_1^-}(s) 
   \nnb \\
   +& 2 \, i\pi \, H_{w_1^+}(s) \, \ln(2) - 
   2 \, H_{w_1^+, w_1^+}(s) \, \ln(2) + \frac{21}{2} \, \zeta_3] + {\cal O}(\ep^4) \Big\} \, .
\end{align}

\subsection{$M_{23}$ -- $M_{25}$}
\label{sec-M23}

Also this topology is quite large and we need nine integrals
\begin{multline}
\vec{M} =\left\{\tilde M_{23}(r,s_1),\tilde M_{24}(r,s_1),\tilde M_{25}(r,s_1),\tilde M_{5}(r),\big[\tilde M_{1}'(\zc)\big]^2,\tilde M_{1}'(\zc)\tilde M_{1}'(\zc=1), \right. \\
\left. \tilde M_{4}'(\zc),\tilde M_{5}'(\zc),\tilde M_{1}'(\zc)\tilde M_{1}(r=i\sqrt{3},s_1)\right\} \, ,
\end{multline}
where $r=i\sqrt{3}$ corresponds to $\zc=1$. Here we choose the set of variables $(r,s_1)$. The fact that the number of integrals is large is not the only complication of this topology. As can be seen from the matrix $\tilde A_{23-25}(r,s_1)$ in Eq.~(\ref{eq:Atilde23}), many factors appear in the differential equations which are irrational in both $r$ and $s_1$. For example,
\begin{align}
\frac{\partial \tilde M_{23}(r,s_1)}{\partial s_1} =& \frac{2\,\eps\,  \tilde M_{23}(r,s_1)  \,s_1 \! \left( 5 - s_1^2 \right)}
   {\left( 1 - s_1^2 \right) \,\left( 3 + s_1^2 \right) } - 
  \frac{\eps\, \tilde M_{24}(r,s_1)  \,\left( 3 - s_1 \right) }
   {4 (1-s_1^2) \, \textstyle{\sqrt{1 + \frac{2 (1-r^2)(1-s_1)}{(1 + s_1)^2} }}} \nnb \\
  +&\frac{\eps\, \tilde M_{25}(r,s_1)  \,\left( 3 + s_1 \right) }
   {4\left( 1 - s_1^2 \right)
     \, \textstyle{\sqrt{1+ \frac{2(1-r^2) (1 + s_1)}{(1-s_1)^2}}}} + \frac{2\, \ep \, \tilde M_{4}'(\zc)\, s_1}{1-s_1^2}\; .
\end{align}
Fortunately, we can still find a form of the differential equations which allows us to apply the formulas for iterated integrals from Section~\ref{sec-def}. There are two reasons why this is possible. First, there exist variable transformations which rationalise either of the square roots, namely
\begin{align}
t = & \frac{1-s_1}{2} + \frac{1+s_1}{2} \, \sqrt{1 + \frac{2 (1-r^2)(1-s_1)}{(1 + s_1)^2} } \quad
\Longrightarrow &\quad s_1 = & \frac{2 t^2-2 t-1+r^2}{r^2-2 t+1} \, , \label{eq:t}
\end{align}
and
\begin{align}
v = & \frac{1+s_1}{2} + \frac{1-s_1}{2} \, \sqrt{1 + \frac{2 (1-r^2)(1+s_1)}{(1 - s_1)^2} } \quad
\Longrightarrow &\quad s_1 = & -\frac{2 v^2-2 v-1+r^2}{r^2-2 v+1} \, . \label{eq:v}
\end{align}
For later convenience we also define
\begin{align}
t_0 = & e^{\frac{i\pi}{3}} \, r + e^{-\frac{i\pi}{3}} \, ,\nnb \\
v_0 = & e^{-\frac{i\pi}{3}} \, r + e^{\frac{i\pi}{3}}\, ,
\end{align}
which correspond to the limit $s_1 \to +i\sqrt{3}$ of $t$ and $v$, respectively. Second, it turns out that we only need the lowest order in the $\ep$-expansion for each of the integrals $M_{23-25}$. This ensures that $M_{24}$ appears only in combination with $t$, whereas $M_{25}$ appears only with $v$, without any admixture of the respective other variable. This does not hold at higher orders in $\ep$, which can be concluded for instance from the appearance of the logarithm $L_{15}$ in $\tilde A_{23-25}(r,s_1)$ in Eq.~(\ref{eq:Atilde23}) which contains both $t$ and $v$. Having said this, we find
\allowdisplaybreaks{
\begin{align}
\tilde M_{23}(r,s_1) = &  \ep^3\,[f^{(1)}(t) + f^{(2)}(t) + f^{(1)}(v) - f^{(2)}(v) + f^{(3)}(v) \nnb \\
 - & f^{(1)}(t_0) - f^{(2)}(t_0) - f^{(1)}(v_0) + f^{(2)}(v_0) - f^{(3)}(v_0) + (H_{w_1^-}(s_1)+2\ln(2)) \nnb \\
 \times & (-\frac{\pi^2}{12} - \frac{1}{2}\,H_{w_1^-, w_1^-}(r) - H_{w_1^-}(r)\,\ln(2) - 
    \ln^2(2) - \frac{1}{2}\,{\rm Li}_2(1-\zc))] + {\cal O}(\ep^4) \, , \\[0.5em]
\tilde M_{24}(r,s_1) = &  \ep^2\,[f^{(4)}(t) + f^{(5)}(t)] + {\cal O}(\ep^3) \, , \\[0.5em]
\tilde M_{25}(r,s_1) = &  \ep^2\,[f^{(4)}(v) - f^{(5)}(v)+ f^{(6)}(v)] + {\cal O}(\ep^3) \, ,
\end{align}
}
with
\allowdisplaybreaks{
\begin{align}
f^{(1)}(x) =& -\frac{5\pi^2}{12}\, H_{w_1^+}(x)-\frac{5\pi^2}{24}\, H_{w_3^-}(x)-
\frac{5\pi^2}{24}\, H_{w_3^+}(x)+H_{w_1^+}(x)\, H_{-1,0}(r^2) \nnb \\ + &
\frac{1}{2}\, H_{w_3^-}(x)\, H_{-1,0}(r^2)+
\frac{1}{2}\, H_{w_3^+}(x)\, H_{-1,0}(r^2)+2\, H_{w_1^+}(x)\, H_{w_1^-,0}(r) \nnb \\ + & H_{w_3^-}(x)\, 
H_{w_1^-,0}(r)+H_{w_3^+}(x)\, H_{w_1^-,0}(r)+
\frac{1}{2}\, H_{w_1^+}(x)\, H_{w_1^-,w_1^-}(r) \nnb \\ + &
\frac{1}{4}\, H_{w_3^-}(x)\, H_{w_1^-,w_1^-}(r)+
\frac{1}{4}\, H_{w_3^+}(x)\, H_{w_1^-,w_1^-}(r)-2\, H_{0}(r)\, H_{w_1^+,w_1^+}(x) \nnb \\ - & H_{w_1^-}(r)\, H_{w_1^+,w_1^+}(x)-
H_{0}(r)\, H_{w_1^+,w_3^-}(x)-H_{0}(r)\, H_{w_1^+,w_3^+}(x)-
H_{0}(r)\, H_{w_3^-,w_1^+}(x) \nnb \\ - & 
\frac{1}{2}\, H_{w_1^-}(r)\, H_{w_3^-,w_1^+}(x)-
\frac{1}{2}\, H_{0}(r)\, H_{w_3^-,w_3^-}(x)-
\frac{1}{2}\, H_{0}(r)\, H_{w_3^-,w_3^+}(x) \nnb \\ - &
H_{0}(r)\, H_{w_3^+,w_1^+}(x)-
\frac{1}{2}\, H_{w_1^-}(r)\, H_{w_3^+,w_1^+}(x)-
\frac{1}{2}\, H_{0}(r)\, H_{w_3^+,w_3^-}(x) \nnb \\ - &
\frac{1}{2}\, H_{0}(r)\, H_{w_3^+,w_3^+}(x)-H_{w_1^+,w_1^+,w_1^-}(x)+
\frac{1}{2}\, H_{w_1^+,w_1^+,w_5^-}(x)+\frac{1}{2}\, H_{w_1^+,w_1^+,w_5^+}(x) \nnb \\ + &
\frac{1}{2}\, H_{w_1^+,w_1^+,w_4^-}(x)+\frac{1}{2}\, H_{w_1^+,w_1^+,w_4^+}(x)-
\frac{1}{2}\, H_{w_1^+,w_3^-,w_1^-}(x)+\frac{1}{4}\, H_{w_1^+,w_3^-,w_5^-}(x) \nnb \\ + &
\frac{1}{4}\, H_{w_1^+,w_3^-,w_5^+}(x)+\frac{1}{4}\, H_{w_1^+,w_3^-,w_4^-}(x)+\frac{1}{4}\, H_{w_1^+,w_3^-,w_4^+}(x)-\frac{1}{2}\, H_{w_1^+,w_3^+,w_1^-}(x) \nnb \\ + & \frac{1}{4}\, H_{w_1^+,w_3^+,w_5^-}(x)+
\frac{1}{4}\, H_{w_1^+,w_3^+,w_5^+}(x)+\frac{1}{4}\, H_{w_1^+,w_3^+,w_4^-}(x)+\frac{1}{4}\, H_{w_1^+,w_3^+,w_4^+}(x) \nnb \\ - &\frac{1}{2}\, H_{w_3^-,w_1^+,w_1^-}(x)+\frac{1}{4}\, H_{w_3^-,w_1^+,w_5^-}(x)+
\frac{1}{4}\, H_{w_3^-,w_1^+,w_5^+}(x)+\frac{1}{4}\, H_{w_3^-,w_1^+,w_4^-}(x) \nnb \\ + & \frac{1}{4}\, H_{w_3^-,w_1^+,w_4^+}(x)-
\frac{1}{4}\, H_{w_3^-,w_3^-,w_1^-}(x)+
\frac{1}{8}\, H_{w_3^-,w_3^-,w_5^-}(x)+
\frac{1}{8}\, H_{w_3^-,w_3^-,w_5^+}(x) \nnb \\ + &
\frac{1}{8}\, H_{w_3^-,w_3^-,w_4^-}(x)+
\frac{1}{8}\, H_{w_3^-,w_3^-,w_4^+}(x)-
\frac{1}{4}\, H_{w_3^-,w_3^+,w_1^-}(x)+
\frac{1}{8}\, H_{w_3^-,w_3^+,w_5^-}(x) \nnb \\ + &
\frac{1}{8}\, H_{w_3^-,w_3^+,w_5^+}(x)+
\frac{1}{8}\, H_{w_3^-,w_3^+,w_4^-}(x)+
\frac{1}{8}\, H_{w_3^-,w_3^+,w_4^+}(x)-\frac{1}{2}\, H_{w_3^+,w_1^+,w_1^-}(x) \nnb \\ + & \frac{1}{4}\, H_{w_3^+,w_1^+,w_5^-}(x)+
\frac{1}{4}\, H_{w_3^+,w_1^+,w_5^+}(x)+\frac{1}{4}\, H_{w_3^+,w_1^+,w_4^-}(x)+\frac{1}{4}\, H_{w_3^+,w_1^+,w_4^+}(x) \nnb \\ - &
\frac{1}{4}\, H_{w_3^+,w_3^-,w_1^-}(x)+
\frac{1}{8}\, H_{w_3^+,w_3^-,w_5^-}(x)+
\frac{1}{8}\, H_{w_3^+,w_3^-,w_5^+}(x)+
\frac{1}{8}\, H_{w_3^+,w_3^-,w_4^-}(x) \nnb \\ + &
\frac{1}{8}\, H_{w_3^+,w_3^-,w_4^+}(x)-\frac{1}{4}\, H_{w_3^+,w_3^+,w_1^-}(x)+\frac{1}{8}\, H_{w_3^+,w_3^+,w_5^-}(x)+
\frac{1}{8}\, H_{w_3^+,w_3^+,w_5^+}(x) \nnb \\ + &
\frac{1}{8}\, H_{w_3^+,w_3^+,w_4^-}(x)+
\frac{1}{8}\, H_{w_3^+,w_3^+,w_4^+}(x)+H_{w_1^-}(r)\, H_{w_1^+}(x)\, \ln(2) \nnb \\ + & \frac{1}{2}\, H_{w_1^-}(r)\, H_{w_3^-}(x)\, \ln(2)+
\frac{1}{2}\, H_{w_1^-}(r)\, H_{w_3^+}(x)\, \ln(2)-2\, H_{w_1^+,w_1^+}(x)\, \ln(2) \nnb \\ - & H_{w_3^-,w_1^+}(x)\, \ln(2)-
H_{w_3^+,w_1^+}(x)\, \ln(2)+H_{w_1^+}(x)\, \ln^2(2)+
\frac{1}{2}\, H_{w_3^-}(x)\, \ln^2(2) \nnb \\ + & \frac{1}{2}\, H_{w_3^+}(x)\, \ln^2(2)+
\frac{1}{2}\, H_{w_1^+}(x)\, {\rm Li}_2(1-\zc)+
\frac{1}{4}\, H_{w_3^-}(x)\, {\rm Li}_2(1-\zc) \nnb \\ + &
\frac{1}{4}\, H_{w_3^+}(x)\, {\rm Li}_2(1-\zc)\, , \\[1em]
f^{(2)}(x) =& \, i\pi \, \big[-H_{-1}(r^2)\, H_{w_1^+}(x)-H_{w_1^-}(r)\, H_{w_1^+}(x)-
\frac{1}{2}\, H_{-1}(r^2)\, H_{w_3^-}(x) \nnb \\ - & \frac{1}{2}\, H_{w_1^-}(r)\, H_{w_3^-}(x)-\frac{1}{2}\, H_{-1}(r^2)\, H_{w_3^+}(x)-
\frac{1}{2}\, H_{w_1^-}(r)\, H_{w_3^+}(x)+H_{w_1^+,w_1^+}(x) \nnb \\ + & 
\frac{1}{2}\, H_{w_1^+,w_3^-}(x)+\frac{1}{2}\, H_{w_1^+,w_3^+}(x)+
\frac{1}{2}\, H_{w_3^-,w_1^+}(x)+\frac{1}{4}\, H_{w_3^-,w_3^-}(x)+
\frac{1}{4}\, H_{w_3^-,w_3^+}(x) \nnb \\ + & \frac{1}{2}\, H_{w_3^+,w_1^+}(x)+
\frac{1}{4}\, H_{w_3^+,w_3^-}(x)+\frac{1}{4}\, H_{w_3^+,w_3^+}(x) \big]\, , \\[1em]
f^{(3)}(x) =& \big[\frac{1}{4}\, H_{0}(r^4)-H_{0}(r)\big] \times \big[4\, H_{-1}(r^2)\, H_{w_1^+}(x)+
4\, H_{w_1^-}(r)\, H_{w_1^+}(x) \nnb \\ + & 2\, H_{-1}(r^2)\, H_{w_3^-}(x)+
2\, H_{w_1^-}(r)\, H_{w_3^-}(x)+2\, H_{-1}(r^2)\, 
H_{w_3^+}(x)+2\, H_{w_1^-}(r)\, H_{w_3^+}(x) \nnb \\ - &
4\, H_{w_1^+,w_1^+}(x)-2\, H_{w_1^+,w_3^-}(x)-
2\, H_{w_1^+,w_3^+}(x)-2\, H_{w_3^-,w_1^+}(x)-
H_{w_3^-,w_3^-}(x) \nnb \\ - & H_{w_3^-,w_3^+}(x)-
2\, H_{w_3^+,w_1^+}(x)-H_{w_3^+,w_3^-}(x)-
H_{w_3^+,w_3^+}(x)\big]\, , \\[1em]
f^{(4)}(x) =& \, \frac{5\, \pi^2}{3}+8\, H_{0}(r)\, H_{w_1^+}(x)+
4\, H_{w_1^-}(r)\, H_{w_1^+}(x)+4\, H_{0}(r)\, H_{w_3^-}(x)+
4\, H_{0}(r)\, H_{w_3^+}(x) \nnb \\ - & 4\, H_{-1,0}(r^2)-
8\, H_{w_1^-,0}(r)-2\, H_{w_1^-,w_1^-}(r)+4\, H_{w_1^+,w_1^-}(x)-
2\, H_{w_1^+,w_5^-}(x) \nnb \\ - & 2\, H_{w_1^+,w_5^+}(x)-
2\, H_{w_1^+,w_4^-}(x)-2\, H_{w_1^+,w_4^+}(x)+
2\, H_{w_3^-,w_1^-}(x)-H_{w_3^-,w_5^-}(x) \nnb \\ - &
H_{w_3^-,w_5^+}(x)-H_{w_3^-,w_4^-}(x)-
H_{w_3^-,w_4^+}(x)+2\, H_{w_3^+,w_1^-}(x)-
H_{w_3^+,w_5^-}(x)-H_{w_3^+,w_5^+}(x) \nnb \\ - &
H_{w_3^+,w_4^-}(x)-H_{w_3^+,w_4^+}(x)-
4\, H_{w_1^-}(r)\, \ln(2)+8\, H_{w_1^+}(x)\, \ln(2)-4\, \ln^2(2) \nnb \\ - &
2\, {\rm Li}_2(1-\zc)\, , \\[1em]
f^{(5)}(x) =&  \, i\pi \, \big[4\, H_{-1}(r^2)+4\, H_{w_1^-}(r)-4\, H_{w_1^+}(x)-
2\, H_{w_3^-}(x)-2\, H_{w_3^+}(x)\big]\, , \\[1em]
f^{(6)}(x) =& \, 2\, \big[4\, H_{0}(r)-H_{0}(r^4)\big]\, \big[2\, H_{-1}(r^2)+2\, H_{w_1^-}(r)-
2\, H_{w_1^+}(x)-H_{w_3^-}(x)-H_{w_3^+}(x)\big]\, .
\end{align}
}

\noindent
For numerical cross-checks, we also present two-fold integral representations over ordinary Feynman parameters. For the relevant coefficients of the $\ep$-expansion of $M_{23-25}(r,s_1)$, they read ($\bar x = 1-x$,~$\hat x=1+x$)
\begin{align}
\tilde M_{23}(r,s_1) = & \int\limits_0^1 \!\! dt_1 \! \int\limits_0^1 \!\! dt_2 \, \frac{\ep^3\left(s_1^2+3\right) \bar t_2 \, \ln\!\left[\frac{\left(1-s_1^2\right)\left(t_1^2 t_2 \bar t_2+\bar t_1 \zc\right)}{t_2\bar t_2\left(\left(1-2t_1\right)^2-s_1^2\right)}\right]}{t_2\bar t_2 \left(\left(s_1^2+3\right) t_1+s_1^2-1\right)+\left(1-s_1^2\right) \zc} + {\cal O}(\ep^4) \, , \nnb \\[1em]
\tilde M_{24}(r,s_1) = &  \int\limits_0^1 \!\! dt_1 \! \int\limits_0^1 \!\! dt_2 \, \frac{\ep^2\hat s_1\sqrt{1+\frac{8 \bar s_1 \zc}{\hat s_1^2}}}{\hat s_1^2 t_2 \bar t_2+2 \bar s_1\zc} \left[\frac{2 \left(\hat s_1 t_1 t_2 \bar t_2-2 \zc\right)}{t_1^2  t_2 \bar t_2  +\bar t_1 \zc}
+\frac{4 s_1 \hat s_1}{s_1^2-\left(1-2 t_1\right)^2}\right] + {\cal O}(\ep^3) \, ,\nnb \\[1em]
\tilde M_{25}(r,s_1) = & \tilde M_{24}(r,-s_1) \; . \label{eq:FP2325}
\end{align}

\subsection{$M_{26}$ and $M_{27}$}
\label{sec-M26}

This topology consists of four integrals, $\vec{M} = \big\{\tilde M_{26}(s_1),\tilde M_{27}(s_1),\tilde M_{6}',\tilde M_{7}'\big\}$, and the matrix $\tilde A_{26,27}(s_1)$. The integrals in this topology depend on a single variable and we only need functions up to weight two. The solution reads
\begin{align}
\tilde M_{26}(s_1) = &  \ep^2\,[-\frac{4\pi^2}{3} - 3\,i\pi\,H_{w_1^+}(s_1) + 3\,H_{w_1^+, w_1^+}(s_1)] + {\cal O}(\ep^3) \, , \\[0.5em]
\tilde M_{27}(s_1) = &  \ep\,[- H_{w_1^+}(s_1)+i\pi] \nnb\\+& 
  \ep^2\,[\frac{1}{2} \, H_{w_1^-, w_1^+}(s_1) + 2 \, H_{0, w_1^+}(s_1) - \frac{i\pi}{2} \, H_{w_1^-}(s_1) - 2 \, i\pi \, H_{0}(s_1) -\pi^2-i\pi\, \ln(2)] \nnb \\ +& {\cal O}(\ep^3) \, .
\end{align}

\subsection{$M_{28}$ and $M_{29}$}
\label{sec-M28}

The integrals in this topology already appeared in the two-loop calculation of the tree amplitudes~\cite{Bell:2007tv,Bell:2009nk,Beneke:2009ek}, where explicit Mellin-Barnes (MB) representations have been used for their numerical evaluation (for a convenient parameterisation cf.~also the appendix of~\cite{Bell:2008ws}). With the current techniques, we are now in the position to compute these integrals analytically.

For this topology, it will be convenient to use the variables $(r,p)$ defined in Section~\ref{sec-def}. We need seven integrals,
\begin{equation}
\vec{M} \!= \!\left\{\tilde M_{28}(r,p),\tilde M_{29}(r,p),\tilde M_{1}'(\zc)\tilde M_{3}'(\ub),\big[\tilde M_{1}'(\zc)\big]^2, \tilde M_{1}'(\zc)\tilde M_{1}'(\zc=1),  \tilde M_{4}'(\zc),\tilde M_{5}'(\zc)\right\}
\end{equation}
and the matrix $\tilde A_{28,29}(r,p)$. The integral $M_{28}$ is required up to functions of weight three, but $M_{29}$ is only needed up to weight two. The solution is again lengthy, and we introduce a short-hand notation for $p_0 = 1-2\sqrt{\zc}$. We find
\allowdisplaybreaks{
\begin{align}
\tilde M_{28}(r,p) = & \ep^3\,[f^{(7)}(p) - f^{(7)}(p_0)] + {\cal O}(\ep^4) \, , \\[0.5em]
\tilde M_{29}(r,p) = &  \ep^2\,[f^{(8)}(p) - f^{(8)}(p_0)] + {\cal O}(\ep^3) \, ,
\end{align}
}
with
\allowdisplaybreaks{
\begin{align}
f^{(7)}(x) = & -i\pi\,H_{w_1^+}(x)\,H_{w_1^+}(p_0)+
2\,H_{0}(r)\,H_{w_1^+}(x)\,H_{w_1^+}(p_0) +
H_{w_1^-}(r)\,H_{w_1^+}(x)\,H_{w_1^+}(p_0) \nnb \\ -&
\frac{\pi^2}{6}\,H_{w_3^-}(x)-
\frac{i\pi}{2}\,H_{w_1^+}(p_0)\,H_{w_3^-}(x)+
H_{0}(r)\,H_{w_1^+}(p_0)\,H_{w_3^-}(x)-\frac{\pi^2}{6}\,H_{w_1^-}(x) \nnb \\ +&
\frac{1}{2}\,H_{w_1^-}(r)\,H_{w_1^+}(p_0)\,H_{w_3^-}(x)-
\frac{i\pi}{2}\,H_{w_1^+}(x)\,H_{w_3^-}(p_0)-
H_{w_1^-}(r)\,H_{w_1^+,w_1^+}(x) \nnb \\ +&
H_{0}(r)\,H_{w_1^+}(x)\,H_{w_3^-}(p_0)-
\frac{i\pi}{4}\,H_{w_3^-}(x)\,H_{w_3^-}(p_0)+
\frac{1}{2}\,H_{0}(r)\,H_{w_3^-}(x)\,H_{w_3^-}(p_0) \nnb \\ -&
\frac{\pi^2}{6}\,H_{w_3^+}(x)-
\frac{i\pi}{2}\,H_{w_1^+}(p_0)\,H_{w_3^+}(x)+
H_{0}(r)\,H_{w_1^+}(p_0)\,H_{w_3^+}(x)-\frac{\pi^2}{2}\,H_{w_1^+}(x) \nnb \\ +&
\frac{1}{2}\,H_{w_1^-}(r)\,H_{w_1^+}(p_0)\,H_{w_3^+}(x)-
\frac{i\pi}{4}\,H_{w_3^-}(p_0)\,H_{w_3^+}(x)+
\frac{1}{2}\,H_{0}(r)\,H_{w_3^-}(p_0)\,H_{w_3^+}(x) \nnb \\ -&
\frac{i\pi}{2}\,H_{w_1^+}(x)\,H_{w_3^+}(p_0)+
H_{0}(r)\,H_{w_1^+}(x)\,H_{w_3^+}(p_0)-H_{0}(r)\,H_{w_1^+,w_3^-}(x) \nnb \\ -&
\frac{i\pi}{4}\,H_{w_3^-}(x)\,H_{w_3^+}(p_0)+
\frac{1}{2}\,H_{0}(r)\,H_{w_3^-}(x)\,H_{w_3^+}(p_0)-
\frac{i\pi}{4}\,H_{w_3^+}(x)\,H_{w_3^+}(p_0) \nnb \\ +&
\frac{1}{2}\,H_{0}(r)\,H_{w_3^+}(x)\,H_{w_3^+}(p_0)-
2\,H_{w_1^-}(x)\,H_{0,0}(\sqrt{\zc}) + H_{w_3^-}(x)\,H_{0,0}(\sqrt{\zc}) \nnb \\+&
H_{w_3^+}(x)\,H_{0,0}(\sqrt{\zc})+
\frac{1}{2}\,H_{w_3^-}(x)\,H_{w_1^-,0}(\sqrt{\zc})+
\frac{1}{2}\,H_{w_3^+}(x)\,H_{w_1^-,0}(\sqrt{\zc}) \nnb \\ -&
H_{w_1^+}(x)\,H_{w_1^+,0}(\sqrt{\zc})-\frac{1}{2}\,H_{w_3^-}(x)\,H_{w_1^+,0}(\sqrt{\zc})-
H_{w_1^-}(x)\,H_{w_1^-,0}(\sqrt{\zc}) \nnb \\ - &
\frac{1}{2}\,H_{w_3^+}(x)\,H_{w_1^+,0}(\sqrt{\zc})+
H_{w_1^+}(x)\,H_{w_1^+,w_1^-}(p_0)+
\frac{1}{2}\,H_{w_3^-}(x)\,H_{w_1^+,w_1^-}(p_0) \nnb \\ +&
\frac{1}{2}\,H_{w_3^+}(x)\,H_{w_1^+,w_1^-}(p_0)-
\frac{1}{2}\,H_{w_1^+}(x)\,H_{w_1^+,w_5^-}(p_0)-
\frac{1}{4}\,H_{w_3^-}(x)\,H_{w_1^+,w_5^-}(p_0) \nnb \\ -&
\frac{1}{4}\,H_{w_3^+}(x)\,H_{w_1^+,w_5^-}(p_0)-
\frac{1}{2}\,H_{w_1^+}(x)\,H_{w_1^+,w_5^+}(p_0)-
\frac{1}{4}\,H_{w_3^-}(x)\,H_{w_1^+,w_5^+}(p_0) \nnb \\ -&
\frac{1}{4}\,H_{w_3^+}(x)\,H_{w_1^+,w_5^+}(p_0)-
\frac{1}{2}\,H_{w_1^+}(x)\,H_{w_1^+,w_4^-}(p_0)-
\frac{1}{4}\,H_{w_3^-}(x)\,H_{w_1^+,w_4^-}(p_0) \nnb \\ -&
\frac{1}{4}\,H_{w_3^+}(x)\,H_{w_1^+,w_4^-}(p_0)-
\frac{1}{2}\,H_{w_1^+}(x)\,H_{w_1^+,w_4^+}(p_0)-
\frac{1}{4}\,H_{w_3^-}(x)\,H_{w_1^+,w_4^+}(p_0) \nnb \\ -&
\frac{1}{4}\,H_{w_3^+}(x)\,H_{w_1^+,w_4^+}(p_0)+
i\pi\,H_{w_1^+,w_1^+}(x)-2\,H_{0}(r)\,H_{w_1^+,w_1^+}(x)+
\frac{i\pi}{2}\,H_{w_1^+,w_3^-}(x) \nnb \\ +&
\frac{1}{2}\,H_{w_1^+}(x)\,H_{w_3^-,w_1^-}(p_0)+
\frac{1}{4}\,H_{w_3^-}(x)\,H_{w_3^-,w_1^-}(p_0)+
\frac{1}{4}\,H_{w_3^+}(x)\,H_{w_3^-,w_1^-}(p_0) \nnb \\ -&
\frac{1}{4}\,H_{w_1^+}(x)\,H_{w_3^-,w_5^-}(p_0)-
\frac{1}{8}\,H_{w_3^-}(x)\,H_{w_3^-,w_5^-}(p_0)-
\frac{1}{8}\,H_{w_3^+}(x)\,H_{w_3^-,w_5^-}(p_0) \nnb \\ -&
\frac{1}{4}\,H_{w_1^+}(x)\,H_{w_3^-,w_5^+}(p_0)-
\frac{1}{8}\,H_{w_3^-}(x)\,H_{w_3^-,w_5^+}(p_0)-
\frac{1}{8}\,H_{w_3^+}(x)\,H_{w_3^-,w_5^+}(p_0) \nnb \\ -&
\frac{1}{4}\,H_{w_1^+}(x)\,H_{w_3^-,w_4^-}(p_0)-
\frac{1}{8}\,H_{w_3^-}(x)\,H_{w_3^-,w_4^-}(p_0)-
\frac{1}{8}\,H_{w_3^+}(x)\,H_{w_3^-,w_4^-}(p_0) \nnb \\ -&
\frac{1}{4}\,H_{w_1^+}(x)\,H_{w_3^-,w_4^+}(p_0)-
\frac{1}{8}\,H_{w_3^-}(x)\,H_{w_3^-,w_4^+}(p_0)-
\frac{1}{8}\,H_{w_3^+}(x)\,H_{w_3^-,w_4^+}(p_0) \nnb \\ +&
\frac{i\pi}{2}\,H_{w_3^-,w_1^+}(x)-H_{0}(r)\,
H_{w_3^-,w_1^+}(x)-\frac{1}{2}\,H_{w_1^-}(r)\,H_{w_3^-,w_1^+}(x)+ 
\frac{i\pi}{4}\,H_{w_3^-,w_3^+}(x) \nnb \\ + &
\frac{i\pi}{4}\,H_{w_3^-,w_3^-}(x)-
\frac{1}{2}\,H_{0}(r)\,H_{w_3^-,w_3^-}(x)+
\frac{i\pi}{2}\,H_{w_1^+,w_3^+}(x)-H_{0}(r)\,H_{w_1^+,w_3^+}(x)  \nnb \\ - &
\frac{1}{2}\,H_{0}(r)\,H_{w_3^-,w_3^+}(x)+
\frac{1}{2}\,H_{w_1^+}(x)\,H_{w_3^+,w_1^-}(p_0)+
\frac{1}{4}\,H_{w_3^-}(x)\,H_{w_3^+,w_1^-}(p_0) \nnb \\ +&
\frac{1}{4}\,H_{w_3^+}(x)\,H_{w_3^+,w_1^-}(p_0)-
\frac{1}{4}\,H_{w_1^+}(x)\,H_{w_3^+,w_5^-}(p_0)-
\frac{1}{8}\,H_{w_3^-}(x)\,H_{w_3^+,w_5^-}(p_0) \nnb \\ -&
\frac{1}{8}\,H_{w_3^+}(x)\,H_{w_3^+,w_5^-}(p_0)-
\frac{1}{4}\,H_{w_1^+}(x)\,H_{w_3^+,w_5^+}(p_0)-
\frac{1}{8}\,H_{w_3^-}(x)\,H_{w_3^+,w_5^+}(p_0) \nnb \\ -&
\frac{1}{8}\,H_{w_3^+}(x)\,H_{w_3^+,w_5^+}(p_0)-
\frac{1}{4}\,H_{w_1^+}(x)\,H_{w_3^+,w_4^-}(p_0)-
\frac{1}{8}\,H_{w_3^-}(x)\,H_{w_3^+,w_4^-}(p_0) \nnb \\ -&
\frac{1}{8}\,H_{w_3^+}(x)\,H_{w_3^+,w_4^-}(p_0)-
\frac{1}{4}\,H_{w_1^+}(x)\,H_{w_3^+,w_4^+}(p_0)-
\frac{1}{8}\,H_{w_3^-}(x)\,H_{w_3^+,w_4^+}(p_0) \nnb \\ -&
\frac{1}{8}\,H_{w_3^+}(x)\,H_{w_3^+,w_4^+}(p_0)+
\frac{i\pi}{2}\,H_{w_3^+,w_1^+}(x)-H_{0}(r)\,H_{w_3^+,w_1^+}(x)+
\frac{i\pi}{4}\,H_{w_3^+,w_3^-}(x) \nnb \\ -&
\frac{1}{2}H_{w_1^-}(r)\,H_{w_3^+,w_1^+}(x) -
\frac{1}{2}\,H_{0}(r)\,H_{w_3^+,w_3^-}(x)+
\frac{i\pi}{4}\,H_{w_3^+,w_3^+}(x) -
\frac{1}{2}\,H_{0}(r)\,H_{w_3^+,w_3^+}(x) \nnb \\ +&
\frac{1}{2}\,H_{w_1^+,w_1^+,w_5^-}(x)+
\frac{1}{2}\,H_{w_1^+,w_1^+,w_5^+}(x)-
\frac{1}{2}\,H_{w_3^-,w_1^+,w_1^-}(x)-
             H_{w_1^+,w_1^+,w_1^-}(x)  \nnb \\ +&
\frac{1}{2}\,H_{w_1^+,w_1^+,w_4^-}(x)+
\frac{1}{2}\,H_{w_1^+,w_1^+,w_4^+}(x)-
\frac{1}{2}\,H_{w_1^+,w_3^-,w_1^-}(x)+
\frac{1}{4}\,H_{w_1^+,w_3^-,w_5^-}(x) \nnb \\ +&
\frac{1}{4}\,H_{w_1^+,w_3^-,w_5^+}(x)+
\frac{1}{4}\,H_{w_1^+,w_3^-,w_4^-}(x)+
\frac{1}{4}\,H_{w_1^+,w_3^-,w_4^+}(x)-
\frac{1}{2}\,H_{w_1^+,w_3^+,w_1^-}(x) \nnb \\ +&
\frac{1}{4}\,H_{w_1^+,w_3^+,w_5^-}(x)+
\frac{1}{4}\,H_{w_1^+,w_3^+,w_5^+}(x)+
\frac{1}{4}\,H_{w_1^+,w_3^+,w_4^-}(x)+
\frac{1}{4}\,H_{w_1^+,w_3^+,w_4^+}(x) \nnb \\ + &
\frac{1}{4}\,H_{w_3^-,w_1^+,w_5^-}(x)+
\frac{1}{4}\,H_{w_3^-,w_1^+,w_5^+}(x)+
\frac{1}{4}\,H_{w_3^-,w_1^+,w_4^-}(x)+
\frac{1}{4}\,H_{w_3^-,w_1^+,w_4^+}(x) \nnb \\ - &
\frac{1}{4}\,H_{w_3^-,w_3^-,w_1^-}(x)+
\frac{1}{8}\,H_{w_3^-,w_3^-,w_5^-}(x)+
\frac{1}{8}\,H_{w_3^-,w_3^-,w_5^+}(x)+
\frac{1}{8}\,H_{w_3^-,w_3^-,w_4^-}(x) \nnb \\ +&
\frac{1}{8}\,H_{w_3^-,w_3^-,w_4^+}(x)-
\frac{1}{4}\,H_{w_3^-,w_3^+,w_1^-}(x)+
\frac{1}{8}\,H_{w_3^-,w_3^+,w_5^-}(x)+
\frac{1}{8}\,H_{w_3^-,w_3^+,w_5^+}(x) \nnb \\ + &
\frac{1}{8}\,H_{w_3^-,w_3^+,w_4^-}(x)+
\frac{1}{8}\,H_{w_3^-,w_3^+,w_4^+}(x)-
\frac{1}{2}\,H_{w_3^+,w_1^+,w_1^-}(x)+
\frac{1}{4}\,H_{w_3^+,w_1^+,w_5^-}(x) \nnb \\ +&
\frac{1}{4}\,H_{w_3^+,w_1^+,w_5^+}(x)+
\frac{1}{4}\,H_{w_3^+,w_1^+,w_4^-}(x)+
\frac{1}{4}\,H_{w_3^+,w_1^+,w_4^+}(x)-
\frac{1}{4}\,H_{w_3^+,w_3^-,w_1^-}(x) \nnb \\ +&
\frac{1}{8}\,H_{w_3^+,w_3^-,w_5^-}(x)+
\frac{1}{8}\,H_{w_3^+,w_3^-,w_5^+}(x)+
\frac{1}{8}\,H_{w_3^+,w_3^-,w_4^-}(x)+
\frac{1}{8}\,H_{w_3^+,w_3^-,w_4^+}(x) \nnb \\ -&
\frac{1}{4}\,H_{w_3^+,w_3^+,w_1^-}(x)+
\frac{1}{8}\,H_{w_3^+,w_3^+,w_5^-}(x)+
\frac{1}{8}\,H_{w_3^+,w_3^+,w_5^+}(x)+
\frac{1}{8}\,H_{w_3^+,w_3^+,w_4^-}(x) \nnb \\ +&
\frac{1}{8}\,H_{w_3^+,w_3^+,w_4^+}(x)+
2H_{w_1^+}(x)\,H_{w_1^+}(p_0)\,\ln(2)-  H_{w_3^-,w_1^+}(x)\,
\ln(2)  - H_{w_3^+,w_1^+}(x)\,\ln(2) \nnb \\ +&
H_{w_1^+}(p_0)\,H_{w_3^-}(x)\,\ln(2)+
H_{w_1^+}(p_0)\,H_{w_3^+}(x)\,\ln(2)-
2\,H_{w_1^+,w_1^+}(x)\,\ln(2)  \, , \\[0.5em]
f^{(8)}(x) = & \, i\pi\,H_{w_1^+}(x)-
2\,H_{0}(r)\,H_{w_1^+}(x)-H_{w_1^-}(r)\,H_{w_1^+}(x)+\frac{i\pi}{2}\,H_{w_3^-}(x) - 
H_{0}(r)\,H_{w_3^-}(x) \nnb \\ + &
\frac{i\pi}{2}\,H_{w_3^+}(x) -  H_{0}(r)\,H_{w_3^+}(x) 
-H_{w_1^+,w_1^-}(x) +\frac{1}{2}\,H_{w_1^+,w_5^-}(x) + \frac{1}{2}\,H_{w_1^+,w_4^-}(x)\nnb \\ +&
\frac{1}{2}\,H_{w_1^+,w_5^+}(x)+ \frac{1}{2}\,H_{w_1^+,w_4^+}(x)
-\frac{1}{2}\,H_{w_3^-,w_1^-}(x)+
\frac{1}{4}\,H_{w_3^-,w_5^-}(x)+\frac{1}{4}\,H_{w_3^-,w_4^-}(x) \nnb \\ +&
\frac{1}{4}\,H_{w_3^-,w_5^+}(x)+
\frac{1}{4}\,H_{w_3^-,w_4^+}(x)-
\frac{1}{2}\,H_{w_3^+,w_1^-}(x)+
\frac{1}{4}\,H_{w_3^+,w_5^-}(x)+
\frac{1}{4}\,H_{w_3^+,w_4^-}(x) \nnb \\ +&
\frac{1}{4}\,H_{w_3^+,w_5^+}(x)+
\frac{1}{4}\,H_{w_3^+,w_4^+}(x) -2\,H_{w_1^+}(x)\,\ln(2) \, .
\end{align}
}


\section{Checks and Validation}
\label{sec-checks}

We performed several cross checks of the analytic results presented in the previous section. First of all, we evaluated the generalised HPLs numerically by rewriting them in terms of Goncharov polylogarithms and evaluating them both with the GiNaC-library~\cite{Bauer:2000cp,Vollinga:2004sn} and an in-house {\tt Mathematica} routine. We also derived MB representations for most of the integrals, where the {\tt AMBRE}-package~\cite{Gluza:2007rt} proved to be useful. Their numerical evaluation with the {\tt MB}-package~\cite{Czakon:2005rk}, however, turned out to be difficult due to highly oscillating integrands related to the presence of the threshold. We therefore used the MB representations to derive ordinary Feynman parameter representations, similar to the ones given in~(\ref{eq:FP2325}). Another purely numerical method is sector decomposition, where we used both the {\tt SecDec}-package~\cite{Carter:2010hi,Borowka:2012yc} as well a {\tt Mathematica}-based in-house routine. For the most complicated coefficients the numerical evaluations confirm the analytic results at the level of $10^{-4}$, and for the simpler coefficients the agreement is of several orders of magnitude better.


\section{Conclusion and Outlook}
\label{sec-conclusion}

We computed the master integrals that arise in the computation of the two-loop correction to the vertex kernel of the leading penguin amplitudes in non-leptonic $B$-decays. The calculation is complicated by the presence of two non-trivial scales ($\ub$ and $\zc=m_f^2/m_b^2$), as well as the kinematic threshold at $\ub =4\zc$. We computed the master integrals in a recently advocated canonical basis, which enabled us to derive analytic results for all master integrals in terms of generalised HPLs. The results are given up to the relevant order in the $\eps$-expansion that is needed to obtain the finite terms of the penguin amplitudes. Our calculation is the first application of a canonical basis to integrals with two different internal masses. Apart from the integral basis, we find that the choice of the kinematic variables is of utmost importance since it renders the logarithms in the matrices $\tilde A_k$ rational and therefore makes the formulas for iterated integrals applicable. 

The results of this paper form the basis to derive fully analytic expressions for the hard-scattering kernels $T_i^I$ in the factorisation formula~(\ref{factorisation}). In phenomenological applications, one has to integrate over the product of the kernels and the Gegenbauer expansion of the light-cone distribution amplitudes. The presence of the charm threshold makes the numerical evaluation of the convolutions delicate. The threshold is much easier to handle in an analytic approach, and the convolutions can now be computed to very high precision.

The integrals presented here are also relevant for other applications such as rare or radiative $B$-meson decays. For example, the two-loop QCD correction to the matrix elements of current-current operators in inclusive $\bar B \to X_s \ell^+\ell^-$ decays have to date only been computed numerically~\cite{Ghinculov:2003qd} or as expansions in the lepton-invariant mass $q^2$~\cite{Asatryan:2001zw,Greub:2008cy}. With the present results, one can now obtain completely analytical expressions for any value of $q^2$. In exclusive $\bar B \to K^{(*)} \ell^+\ell^-$ decays, one can study non-factorisable corrections to charm-loop effects. 


\section*{Acknowledgments}
We thank Sophia Borowka and Gudrun Heinrich for assistance on the program SecDec and for performing cross-checks on some of the integrals. We thank Claude Duhr, Johannes Henn, Andreas von Manteuffel and Stefan Weinzierl for useful discussions. The work of TH is supported by DFG research unit FOR 1873 ``Quark Flavour Physics and Effective Field Theories''. GB gratefully acknowledges the support of a University Research Fellowship by the Royal Society.

\appendix


\newpage
\section{Matrices $\tilde A_k$}
\label{app:Atilde}

In this appendix we list the matrices $\tilde A_k$ for the different subtopologies. To this end, we define the following logarithms,
\allowdisplaybreaks{
\begin{align}
L_1^x =& \ln(x) \; , & L_{11}^x =& \ln\left(\frac{1-\sqrt{1-r^2}+x}{1-\sqrt{1-r^2}-x}\right) \; , \nnb \\
L_2^x =& \ln(1-x^2) \; , & L_{12} =& \ln\left(\frac{2+\sqrt{1-r^2}}{2-\sqrt{1-r^2}}\right) \; , \nnb \\
L_3^x =& \ln\left(\frac{1+x}{1-x}\right) \; , & L_{13}^x =& \ln\left(\frac{x^2+3}{4}\right) \; , \nnb  \\
L_4^x =& \ln(r^2-x^2) \; , & L_{14}^x =& \ln\left(x^2-x\,(r^2+1)+1\right) \; , \nnb \\
L_5^x =& \ln\left(\frac{r+x}{r-x}\right)\; , & L_{15} =&  \ln\left((1-s_1)(1-t)-(1+s_1)(1-v)\right) \; , \nnb \\
L_6^x =& \ln\left(\left(\frac{r^2+1}{2}\right)^2-x^2\right) \; , & L_{16} =& \ln\left((1+s_1)(1+r)-2(1-t)\right) \; , \nnb \\
L_7^x =& \ln\left(\frac{r^2+2x+1}{r^2-2x+1}\right) \; , & L_{17} =& \ln\left((1-s_1)(1+r)-2(1-v)\right) \; , \nnb \\
L_8^x =& \ln\left(\left(1+\sqrt{1-r^2}\right)^2-x^2\right) \; , & L_{18} =& \ln\left(\sqrt{1-r^2}-2t+(1-s_1)(1+\frac{1}{2}\sqrt{1-r^2})\right) \, , \nnb \\
L_9^x =& \ln\left(\frac{1+\sqrt{1-r^2}+x}{1+\sqrt{1-r^2}-x}\right) \; , & L_{19} =& \ln\left(\sqrt{1-r^2}-2v+(1+s_1)(1+\frac{1}{2}\sqrt{1-r^2})\right) \, . \nnb \\
L_{10}^x =& \ln\left(\left(1-\sqrt{1-r^2}\right)^2-x^2\right) \; , & &
\end{align}}
The matrices $\tilde A_k$ now assume a compact form,
\allowdisplaybreaks{
\begin{align}
\tilde A_{3,4} = &
\left(
\begin{array}{ccc}
 -L_2^s-2 L_2^r & -L_3^s & 0 \\
 6 L_3^s & -6 L_1^s+4 L_2^s-2 L_2^r & -2 L_3^s \\
 0 & 0 & -2 L_2^r
\end{array}
\right) \; , \\[1em]
\tilde A_{5} = &
\left(
\begin{array}{ccc}
 -2 L_1^r & -2 L_3^r & 2 L_3^r \\
 0 & -2 L_2^r & 0 \\
 0 & 0 & -L_2^r
\end{array}
\right)\; , \\[1em]
\tilde A_{6,7} = &\!
\left(\!\!
\begin{array}{cccccc}
 -L_2^s-L_2^r-L_4^s & -L_3^s & \frac{L_2^s}{2}-\frac{L_4^s}{2} & 0 & 0 & 0 \\
 3 L_3^s & -4 L_1^s+3 L_2^s-L_2^r-L_4^s & -\frac{3 L_3^s}{2} & \frac{L_5^s}{2} & L_3^s & L_3^s \\
 0 & 0 & -3 L_2^r & -L_3^r & 0 & 0 \\
 0 & 0 & 6 L_3^r & 2 L_2^r-6 L_1^r & 0 & -2 L_3^r \\
 0 & 0 & 0 & 0 & L_2^s-L_2^r-L_4^s & 0 \\
 0 & 0 & 0 & 0 & 0 & -2 L_2^r
\end{array}
\!\right) \\
\tilde A_{8,9} = &
\left(
\begin{array}{cccccc}
 -L_2^s-3 L_2^r+L_4^s & L_3^s & \frac{L_3^r}{2} & 0 & 0 & 0 \\
 -3 L_3^s & -4 L_1^s+3 L_2^s-3 L_2^r+L_4^s & -\frac{L_5^s}{2} & L_3^s & L_3^s & -L_3^s \\
 0 & 0 & -2 L_1^r & 0 & -2 L_3^r & 2 L_3^r \\
 0 & 0 & 0 & L_2^s-3 L_2^r+L_4^s & 0 & L_2^r-L_2^s \\
 0 & 0 & 0 & 0 & -2 L_2^r & 0 \\
 0 & 0 & 0 & 0 & 0 & -L_2^r
\end{array}
\right)
\end{align}
}
\begin{align}
\tilde A_{10,11} = &
\left(
\begin{array}{ccc}
 -2 L_2^s-3 L_2^r+2 L_4^s & -L_3^s & \frac{L_2^s}{2}-\frac{L_2^r}{2} \\
 6 L_3^s & -6 L_1^s+3 L_2^s-3 L_2^r+2 L_4^s & -3 L_3^s \\
 0 & 0 & -L_2^s-2 L_2^r \\
 0 & 0 & 6 L_3^s \\
 0 & 0 & 0 \\
 0 & 0 & 0 \\
 0 & 0 & 0
\end{array}
\right. \nnb \\
&\hspace*{188pt} \left.
\begin{array}{cccc}
  -\frac{L_3^s}{2} & \frac{L_3^r}{2} & 0 & 0 \\
  L_4^s-L_2^r & L_5^s & 0 & L_3^s \\
  -L_3^s & 0 & 0 & 0 \\
 -6 L_1^s+4 L_2^s-2 L_2^r & 0 & -2 L_3^s & 0 \\
 0 & -2 L_1^r & -2 L_3^r & 2 L_3^r \\
 0 & 0 & -2 L_2^r & 0 \\
 0 & 0 & 0 & -L_2^r
\end{array}
\right)  \\[2em]
\tilde A_{12-14} = &
\left(
\begin{array}{ccc}
 -L_2^s-2 L_2^r+2 L_4^s-L_6^s & L_2^r-\frac{L_6^s}{2} & \frac{L_7^s}{2}-L_3^s \\
 L_2^s-2 L_4^s+L_6^s & L_2^s-L_2^r-2 L_4^s+\frac{L_6^s}{2} & -\frac{L_7^s}{2} \\
 L_3^s-L_7^s & -L_3^s-\frac{L_7^s}{2} & -2 L_1^s+2 L_2^s-L_2^r-2 L_4^s+\frac{L_6^s}{2} \\
 0 & 0 & 0  \\
 0 & 0 & 0  \\
 0 & 0 & 0  \\
 0 & 0 & 0
\end{array}
\right. \nnb \\
&\hspace*{109pt}
\left.
\begin{array}{cccc}
  -L_2^s-\frac{L_2^r}{2}+\frac{3 L_6^s}{4} & 0 & 0 & 0 \\
  \frac{L_2^s}{2}+L_2^r-\frac{3 L_6^s}{4} & \frac{L_3^r}{2} & 0 & L_3^s \\
 \frac{3 L_3^s}{2}+\frac{3 L_7^s}{4} & -L_5^s & 0 & -L_2^s-2 L_2^r+2 L_4^s \\
  -3 L_2^r & -L_3^r & 0 & 0 \\
  6 L_3^r & 2 L_2^r-6 L_1^r & -2 L_3^r & 0 \\
  0 & 0 & -2 L_2^r & 0 \\
  0 & 0 & L_3^s & -2 L_1^s+L_2^s-2 L_2^r
\end{array}
\right) 
\end{align}
\allowdisplaybreaks{
\begin{align}
\tilde A_{15-17} = &
\left(
\begin{array}{ccccc}
 -L_2^s-2 L_2^r & 0 & -L_3^s & 0 & 0 \\
 L_2^s & L_2^s-2 L_2^r & 0 & 0 & -L_3^s \\
 L_3^s & -L_3^s & -2 L_1^s+2 L_2^s-2 L_2^r & 0 & L_2^s \\
 0 & 0 & 0 & -2 L_2^r & 0 \\
 0 & 0 & 0 & L_3^s & -2 L_1^s+L_2^s-2 L_2^r
\end{array}
\right) \; ,
\end{align}
}
\begin{align}
\tilde A_{18-21} = &
\left(
\begin{array}{cc}
 -L_2^s-2 L_2^r+2 L_4^s-L_6^s & 0 \\
 L_6^s-L_2^s & -2 L_2^s-L_2^r+2 L_4^s \\
 -3 L_3^s-L_7^s+2 L_9^s+2 L_{11}^s & -2 L_3^s+2 L_9^s+2 L_{11}^s \\
 -10 L_2^s+4 L_4^s-2 L_6^s+4 L_8^s+4 L_{10}^s & -12 L_2^s+4 L_4^s+4 L_8^s+4 L_{10}^s \\
 0 & 0 \\
 0 & 0 \\
 0 & 0 \\
 0 & 0 \\
 0 & 0 \\
 0 & 0 \\
 0 & 0 
\end{array}
\right.
 \nnb \\
& \hspace*{-49pt}
\left.
\begin{array}{cccc}
  L_3^s+\frac{L_7^s}{2} & \frac{L_6^s}{4}-\frac{L_2^r}{2} & L_3^r & 0 \\
  -\frac{L_7^s}{2} & \frac{L_2^s}{2}-\frac{L_6^s}{4} & -\frac{L_3^r}{2} & 0 \\
 -2 L_1^s+2 L_2^s-L_2^r+2 L_4^s+\frac{L_6^s}{2}-2 L_8^s-2 L_{10}^s & \frac{L_3^s}{2}+\frac{L_7^s}{4}-L_9^s-L_{11}^s & -L_5^s & L_3^s \\
  4 L_3^s+L_7^s-4 L_9^s-4 L_{11}^s & 3 L_2^s-L_2^r+\frac{L_6^s}{2}-2 L_8^s-2 L_{10}^s & 0 & 0 \\
  0 & 0 & -2 L_1^r & -2 L_3^r \\
  0 & 0 & 0 & -2 L_2^r \\
  0 & 0 & 0 & 0 \\
  0 & 0 & 0 & L_3^s \\
  0 & 0 & 0 & 0 \\
  0 & 0 & 0 & \frac{L_2^r}{2}-\frac{L_{13}^r}{2} \\
  0 & 0 & 0 & \frac{L_{12}}{2}
\end{array}
\right.
 \nnb \\
& \hspace*{-43pt}
\left.
\begin{array}{ccccc}
  0 & \frac{L_7^s}{2}-L_3^s & L_3^s-\frac{L_7^s}{2} & 2 L_2^s+2 L_2^r-2 L_6^s & 0 \\
  0 & L_3^s-\frac{L_7^s}{2} & \frac{L_7^s}{2}-L_3^s & -2 L_2^s-2 L_2^r+2 L_6^s & 0 \\
  -L_3^s & \frac{L_6^s}{2}-L_2^r & L_2^s-\frac{L_6^s}{2} & 2 L_3^s-2 L_7^s+2 L_9^s+2 L_{11}^s & 2 L_{11}^s-2 L_9^s \\
  0 & 2 L_3^s+L_7^s & -L_7^s & -4 L_2^s-4 L_6^s+4 L_8^s+4 L_{10}^s & 4 L_{10}^s-4 L_8^s \\
  2 L_3^r & 0 & 0 & 0 & 0 \\
  0 & 0 & 0 & 0 & 0 \\
  -L_2^r & 0 & 0 & 0 & 0 \\
  0 & -2 L_1^s+L_2^s-2 L_2^r & 0 & 0 & 0 \\
  L_3^s & 0 & -2 L_1^s+L_2^s-L_2^r & 0 & 0 \\
  \frac{L_{13}^r}{2}-\frac{L_2^r}{2} & 0 & 0 & -3 L_{13}^r & -L_{12} \\
  -\frac{L_{12}}{2} & 0 & 0 & 3 L_{12} & L_{13}^r-2 L_2^r
\end{array}
\right) 
\end{align}

\allowdisplaybreaks{
\begin{align}
\tilde A_{22} = &
\left(
\begin{array}{ccccccc}
 L_2^s & -L_2^s & 0 & 0 & 0 & -L_3^s & -L_3^s \\
 0 & -L_2^s & -L_3^s & 0 & 0 & 0 & 0 \\
 0 & 6 L_3^s & 4 L_2^s-6 L_1^s & 0 & -2 L_3^s & 0 & 0 \\
 0 & 0 & 0 & L_2^s & 0 & 0 & 0 \\
 0 & 0 & 0 & 0 & 0 & 0 & 0 \\
 0 & 0 & 0 & 0 & L_3^s & L_2^s-2 L_1^s & 0 \\
 0 & 0 & 0 & L_3^s & 0 & 0 & 2 L_2^s-2 L_1^s
\end{array}
\right) - 2 L_2^r \, \mathds{1}_{7 \times 7} \, ,
\end{align}
}
\allowdisplaybreaks{
\begin{align}
\tilde A_{23-25} = &
\left(
\begin{array}{cc}
-L_2^{s_1}-3 L_2^r+2 L_{13}^{s_1} & -\frac{L_3^t}{4}-\frac{L_2^r}{4}+\frac{L_6^t}{8}-\frac{L_7^t}{8} \\
 12 L_3^t+12 L_2^r-6 L_6^t+6 L_7^t  & -L_1^{s_1}+\frac{L_2^{s_1}}{2}-\frac{5 L_3^{s_1}}{2}+L_2^r+L_{13}^{s_1}+2 L_6^t-2 L_7^t-4 L_{14}^t \\
 12 L_3^v+12 L_2^r-6 L_6^v+6 L_7^v  & -L_1^{s_1}-L_2^r-L_{13}^{s_1}+2 L_{15}  \\
 0 & 0  \\
 0 & 0  \\
 0 & 0  \\
 0 & 0  \\
 0 & 0  \\
 0 & 0 
\end{array} \right. \nnb \\
& \hspace*{-43pt}
\left.
\begin{array}{ccc}
 -\frac{L_3^v}{4}-\frac{L_2^r}{4}+\frac{L_6^v}{8}-\frac{L_7^v}{8} & -\frac{L_3^r}{2} & 0 \\
 -L_1^{s_1}-L_2^r-L_{13}^{s_1}+2 L_{15}  & 2 L_2^r+2 L_{13}^{s_1}-4 L_{16}  & 4 L_3^t+2 L_2^r-2 L_{13}^r \\
 -L_1^{s_1}+\frac{L_2^{s_1}}{2}+\frac{5 L_3^{s_1}}{2}+L_2^r+L_{13}^{s_1}+2 L_6^v-2 L_7^v-4 L_{14}^v  & 2 L_2^r+2 L_{13}^{s_1}-4 L_{17}  & 4 L_3^v+2 L_2^r-2 L_{13}^r  \\
  0 & -2 L_1^r & -2 L_3^r \\
  0 & 0 & -2 L_2^r  \\
  0 & 0 & 0 \\
  0 & 0 & \frac{L_2^r}{2}-\frac{L_{13}^r}{2} \\
  0 & 0 & \frac{L_{12}}{2} \\
  0 & 0 & 0
\end{array} \right. \nnb \\
& \hspace*{-15pt}
\left.
\begin{array}{ccc}
  0 & -L_2^{s_1} & 0\\
 -4 L_3^t-2 L_2^r+2 L_{13}^r &  12 L_3^t+6 L_6^t-6 L_7^t-12 L_{13}^r &  -4 L_2^{s_1}-4 L_3^{s_1}-4 L_{13}^r+8 L_{18} \\
  -4 L_3^v-2 L_2^r+2 L_{13}^r  & 12 L_3^v+6 L_6^v-6 L_7^v-12 L_{13}^r  &  -4 L_2^{s_1}+4 L_3^{s_1}-4 L_{13}^r+8 L_{19} \\
  2 L_3^r & 0 & 0 \\
   0 & 0 & 0 \\
   -L_2^r & 0 & 0 \\
   \frac{L_{13}^r}{2}-\frac{L_2^r}{2} & -3 L_{13}^r & -L_{12} \\
   -\frac{L_{12}}{2} & 3 L_{12} & L_{13}^r-2 L_2^r \\
  L_3^{s_1} & 0 & 0 
\end{array} \right. \nnb \\
& \hspace*{193pt}
\left.
\begin{array}{c}
   0\\
   4 L_3^t+4 L_2^r-2 L_6^t+2 L_7^t \\
  -4 L_3^v-4 L_2^r+2 L_6^v-2 L_7^v  \\
  0 \\
   0 \\
   0 \\
    0 \\
    0 \\
   -2 L_1^{s_1}+L_2^{s_1}-L_2^r
\end{array} \right)\label{eq:Atilde23} \; , \\[4em]
\tilde A_{26,27} = &
\left(
\begin{array}{cccc}
 2 L_{13}^{s_1}-\frac{5 L_2^{s_1}}{2} & -3 L_3^{s_1} & \frac{L_2^{s_1}}{4} & -L_2^{s_1} \\
 -\frac{9 L_3^{s_1}}{4} & \frac{L_2^{s_1}}{2}-2 L_1^{s_1} & \frac{5 L_3^{s_1}}{8} & \frac{3 L_3^{s_1}}{2} \\
 0 & 0 & 0 & 0 \\
 0 & 0 & 0 & 0
\end{array}
\right)
 \; ,
\end{align}
}
\allowdisplaybreaks{
\begin{align}
\tilde A_{28,29} = &\left(
\begin{array}{c}
 -L_2^p-3 L_2^r-\frac{L_6^p}{2}+\frac{L_7^p}{2}+L_8^p-L_9^p+L_{10}^p-L_{11}^p \\
 -3 L_3^p-3 L_2^r+\frac{3 L_6^p}{2}-\frac{3 L_7^p}{2}  \\
 0 \\
 0 \\
 0 \\
 0 \\
 0 
\end{array}
\right. \nnb \\
& \hspace*{-70pt}
\left.
\begin{array}{cc}
   L_3^p+L_2^r-\frac{L_6^p}{2}+\frac{L_7^p}{2} & 0 \\
   3 L_2^p+L_2^r+\frac{3 L_6^p}{2}-\frac{3
   L_7^p}{2}-L_8^p+L_9^p-L_{10}^p+L_{11}^p-4 L_{14}^p & -L_3^p-L_2^r+\frac{L_6^p}{2}-\frac{L_7^p}{2} \\
  0 & L_2^p-L_2^r+\frac{L_6^p}{2}-\frac{L_7^p}{2}-L_8^p+L_9^p-L_{10}^p+L_{11}^p \\
  0 & 0  \\
  0 & 0  \\
  0 & 0  \\
  0 & 0 
\end{array}
\right. \nnb \\
& \hspace*{-35pt}
\left.
\begin{array}{ccc}
    0 & 0 & -L_2^p+\frac{L_6^p}{2}-\frac{L_7^p}{2}  \\
   -L_3^p-\frac{L_2^r}{2}-\frac{L_{12}}{2}+\frac{L_{13}^r}{2} &
   L_3^p+\frac{L_2^r}{2}+\frac{L_{12}}{2}-\frac{L_{13}^r}{2} & -3 L_3^p-\frac{3 L_6^p}{2}+\frac{3 L_7^p}{2}-3
   L_{12}+3 L_{13}^r \\
  0 & -L_2^p+\frac{L_8^p}{2}-\frac{L_9^p}{2}+\frac{L_{10}^p}{2}-\frac{L_{11}^p}{2} & 0  \\
   -2 L_2^r & 0 & 0  \\
   0 & -L_2^r & 0  \\
   \frac{L_2^r}{2}-\frac{L_{13}^r}{2} & \frac{L_{13}^r}{2}-\frac{L_2^r}{2} & -3 L_{13}^r \\
   \frac{L_{12}}{2} & -\frac{L_{12}}{2} & 3 L_{12}
\end{array}
\right. \nnb \\
& \hspace*{70pt}
\left.
\begin{array}{c}
    L_3^p+L_2^r-\frac{L_6^p}{2}+\frac{L_7^p}{2} \\
   3 L_2^p+3 L_2^r+\frac{3 L_6^p}{2}-\frac{3 L_7^p}{2}-2 L_8^p+2 L_9^p+L_{12}-L_{13}^r-4 L_{14}^p \\
   0 \\
   0 \\
   0 \\
   -L_{12} \\
   L_{13}^r-2 L_2^r
\end{array}
\right) \; .
\end{align}
}

\section{Auxiliary Integrals}
\label{app:auxints}

Here we collect the results of the integrals that are already known from previous calculations, but which appear as subtopologies of the master integrals discussed in the main text and are needed in order to make the system of differential equations complete. In terms of the integrals defined in Fig.~\ref{fig:auxintegrals}, they read
\begin{align} 
M_1'(\zc) &= \ep \, I_1'(z_f) \,, &
M_5'(\zc) &= \ep^2 \,\sqrt{\zc} \, 
\big( I_5'(z_f) + 2 I_4'(z_f) \big)\,, 
\label{eq:M1p} \\[0.6em]
M_2'(x) &= \ep\, x \, I_2'(x) \,, &
M_6' &= \ep^2 \, 
\big( I_6' + 2 I_7' \big)\,, 
\\[0.6em]
M_3'(x) &= \ep \, x \, I_3'(x) \,, &
M_7' &=\ep^2 \, I_8' \,.
\\[0.6em]
M_4'(\zc) &= \ep^2 \,I_4'(\zc) \,,
\label{eq:M4p}
\end{align}
Normalizing these integrals according to the definition in \eqref{eq:mastergeneric}, the results become
\begin{align}
\tilde M_{1}'(\zc) = & \; \zc^{-\ep} \;
\Gamma(1-\ep) \, \Gamma(1+\ep)  \, , \\[0.6em]
\tilde M_{2}'(x) = & 
- e^{i \pi  \ep} \, x^{-\ep} \, \frac{\Gamma^3(1-\ep) \Gamma(1+\ep)}{\Gamma(1-2\ep)} 
\, , \\[0.6em]
\tilde M_{3}'(x) = & 
- \frac{\ep \, x}{1-\ep} \,\Gamma(1-\ep) \, \Gamma(1+\ep)  \; \pFq{2}{1}{1,1+\ep}{2-\ep}{x} \nnb \\[0.6em]
          = & \; 
          \ep \, \ln (1-x) -\ep^2 \, [{\rm Li}_2(x) + \ln^2(1-x)] 
	  + \ep^3 [-2 \, {\rm Li}_3(1-x)-{\rm Li}_3(x) 
\nnb \\[0.6em]
	  + & \;\frac{2}{3} \, \ln^3(1-x)-\ln (x)\ln^2(1-x)
	  +  \frac{1}{2} \, \pi ^2 \ln(1-x)+2 \zeta_3] + {\cal O}(\ep^4)\, , \\[0.6em]
\tilde  M_{4}'(\zc) = & \;
\ep^2\,[-\frac{\pi^2}{12} - \frac{1}{2}\,H_{w_1^-, w_1^-}(r) - H_{w_1^-}(r)\,\ln(2) - 
    \ln^2(2) - \frac{1}{2}\,{\rm Li}_2(1-\zc)] 
    \nnb\\[0.6em]
    + & \; 
  \ep^3\,[\frac{\pi^2}{2}\,H_{w_1^+}(\sqrt{\zc}) - \frac{3}{2}\,H_{w_1^-, w_1^-, w_1^-}(r)
    + H_{w_1^+, w_1^+, 0}(\sqrt{\zc}) - 3\,H_{w_1^-, w_1^-}(r)\,\ln(2) 
\nnb \\[0.6em]
    - & \;  
    3\,H_{w_1^-}(r)\,\ln^2(2) - 2\,\ln^3(2) + \frac{\pi^2}{4}\,\ln(1 - \zc) + 
    \frac{3}{2}\,{\rm Li}_3(1-\zc) - \frac{7}{2}\,\zeta_3] + {\cal O}(\ep^4) \, , 
 \\[0.6em]
\tilde  M_{5}'(\zc) = & \;
 \ep^2\,[-\frac{\pi^2}{2} - H_{w_1^+, 0}(\sqrt{\zc})] +
 \ep^3\,[2\,\pi^2\,H_{0}(\sqrt{\zc}) + \frac{\pi^2}{2}\,H_{w_1^-}(\sqrt{\zc}) - 
   \frac{\pi^2}{2}\,H_{w_1^+}(\sqrt{\zc}) 
   \nnb \\[0.6em]
   +& \; 4\,H_{0, w_1^+, 0}(\sqrt{\zc}) +
   H_{w_1^-, w_1^+, 0}(\sqrt{\zc}) - 3\,H_{w_1^+, w_1^-, 0}(\sqrt{\zc}) + 
   4\,\pi^2\,\ln(2)]+ {\cal O}(\ep^4) \, , \\[0.6em]
\tilde  M_{6}' = & - \Gamma^3(1-\ep) \, \Gamma(1+\ep) \, \Gamma(1+2 \ep) \, , \\[0.6em]
\tilde  M_{7}' = & - \frac{\Gamma(1-4 \ep) \, \Gamma^4(1-\ep) \, \Gamma (1+\ep) \, \Gamma (1+2 \ep)}{4 \, 
   \Gamma(1-3 \ep) \, \Gamma (1-2\ep)}\, . 
\end{align}

\FIGURE[t]{
\includegraphics[width=1\textwidth]{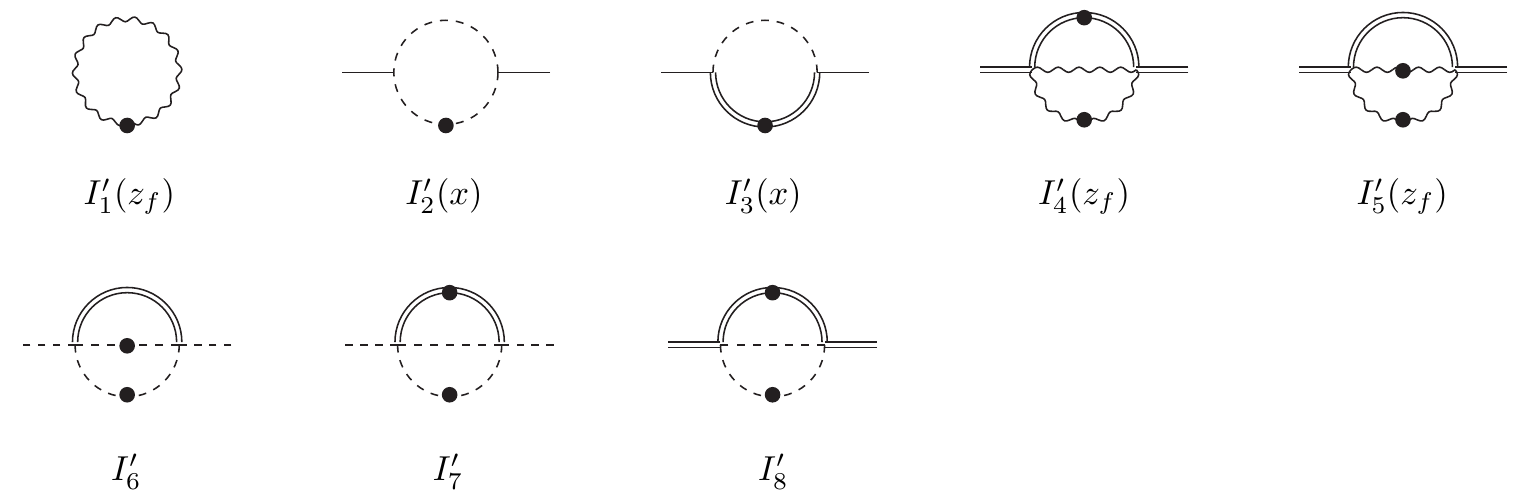}
\caption{Integrals required to define the auxiliary integrals in \eqref{eq:M1p}--\eqref{eq:M4p}. The notation has been introduced in the caption of Fig.~\ref{fig:integrals}.}\label{fig:auxintegrals}}

\section{$\tilde M_{18} + \tilde M_{19}$ to ${\cal O}(\ep^4)$}
\label{app:M1819ep4}

Here we present the result of $\tilde M_{18}(r,s) + \tilde M_{19}(r,s)$ to order ${\cal O}(\ep^4)$. This result is needed in the final result of the QCD amplitude but due to its length was relegated to this appendix.

\allowdisplaybreaks{
\begin{eqnarray}
&&\tilde M_{18}(r,s)+ \tilde M_{19}(r,s)_{\big|\ep^4}= - 2\,\pi^2\,H_{0}\left(\textstyle\frac{1}{1+2\,\sqrt{\zc}}\displaystyle\right)\,
     H_{w_1^-}(r) + 2\,\pi^2\,H_{0}\left(\textstyle\frac{1}{1+2\,\sqrt{\zc}}\displaystyle\right)\,
     H_{w_1^-}(s) \nnb \\&&- \frac{\pi^2}{2}\,H_{w_1^-}(r)\,H_{w_1^-}(s) + 
    \pi^2\,H_{w_1^-}(r)\,H_{w_1^+}(s) + \pi^2\,H_{w_1^-}(r)\,
     H_{w_2^-}(s) + 2\,i\pi\,H_{w_1^+}(r)\,H_{-1, 1}(r^2)\nnb \\&& - 
    2\,i\pi\,H_{w_1^+}(s)\,H_{-1, 1}(r^2) + 4\,i\pi\,H_{w_1^+}(s)\,
     H_{0, w_1^-}(r) + \frac{4}{3}\,\pi^2\,H_{w_1^-, w_1^-}(r) \nnb \\&&- 
    3\,i\pi\,H_{w_1^+}(s)\,H_{w_1^-, w_1^-}(r) - 
    \frac{\pi^2}{3}\,H_{w_1^-, w_1^-}(s) + H_{w_1^-, w_1^-}(r)\,
     H_{w_1^-, w_1^-}(s) + \frac{\pi^2}{6}\,H_{w_1^-, w_5^-}(r) \nnb \\&& - 
    \frac{\pi^2}{6}\,H_{w_1^-, w_5^-}(s)- H_{w_1^-, w_1^-}(r)\,
     H_{w_1^-, w_5^-}(s) - 2\,i\pi\,H_{w_1^-}(r)\,
     H_{w_1^-, w_5^+}(s) - \frac{11}{6}\,\pi^2\,H_{w_1^-, w_4^-}(r) \nnb \\&&+ 
    \frac{11}{6}\,\pi^2\,H_{w_1^-, w_4^-}(s) - H_{w_1^-, w_1^-}(r)\,
     H_{w_1^-, w_4^-}(s) - 2\,i\pi\,H_{w_1^-}(r)\,
     H_{w_1^-, w_4^+}(s) - \pi^2\,H_{w_1^-, w_1^+}(s) \nnb \\&&+ 
    2\,i\pi\,H_{w_1^-}(r)\,H_{w_1^-, w_1^+}(s) + 
    \frac{\pi^2}{12}\,H_{w_1^-, w_3^-}(r) - \frac{\pi^2}{12}\,H_{w_1^-, w_3^-}(s)
    + H_{w_1^-, w_1^-}(r)\,H_{w_1^-, w_3^-}(s) \nnb \\&&+ 
    i\pi\,H_{w_1^-}(r)\,H_{w_1^-, w_3^+}(s) - 
    \pi^2\,H_{w_1^-, w_2^-}(r) + 2\,i\pi\,H_{w_1^-}(r)\,H_{w_1^+, 0}(s) \nnb \\&&+ 
    2\,H_{w_1^-, w_5^-}(r)\,H_{w_1^+, 0}(\sqrt{\zc}) - 
    2\,H_{w_1^-, w_5^-}(s)\,H_{w_1^+, 0}(\sqrt{\zc}) - 
    2\,H_{w_1^-, w_4^-}(r)\,H_{w_1^+, 0}(\sqrt{\zc}) \nnb \\&&+ 
    2\,H_{w_1^-, w_4^-}(s)\,H_{w_1^+, 0}(\sqrt{\zc}) - 
    \pi^2\,H_{w_1^+, w_1^-}(s) + 3\,i\pi\,H_{w_1^-}(r)\,
     H_{w_1^+, w_1^-}(s) \nnb \\&&- 2\,i\pi\,H_{w_1^-}(r)\,
     H_{w_1^+, w_5^-}(s) + \frac{\pi^2}{6}\,H_{w_1^+, w_5^+}(r) + 
    2\,H_{w_1^+, 0}(\sqrt{\zc})\,H_{w_1^+, w_5^+}(r) - 
    \frac{\pi^2}{6}\,H_{w_1^+, w_5^+}(s) \nnb \\&&- H_{w_1^-, w_1^-}(r)\,
     H_{w_1^+, w_5^+}(s) - 2\,H_{w_1^+, 0}(\sqrt{\zc})\,
     H_{w_1^+, w_5^+}(s) - 2\,i\pi\,H_{w_1^-}(r)\,
     H_{w_1^+, w_4^-}(s) \nnb \\&&- \frac{11}{6}\,\pi^2\,H_{w_1^+, w_4^+}(r) - 
    2\,H_{w_1^+, 0}(\sqrt{\zc})\,H_{w_1^+, w_4^+}(r) + 
    \frac{11}{6}\,\pi^2\,H_{w_1^+, w_4^+}(s) \nnb \\&&- H_{w_1^-, w_1^-}(r)\,
     H_{w_1^+, w_4^+}(s) + 2\,H_{w_1^+, 0}(\sqrt{\zc})\,
     H_{w_1^+, w_4^+}(s) + \frac{2}{3}\,\pi^2\,H_{w_1^+, w_1^+}(r) - 
    \frac{2}{3}\,\pi^2\,H_{w_1^+, w_1^+}(s) \nnb \\&&+ 2\,H_{w_1^-, w_1^-}(r)\,
     H_{w_1^+, w_1^+}(s) + i\pi\,H_{w_1^-}(r)\,H_{w_1^+, w_3^-}(s) + 
    \frac{\pi^2}{12}\,H_{w_1^+, w_3^+}(r) - \frac{\pi^2}{12}\,H_{w_1^+, w_3^+}(s) \nnb \\&&+ 
    H_{w_1^-, w_1^-}(r)\,H_{w_1^+, w_3^+}(s) +
    2\,i\pi\,H_{w_1^-}(r)\,H_{w_1^+, w_2^-}(s) + 
    2\,H_{w_1^+, w_1^-}(r)\,H_{w_1^+, w_2^+}(s) \nnb \\&&- 
    \pi^2\,H_{w_2^-, w_1^-}(s) + 2\,i\pi\,H_{w_1^-}(r)\,
     H_{w_2^-, w_1^+}(s) - 4\,i\pi\,H_{0, w_1^-, w_1^+}(r) \nnb \\&&+ 
    8\,H_{w_1^-}(r)\,H_{0, w_1^+, 0}\left(\textstyle\frac{1}{1+2\,\sqrt{\zc}}\displaystyle\right) - 
    8\,H_{w_1^-}(s)\,H_{0, w_1^+, 0}\left(\textstyle\frac{1}{1+2\,\sqrt{\zc}}\displaystyle\right) - 
    4\,i\pi\,H_{0, w_1^+, w_1^-}(r) \nnb \\&&- 4\,H_{w_1^-}(r)\,
     H_{0, w_1^+, w_1^-}(1-2\,\sqrt{\zc}) + 4\,H_{w_1^-}(s)\,
     H_{0, w_1^+, w_1^-}(1-2\,\sqrt{\zc}) \nnb \\&&+ 4\,H_{w_1^-}(r)\,
     H_{0, w_1^+, w_1^-}\left(\textstyle\frac{1}{1+2\,\sqrt{\zc}}\displaystyle\right) - 
    4\,H_{w_1^-}(s)\,H_{0, w_1^+, w_1^-}\left(\textstyle\frac{1}{1+2\,\sqrt{\zc}}\displaystyle\right) \nnb \\&&+ 
    H_{w_1^-}(r)\,H_{0, w_1^+, w_1^-}(1-2\,\zc) - 
    H_{w_1^-}(s)\,H_{0, w_1^+, w_1^-}(1-2\,\zc) \nnb \\&&- 
    4\,H_{w_1^-}(r)\,H_{0, w_1^+, w_1^+}(1-2\,\sqrt{\zc}) + 
    4\,H_{w_1^-}(s)\,H_{0, w_1^+, w_1^+}(1-2\,\sqrt{\zc}) \nnb \\&&+ 
    4\,H_{w_1^-}(r)\,H_{0, w_1^+, w_1^+}\left(\textstyle\frac{1}{1+2\,\sqrt{\zc}}\displaystyle\right) - 
    4\,H_{w_1^-}(s)\,H_{0, w_1^+, w_1^+}\left(\textstyle\frac{1}{1+2\,\sqrt{\zc}}\displaystyle\right) \nnb \\&&+ 
    H_{w_1^-}(r)\,H_{0, w_1^+, w_1^+}(1-2\,\zc) - 
    H_{w_1^-}(s)\,H_{0, w_1^+, w_1^+}(1-2\,\zc) + 
    4\,i\pi\,H_{w_1^-, w_1^-, w_5^+}(r) \nnb \\&&+ 
    4\,i\pi\,H_{w_1^-, w_1^-, w_4^+}(r) - 
    i\pi\,H_{w_1^-, w_1^-, w_1^+}(s) - 
    2\,i\pi\,H_{w_1^-, w_1^-, w_3^+}(r) - 
    2\,i\pi\,H_{w_1^-, w_5^-, w_1^+}(r) \nnb \\&&+ 
    2\,i\pi\,H_{w_1^-, w_5^-, w_1^+}(s) +
    2\,i\pi\,H_{w_1^-, w_5^+, w_1^-}(s) + 2\,H_{w_1^-}(r)\,
     H_{w_1^-, w_5^+, w_1^+}(s) - 2\,i\pi\,H_{w_1^-, w_4^-, w_1^+}(r)\nnb \\&& 
    + 2\,i\pi\,H_{w_1^-, w_4^-, w_1^+}(s) + 
    2\,i\pi\,H_{w_1^-, w_4^+, w_1^-}(s) + 2\,H_{w_1^-}(r)\,
     H_{w_1^-, w_4^+, w_1^+}(s) - 2\,i\pi\,H_{w_1^-, w_1^+, 0}(s)\nnb \\&& - 
    i\pi\,H_{w_1^-, w_1^+, w_1^-}(r) - i\pi\,H_{w_1^-, w_1^+, w_1^-}(s) + 
    2\,i\pi\,H_{w_1^-, w_1^+, w_5^-}(r) + 
    2\,i\pi\,H_{w_1^-, w_1^+, w_4^-}(r) \nnb \\&&- 2\,H_{w_1^-}(r)\,
     H_{w_1^-, w_1^+, w_1^+}(s) - i\pi\,H_{w_1^-, w_1^+, w_3^-}(r) - 
    2\,i\pi\,H_{w_1^-, w_1^+, w_2^-}(r) + 
    \frac{1}{2}\,i\pi\,H_{w_1^-, w_3^-, w_1^+}(r) \nnb \\&&- 
    \frac{1}{2}\,i\pi\,H_{w_1^-, w_3^-, w_1^+}(s) - 
    \frac{1}{2}\,i\pi\,H_{w_1^-, w_3^+, w_1^-}(r) - 
    \frac{1}{2}\,i\pi\,H_{w_1^-, w_3^+, w_1^-}(s) - 
    H_{w_1^-}(r)\,H_{w_1^-, w_3^+, w_1^+}(s) \nnb \\&&- 
    2\,i\pi\,H_{w_1^-, w_2^-, w_1^+}(r) - 4\,i\pi\,H_{w_1^+, 0, w_1^-}(r) - 
    2\,i\pi\,H_{w_1^+, 0, w_1^-}(s) - 2\,H_{w_1^-}(r)\,
     H_{w_1^+, 0, w_1^+}(s) \nnb \\&&- 2\,i\pi\,H_{w_1^+, w_1^-, 0}(s) - 
    3\,i\pi\,H_{w_1^+, w_1^-, w_1^-}(s) + 
    2\,i\pi\,H_{w_1^+, w_1^-, w_5^-}(r) + 
    2\,i\pi\,H_{w_1^+, w_1^-, w_4^-}(r) \nnb \\&&- 3\,H_{w_1^-}(r)\,
     H_{w_1^+, w_1^-, w_1^+}(s) - i\pi\,H_{w_1^+, w_1^-, w_3^-}(r) - 
    2\,i\pi\,H_{w_1^+, w_1^-, w_2^-}(r) + 
    2\,i\pi\,H_{w_1^+, w_5^-, w_1^-}(s) \nnb \\&&+ 2\,H_{w_1^-}(r)\,
     H_{w_1^+, w_5^-, w_1^+}(s) - 
    2\,i\pi\,H_{w_1^+, w_5^+, w_1^+}(r) + 
    2\,i\pi\,H_{w_1^+, w_5^+, w_1^+}(s) +
    2\,i\pi\,H_{w_1^+, w_4^-, w_1^-}(s) \nnb \\&&+ 2\,H_{w_1^-}(r)\,
     H_{w_1^+, w_4^-, w_1^+}(s) - 
    2\,i\pi\,H_{w_1^+, w_4^+, w_1^+}(r) + 
    2\,i\pi\,H_{w_1^+, w_4^+, w_1^+}(s) \nnb \\&&- 2\,H_{w_1^-}(s)\,
     H_{w_1^+, w_1^+, w_1^-}(r) + 2\,H_{w_2^-}(s)\,
     H_{w_1^+, w_1^+, w_1^-}(r) + i\pi\,H_{w_1^+, w_1^+, w_1^+}(r) \nnb \\&&- 
    i\pi\,H_{w_1^+, w_1^+, w_1^+}(s) - \frac{1}{2}\,i\pi\,H_{w_1^+, w_3^-, w_1^-}(r)
    - \frac{1}{2}\,i\pi\,H_{w_1^+, w_3^-, w_1^-}(s) - 
    H_{w_1^-}(r)\,H_{w_1^+, w_3^-, w_1^+}(s) \nnb \\&&+ 
    \frac{1}{2}\,i\pi\,H_{w_1^+, w_3^+, w_1^+}(r) - 
    \frac{1}{2}\,i\pi\,H_{w_1^+, w_3^+, w_1^+}(s) - 
    2\,i\pi\,H_{w_1^+, w_2^-, w_1^-}(s) - 2\,H_{w_1^-}(r)\,
     H_{w_1^+, w_2^-, w_1^+}(s) \nnb \\&&- 2\,i\pi\,H_{w_2^-, w_1^-, w_1^+}(s) - 
    2\,i\pi\,H_{w_2^-, w_1^+, w_1^-}(s) - 2\,H_{w_1^-}(r)\,
     H_{w_2^-, w_1^+, w_1^+}(s) - 6\,H_{w_1^-, w_1^-, w_1^-, w_1^-}(r) \nnb \\&&+ 
    3\,H_{w_1^-, w_1^-, w_1^-, w_5^-}(r) + 
    3\,H_{w_1^-, w_1^-, w_1^-, w_4^-}(r) - 
    3\,H_{w_1^-, w_1^-, w_1^-, w_3^-}(r) +
    2\,H_{w_1^-, w_1^-, w_5^-, w_1^-}(r) \nnb \\&&- 
    4\,H_{w_1^-, w_1^-, w_5^+, w_1^+}(r) + 
    2\,H_{w_1^-, w_1^-, w_4^-, w_1^-}(r) - 
    4\,H_{w_1^-, w_1^-, w_4^+, w_1^+}(r) + 
    H_{w_1^-, w_1^-, w_1^+, w_5^+}(r) \nnb \\&&+
    H_{w_1^-, w_1^-, w_1^+, w_4^+}(r) + H_{w_1^-, w_1^-, w_1^+, w_1^+}(r) 
    + H_{w_1^-, w_1^-, w_1^+, w_1^+}(s) - 
    H_{w_1^-, w_1^-, w_1^+, w_3^+}(r) \nnb \\&&- 
    2\,H_{w_1^-, w_1^-, w_3^-, w_1^-}(r) + 
    2\,H_{w_1^-, w_1^-, w_3^+, w_1^+}(r) + 
    H_{w_1^-, w_5^-, w_1^-, w_1^-}(r) + 
    2\,H_{w_1^-, w_5^-, w_1^+, w_1^+}(r) \nnb \\&&-
    2\,H_{w_1^-, w_5^-, w_1^+, w_1^+}(s) -
    2\,H_{w_1^-, w_5^+, w_1^-, w_1^+}(s) - 
    2\,H_{w_1^-, w_5^+, w_1^+, w_1^-}(r) + 
    H_{w_1^-, w_4^-, w_1^-, w_1^-}(r) \nnb \\&&+
    2\,H_{w_1^-, w_4^-, w_1^+, w_1^+}(r) -
    2\,H_{w_1^-, w_4^-, w_1^+, w_1^+}(s) - 
    2\,H_{w_1^-, w_4^+, w_1^-, w_1^+}(s) - 
    2\,H_{w_1^-, w_4^+, w_1^+, w_1^-}(r) \nnb \\&&+
    2\,H_{w_1^-, w_1^+, 0, w_1^+}(s) + H_{w_1^-, w_1^+, w_1^-, w_5^+}(r) 
    + H_{w_1^-, w_1^+, w_1^-, w_4^+}(r) + 
    2\,H_{w_1^-, w_1^+, w_1^-, w_1^+}(r) \nnb \\&&+ H_{w_1^-, w_1^+, w_1^-, w_1^+}(s) 
    - H_{w_1^-, w_1^+, w_1^-, w_3^+}(r) - 
    2\,H_{w_1^-, w_1^+, w_5^-, w_1^+}(r) + 
    H_{w_1^-, w_1^+, w_5^+, w_1^-}(r) \nnb \\&&- 
    2\,H_{w_1^-, w_1^+, w_4^-, w_1^+}(r) + 
    H_{w_1^-, w_1^+, w_4^+, w_1^-}(r) + 
    2\,H_{w_1^-, w_1^+, w_1^+, w_1^-}(r) + H_{w_1^-, w_1^+, w_3^-, w_1^+}(r)\nnb \\&&
     - H_{w_1^-, w_1^+, w_3^+, w_1^-}(r) + 
    2\,H_{w_1^-, w_1^+, w_2^-, w_1^+}(r) - 
    H_{w_1^-, w_3^-, w_1^-, w_1^-}(r) - 
    \frac{1}{2}\,H_{w_1^-, w_3^-, w_1^+, w_1^+}(r) \nnb \\&&+ 
    \frac{1}{2}\,H_{w_1^-, w_3^-, w_1^+, w_1^+}(s) + 
    \frac{1}{2}\,H_{w_1^-, w_3^+, w_1^-, w_1^+}(r) + 
    \frac{1}{2}\,H_{w_1^-, w_3^+, w_1^-, w_1^+}(s) + 
    H_{w_1^-, w_3^+, w_1^+, w_1^-}(r) \nnb \\&&+ 
    2\,H_{w_1^-, w_2^-, w_1^+, w_1^+}(r) + 
    2\,H_{w_1^+, 0, w_1^-, w_1^+}(s) + 2\,H_{w_1^+, 0, w_1^+, w_1^-}(r) + 
    2\,H_{w_1^+, w_1^-, 0, w_1^+}(s) \nnb \\&&+ H_{w_1^+, w_1^-, w_1^-, w_5^+}(r) 
    + H_{w_1^+, w_1^-, w_1^-, w_4^+}(r) + 
    H_{w_1^+, w_1^-, w_1^-, w_1^+}(r) + 
    3\,H_{w_1^+, w_1^-, w_1^-, w_1^+}(s) \nnb \\&&- H_{w_1^+, w_1^-, w_1^-, w_3^+}(r) 
    - 2\,H_{w_1^+, w_1^-, w_5^-, w_1^+}(r) + 
    H_{w_1^+, w_1^-, w_5^+, w_1^-}(r) - 
    2\,H_{w_1^+, w_1^-, w_4^-, w_1^+}(r) \nnb \\&&+ 
    H_{w_1^+, w_1^-, w_4^+, w_1^-}(r) + 
    3\,H_{w_1^+, w_1^-, w_1^+, w_1^-}(r) - 
    2\,H_{w_1^+, w_1^-, w_1^+, w_2^+}(r) + H_{w_1^+, w_1^-, w_3^-, w_1^+}(r)\nnb \\&&
    - H_{w_1^+, w_1^-, w_3^+, w_1^-}(r) + 
    2\,H_{w_1^+, w_1^-, w_2^-, w_1^+}(r) - 
    2\,H_{w_1^+, w_5^-, w_1^-, w_1^+}(s) - 
    2\,H_{w_1^+, w_5^-, w_1^+, w_1^-}(r) \nnb \\&&+ 
    H_{w_1^+, w_5^+, w_1^-, w_1^-}(r) + 
    2\,H_{w_1^+, w_5^+, w_1^+, w_1^+}(r) - 
    2\,H_{w_1^+, w_5^+, w_1^+, w_1^+}(s) - 
    2\,H_{w_1^+, w_4^-, w_1^-, w_1^+}(s) \nnb \\&&- 
    2\,H_{w_1^+, w_4^-, w_1^+, w_1^-}(r) + 
    H_{w_1^+, w_4^+, w_1^-, w_1^-}(r) + 
    2\,H_{w_1^+, w_4^+, w_1^+, w_1^+}(r) - 
    2\,H_{w_1^+, w_4^+, w_1^+, w_1^+}(s) \nnb \\&&+ 
    2\,H_{w_1^+, w_1^+, w_1^-, w_1^-}(r) - 
    2\,H_{w_1^+, w_1^+, w_1^-, w_2^-}(r) - 
    4\,H_{w_1^+, w_1^+, w_1^-, w_2^+}(r) - H_{w_1^+, w_1^+, w_1^+, w_1^+}(r) \nnb \\&&+ 
    H_{w_1^+, w_1^+, w_1^+, w_1^+}(s) - 2\,H_{w_1^+, w_1^+, w_2^-, w_1^-}(r)
    - 4\,H_{w_1^+, w_1^+, w_2^+, w_1^-}(r) + 
    \frac{1}{2}\,H_{w_1^+, w_3^-, w_1^-, w_1^+}(r) \nnb \\&&+ 
    \frac{1}{2}\,H_{w_1^+, w_3^-, w_1^-, w_1^+}(s) + 
    H_{w_1^+, w_3^-, w_1^+, w_1^-}(r) - 
    H_{w_1^+, w_3^+, w_1^-, w_1^-}(r) - 
    \frac{1}{2}\,H_{w_1^+, w_3^+, w_1^+, w_1^+}(r) \nnb \\&&+ 
    \frac{1}{2}\,H_{w_1^+, w_3^+, w_1^+, w_1^+}(s) + 
    2\,H_{w_1^+, w_2^-, w_1^-, w_1^+}(s) - 
    2\,H_{w_1^+, w_2^+, w_1^+, w_1^-}(r) + 
    2\,H_{w_2^-, w_1^-, w_1^+, w_1^+}(s) \nnb \\&&+ 
    2\,H_{w_2^-, w_1^+, w_1^-, w_1^+}(s) - \frac{3}{2}\,\pi^2\,H_{w_1^-}(r)\,\ln(2)
    + \frac{3}{2}\,\pi^2\,H_{w_1^-}(s)\,\ln(2) \nnb \\&&+ 2\,i\pi\,H_{-1}(r^2)\,
     H_{w_1^+}(r)\,\ln(2) - 2\,i\pi\,H_{-1}(r^2)\,H_{w_1^+}(s)\,\ln(2) - 
    2\,i\pi\,H_{w_1^-}(r)\,H_{w_1^+}(s)\,\ln(2) \nnb \\&&- 
    8\,H_{w_1^-}(r)\,H_{0, w_1^+}(1-2\,\sqrt{\zc})\,\ln(2) + 
    8\,H_{w_1^-}(s)\,H_{0, w_1^+}(1-2\,\sqrt{\zc})\,\ln(2) \nnb \\&&+ 
    8\,H_{w_1^-}(r)\,H_{0, w_1^+}\left(\textstyle\frac{1}{1+2\,\sqrt{\zc}}\displaystyle\right)\,\ln(2) - 
    8\,H_{w_1^-}(s)\,H_{0, w_1^+}\left(\textstyle\frac{1}{1+2\,\sqrt{\zc}}\displaystyle\right)\,\ln(2) \nnb \\&&+ 
    2\,H_{w_1^-}(r)\,H_{0, w_1^+}(1-2\,\zc)\,\ln(2) - 
    2\,H_{w_1^-}(s)\,H_{0, w_1^+}(1-2\,\zc)\,\ln(2) \nnb \\&&+ 
    2\,H_{w_1^-}(r)\,H_{w_1^-, w_1^-}(s)\,\ln(2) - 
    2\,H_{w_1^-}(r)\,H_{w_1^-, w_5^-}(s)\,\ln(2) - 
    2\,H_{w_1^-}(r)\,H_{w_1^-, w_4^-}(s)\,\ln(2) \nnb \\&&+ 
    2\,i\pi\,H_{w_1^-, w_1^+}(s)\,\ln(2) + 2\,H_{w_1^-}(r)\,
     H_{w_1^-, w_3^-}(s)\,\ln(2) - i\pi\,H_{w_1^-, w_3^+}(r)\,
     \ln(2) \nnb \\&&+ i\pi\,H_{w_1^-, w_3^+}(s)\,\ln(2) + 
    2\,i\pi\,H_{w_1^+, w_1^-}(r)\,\ln(2) - 2\,H_{w_1^-}(r)\,
     H_{w_1^+, w_5^+}(s)\,\ln(2) \nnb \\&&- 2\,H_{w_1^-}(r)\,
     H_{w_1^+, w_4^+}(s)\,\ln(2) - 4\,H_{w_1^-}(s)\,
     H_{w_1^+, w_1^+}(r)\,\ln(2) + 4\,H_{w_2^-}(s)\,H_{w_1^+, w_1^+}(r)\,
     \ln(2) \nnb \\&&+ 4\,H_{w_1^-}(r)\,H_{w_1^+, w_1^+}(s)\,\ln(2) - 
    i\pi\,H_{w_1^+, w_3^-}(r)\,\ln(2) + i\pi\,H_{w_1^+, w_3^-}(s)\,
     \ln(2) \nnb \\&&+ 2\,H_{w_1^-}(r)\,H_{w_1^+, w_3^+}(s)\,\ln(2) + 
    4\,H_{w_1^+}(r)\,H_{w_1^+, w_2^+}(s)\,\ln(2) - 
    6\,H_{w_1^-, w_1^-, w_1^-}(r)\,\ln(2) \nnb \\&&+ 
    4\,H_{w_1^-, w_1^-, w_5^-}(r)\,\ln(2) + 
    4\,H_{w_1^-, w_1^-, w_4^-}(r)\,\ln(2) - 
    4\,H_{w_1^-, w_1^-, w_3^-}(r)\,\ln(2) \nnb \\&&+ 
    2\,H_{w_1^-, w_5^-, w_1^-}(r)\,\ln(2) - 
    4\,H_{w_1^-, w_5^+, w_1^+}(r)\,\ln(2) + 
    4\,H_{w_1^-, w_5^+, w_1^+}(s)\,\ln(2) \nnb \\&&+ 
    2\,H_{w_1^-, w_4^-, w_1^-}(r)\,\ln(2) - 
    4\,H_{w_1^-, w_4^+, w_1^+}(r)\,\ln(2) + 
    4\,H_{w_1^-, w_4^+, w_1^+}(s)\,\ln(2) \nnb \\&&+ 
    2\,H_{w_1^-, w_1^+, w_5^+}(r)\,\ln(2) + 
    2\,H_{w_1^-, w_1^+, w_4^+}(r)\,\ln(2) + 4\,H_{w_1^-, w_1^+, w_1^+}(r)\,
     \ln(2) \nnb \\&&- 4\,H_{w_1^-, w_1^+, w_1^+}(s)\,\ln(2) - 
    2\,H_{w_1^-, w_1^+, w_3^+}(r)\,\ln(2) - 
    2\,H_{w_1^-, w_3^-, w_1^-}(r)\,\ln(2) \nnb \\&&+ 
    2\,H_{w_1^-, w_3^+, w_1^+}(r)\,\ln(2) - 
    2\,H_{w_1^-, w_3^+, w_1^+}(s)\,\ln(2) + 4\,H_{w_1^+, 0, w_1^+}(r)\,
     \ln(2) \nnb \\&&- 4\,H_{w_1^+, 0, w_1^+}(s)\,\ln(2) + 
    2\,H_{w_1^+, w_1^-, w_5^+}(r)\,\ln(2) + 
    2\,H_{w_1^+, w_1^-, w_4^+}(r)\,\ln(2)\nnb \\&& + 6\,H_{w_1^+, w_1^-, w_1^+}(r)\,
     \ln(2) - 6\,H_{w_1^+, w_1^-, w_1^+}(s)\,\ln(2) - 
    2\,H_{w_1^+, w_1^-, w_3^+}(r)\,\ln(2) \nnb \\&&- 
    4\,H_{w_1^+, w_5^-, w_1^+}(r)\,\ln(2) + 
    4\,H_{w_1^+, w_5^-, w_1^+}(s)\,\ln(2) + 
    2\,H_{w_1^+, w_5^+, w_1^-}(r)\,\ln(2) \nnb \\&&- 
    4\,H_{w_1^+, w_4^-, w_1^+}(r)\,\ln(2) + 
    4\,H_{w_1^+, w_4^-, w_1^+}(s)\,\ln(2) + 
    2\,H_{w_1^+, w_4^+, w_1^-}(r)\,\ln(2) \nnb \\&&- 
    4\,H_{w_1^+, w_1^+, w_2^-}(r)\,\ln(2) - 8\,H_{w_1^+, w_1^+, w_2^+}(r)\,
     \ln(2) + 2\,H_{w_1^+, w_3^-, w_1^+}(r)\,\ln(2) \nnb \\&&- 
    2\,H_{w_1^+, w_3^-, w_1^+}(s)\,\ln(2) - 
    2\,H_{w_1^+, w_3^+, w_1^-}(r)\,\ln(2) - 4\,H_{w_1^+, w_2^-, w_1^+}(s)\,
     \ln(2) \nnb \\&&- 4\,H_{w_1^+, w_2^+, w_1^+}(r)\,\ln(2) - 
    4\,H_{w_2^-, w_1^+, w_1^+}(s)\,\ln(2) - 2\,H_{w_1^-, w_1^-}(r)\,
     \ln^2(2) \nnb \\&&+ 2\,H_{w_1^-, w_1^-}(s)\,\ln^2(2) + 
    2\,H_{w_1^-, w_5^-}(r)\,\ln^2(2) - 2\,H_{w_1^-, w_5^-}(s)\,
     \ln^2(2) + 2\,H_{w_1^-, w_4^-}(r)\,\ln^2(2) \nnb \\&&- 
    2\,H_{w_1^-, w_4^-}(s)\,\ln^2(2) - 2\,H_{w_1^-, w_3^-}(r)\,
     \ln^2(2) + 2\,H_{w_1^-, w_3^-}(s)\,\ln^2(2) + 
    2\,H_{w_1^+, w_5^+}(r)\,\ln^2(2) \nnb \\&&- 2\,H_{w_1^+, w_5^+}(s)\,
     \ln^2(2) + 2\,H_{w_1^+, w_4^+}(r)\,\ln^2(2) - 
    2\,H_{w_1^+, w_4^+}(s)\,\ln^2(2) - 4\,H_{w_1^+, w_1^+}(r)\,\ln^2(2) \nnb \\&&+ 
    4\,H_{w_1^+, w_1^+}(s)\,\ln^2(2) - 2\,H_{w_1^+, w_3^+}(r)\,\ln^2(2) + 
    2\,H_{w_1^+, w_3^+}(s)\,\ln^2(2) - H_{w_1^-, w_1^-}(r)\,
     {\rm Li}_2(1-\zc) \nnb \\&&+ H_{w_1^-, w_1^-}(s)\,{\rm Li}_2(1-\zc) + 
    H_{w_1^-, w_5^-}(r)\,{\rm Li}_2(1-\zc) - 
    H_{w_1^-, w_5^-}(s)\,{\rm Li}_2(1-\zc) \nnb \\&&+ 
    H_{w_1^-, w_4^-}(r)\,{\rm Li}_2(1-\zc) - 
    H_{w_1^-, w_4^-}(s)\,{\rm Li}_2(1-\zc) - 
    H_{w_1^-, w_3^-}(r)\,{\rm Li}_2(1-\zc) \nnb \\&&+ 
    H_{w_1^-, w_3^-}(s)\,{\rm Li}_2(1-\zc) + 
    H_{w_1^+, w_5^+}(r)\,{\rm Li}_2(1-\zc) - 
    H_{w_1^+, w_5^+}(s)\,{\rm Li}_2(1-\zc) \nnb \\&&+ 
    H_{w_1^+, w_4^+}(r)\,{\rm Li}_2(1-\zc) - 
    H_{w_1^+, w_4^+}(s)\,{\rm Li}_2(1-\zc) + 
    H_{w_1^+, w_1^+}(r)\,{\rm Li}_2(1-\zc) \nnb \\&&- H_{w_1^+, w_1^+}(s)\,
     {\rm Li}_2(1-\zc) - H_{w_1^+, w_3^+}(r)\,{\rm Li}_2(1-\zc) + 
    H_{w_1^+, w_3^+}(s)\,{\rm Li}_2(1-\zc) \nnb \\&&+ 7\,H_{w_1^-}(r)\,\zeta_3 - 
    7\,H_{w_1^-}(s)\,\zeta_3 \; .
\end{eqnarray}
}

\newpage
\bibliographystyle{JHEP}

\providecommand{\href}[2]{#2}\begingroup\raggedright\endgroup

\end{document}